\documentclass[aps, prd, amsmath, floats, floatfix, twocolumn, nofootinbib, superscriptaddress, showpacs]{revtex4-1}
\usepackage{graphicx}
\usepackage{color}
\usepackage{latexsym}
\usepackage{amsfonts}
\usepackage{multirow}
\usepackage{enumerate}

\newcommand{\be}{\begin{eqnarray}}
\newcommand{\ee}{\end{eqnarray}}

\usepackage{longtable}
\usepackage[table]{xcolor}
\usepackage{hyperref}
\hypersetup{
	colorlinks=true, 
	linkcolor=blue,  
	citecolor=cyan,  
}

\begin{document}

\title{Rotating and non-linear magnetic-charged black hole surrounded by quintessence}

\author{Carlos~A.~Benavides-Gallego}
\email{abgcarlos17@fudan.edu.cn}
\affiliation{Center for Field Theory and Particle Physics and Department of Physics, Fudan University, 200438 Shanghai, China}

\author{Ahmadjon Abdujabbarov}
\email{ahmadjon@astrin.uz}
\affiliation{Shanghai Astronomical Observatory, 80 Nandan Road, Shanghai 
	200030, P. R. China}
\affiliation{Ulugh Beg Astronomical Institute, Astronomicheskaya 33, 
	Tashkent 100052, Uzbekistan}
\affiliation{National University of Uzbekistan, Tashkent 100174, 
	Uzbekistan}
\affiliation{Tashkent Institute of Irrigation and Agricultural Mechanization Engineers, Kori Niyoziy, 39, Tashkent 100000, Uzbekistan}

\author{Cosimo~Bambi}
\email[Corresponding author: ]{bambi@fudan.edu.cn}
\affiliation{Center for Field Theory and Particle Physics and Department of Physics, Fudan University, 200438 Shanghai, China}

\date{\today}

\begin{abstract}
\noindent In this work we derived a rotating and non-linear magnetic-charged black hole surrounded by quintessence using the Newman-Janis algorithm. Considering the state parameter $\omega_q=-3/2$, we studied the event horizons, the ergosphere, and the ZAMO. We found that the existence of the outer horizon is constrained by the values of the charge $Q$. Furthermore, we found that the ergo-region increases when both the charge $Q$ and the spin parameter $a$ are increased. On the other hand, regarding  equatorial circular orbits, we studied the limit given by the static radius on the existence of circular geodesics, the photon circular geodesics, and the innermost stable circular orbits (ISCO). We show that photon circular orbits do not depend strongly on $Q$, and $r_{ISCO}$ is constrained by the values of charge. 
\end{abstract}

\maketitle

\section{Introduction \label{seccionI}}

Nowadays, the Standard Model of Cosmology (SMC) is the best theory available for the description of the Universe. It is based on two fundamental ingredients: the standard model of particle physics, and the general theory of relativity \cite{Bambi16a}. The success of the SMC lies on the fact that it agrees with three observational pillars in cosmology: the Hubble's law, the primordial abundances of light elements (Big Bang nucleosynthesis), and the cosmic microwave background: \textit{``the black body radiation left over from the first few hundred thousand years''} \cite{Dodelson03}. Nevertheless, the SMC faces some problems if it is based only on these two fundamental ingredients \cite{Bambi16a}:
\begin{enumerate} [(i)]
\item According to the small angular anisotropies detected in the CMB \cite{Planck14}, different parts of the sky were causally connected at the time of the last scattering. Nevertheless, this is not what one would expect from the Friedmann equations if we consider a universe dominated by matter or radiation. This is known as the horizon problem. 

\item In the case of usual matter, Friedmann equations show that a strong departure from flatness is developed by time. Nevertheless, observations indicate that the universe is close to the geometrically flat case. As a result, the universe must be extremely fine-tuned to a flat one at the very beginning creating the so-called flatness problem.  
\item A mechanism to generate primordial inhomogeneities at cosmological scales is needed. These inhomogeneities are the seeds of future galaxies. 
\item Finally, beyond the framework of the standard model of particle physics, the formation of dangerous relics are expected at early times. These relics, in the form of heavy particles or topological defects, would force the universe to recollapse too soon.
\end{enumerate}

Inflation is one way to solve the problems of the horizon, flatness and dangerous relics. This idea was suggested by Kazanas, Starobinsky, and  Guth~\cite{Kazanas80,Starobinsky80,Guth81,Guth82}. In this scenario, if the universe is supercooled to temperatures below the critical temperature for some phase transition in its early history, then a huge expansion factor would result from a period of exponential growth. This paradigm, according to Guth, is completely natural in the context of grand unified models of elementary particle interactions \cite{Guth81}. However, since the concrete mechanism of inflation suggested by Guth was not realistic, it became necessary to suggest a new inflationary model. In this sense, by introducing a dynamical inflation field, responsible for the exponential expansion of the universe, Linde and, independently, Albrecht and Steinhardt were able to solve this problem \cite{Linde82, Albrecht82}. Now many different scenarios of inflation are worked out and there is common agreement that a sufficiently long period of inflation could solve all four problems mentioned above \cite{Bambi16a}.

On the other hand, observations suggest that only $5\%$ of the Universe is made of protons and neutrons, while $25\%$  seems to be made of some weakly interactive particles known as dark matter. The remaining $70\%$ seems to be a uniformly distributed hypothetical fluid with an unusual equation of state $p\approx -\rho$ (where $p$ is the pressure and $\rho$ is the energy density). Scientists believe, based on the observation of the luminosity distances of a type-I supernovae \cite{Riess98,Perlmutter99} that the current acceleration is an effect of this hypothetical fluid \cite{Yang18}. This substance is known as dark energy and recent models suggest two possibilities for it: the cosmological constant \cite{Krauss95,Weinberg89,Peebles03,Padmanabhan03}, and the quintessence \cite{Jassal05,Jassal05a,Samushia06,Xia08,Zhao07,Xia06}.

The quintessence is a scalar field with the state parameter $\omega_q$ in the range $-1\leq\omega_q\leq -1/3$, where the case $\omega_q=-1$ covers the cosmological constant term \cite{Kiselev03a}. In the case of black holes, for example, the quintessence perturbs the black hole background metric and also modifies the curves of space-time around the singularity. Moreover, cosmological horizons exist in the space-time of black holes or other compact objects immersed in this kind of scalar field. The quintessence scenario can be realized in a large variety of models: scalar field in modified $f(R)$ gravity, string-like cosmology, k-essence, extra dimensions, braneworld models \cite{Toshmatov17}. 

Interesting work considering the quintessence scenario has been done. In reference \cite{Kiselev03a}, for example, the general solutions to the spherically symmetric Einstein equations describing the Schwarzschild black hole surrounded by  dark energy, in the form of a quintessential field, was found by Kiselev. Using this approach he was able to explain the asymptotic behavior of the rotation curves in spiral galaxies. Furthermore, in reference \cite{Kiselev03}, the same author presents a new static spherically symmetric exact solutions of the Einstein field equations for quintessential matter surrounding a black hole (charged or uncharged) and discusses the case in which $w_q= -2/3$.

On the other hand, and motivated by the importance of the geodesic structure in a space-time surrounded by quintessence, a detailed study of the photon trajectories was done in reference \cite{Fernando15}. In~ \cite{Fernando15} the authors obtained an exact solution for the trajectories in terms of the Jacobi elliptic integrals. A study of the motion and collision of particles in the gravitational field of a rotating black hole immersed in quintessential dark energy was done in reference \cite{Oteev16}. The authors of Ref.~\cite{Oteev16} focus on the acceleration of particles due to collisional processes and show how the centre of mass energy depends on the quintessential field parameter $C$. They also study the dependence of the maximal value of the efficiency of energy extraction through the Penrose process for rotating black holes with quintessential field parameter $C$. The authors found that the quintessence field decreases the energy extraction efficiency through the Penrose process and when the parameter $C$ vanishes it is possible to obtain the standard value of the efficiency coefficient for the Kerr black hole.

In reference \cite{Abdujabbarov17b}, the shadow of a rotating black hole with quintessential energy was studied in vacuum, and in the presence of plasma. In~\cite{Abdujabbarov17b} the authors found that, in the vacuum case, the quintessential parameter changes the shape of the shadow considerably. On the other hand, in the case of a quintessence black hole surrounded by plasma, the shape of the shadow depends on the plasma parameters, black hole spin, and the quintessential field parameter. However, for larger values of the quintessential field parameter, the shape of the shadow does not change significantly. 

In this paper, we study a rotating and non-linear magnetic-charged black hole surrounded by quintessence. The manuscript is organized as follow. In section \ref{sectionII}, we review the geometry of quintessential rotating black holes discussed in reference \cite{Toshmatov17}. Then, in section \ref{sectionIII}, we discuss the non-linear magnetic-charged black hole solution surrounded by quintessence. In section \ref{sectionIV}, we obtain the rotating solution of a non-linear magnetic-charger black hole surrounded by quintessence, the Einstein and energy-momentum tensors, and the equation of state. In section \ref{sectionV} we consider the case $w_q=-2/3$ to study the event horizon, the ergosphere, and the angular velocity of a zero angular momentum observer (ZAMO). Finally, in section \ref{sectionVI}, we study the equatorial circular orbits, the static radius, the photon circular orbits, and the innermost stable circular orbits (ISCO).

Along the manuscript geometrized units ($G=1$ and $c=1$) and the signature $(-+++)$ were used. Moreover, in section~\ref{sectionV}, we set the mass of the black hole $M=1$ for all the figures and calculations.  

\section{Quintessential Rotating Black Hole solutions \label{sectionII}}
The Newman-Janis algorithm (NJA) is used to obtain rotating space-times from a spherically symmetric background without solving Einstein's field equations. This mathematical idea was first proposed by E.~T. Newman and A.~I. Janis in 1965~\cite{Newman65b}. Using this ``\textit{curious derivation}'', they were able to obtain Kerr's metric by performing a complex coordinate transformation on the Schwarzschild line element \cite{Newman65b}. The key point behind the NJA lies on the fact that the contravariant tensor $g^{\mu\nu}$, expressed in Eddington-Finkelstein (EF) coordinates, can be defined in terms of a null tetrad.

The NJA has been widely used in the scientific community (see, e.g.~\cite{Broccoli18,Erbin17,Gonzalez15,Erbin15,Keane14,Bambi13,Lombardo04} and references therein). Nevertheless, in order to obtain the general equations to find quintessential rotating black hole solutions, the authors of Ref.~\cite{Toshmatov17} used a modified version obtained in~\cite{Azreg14}. As the authors explain, the only difference from the original NJA is that they skipped the complexification of coordinates~\cite{Azreg14,Azreg-Ainou14,Azreg-Ainou11,Toshmatov14}.

In this section, we review the modified algorithm used in~\cite{Toshmatov17,Azreg14} to obtain the rotating solution of a black hole with quintessence energy. This algorithm can be divided into three important steps. First, it is important to transform the spherically symmetric line element from the Boyer-Lindquist (BL) to Eddington-Finkelstein (EF) coordinates. Then, since the contravariant components of the metric tensor can be expressed by a null tetrad, the next step is to perform a complex coordinate transformation. This transformation enables us to obtain the form of the contravariant tensor $g^{\mu\nu}$ in terms of the tetrad. Finally, once the contravariant components of the metric are obtained, we turn back from EF to BL coordinates. This transformation involves two functions  of $r$: $\lambda(r)$ and $\chi(r)$. These functions can be found under the condition that all non-diagonal components of the metric, with the exception of $g_{t\phi}$, vanish.

A spherically symmetric space-time, expressed in Boyer-Lindquist (BL) coordinates $(t,r,\theta,\phi)$, can be described by the line element \cite{Toshmatov17} 
\begin{equation}
\label{spherically}
ds^2=-f(r)dt^2+g^{-1}(r)dr^2+h(r)d\Omega^2,
\end{equation}
where 
\begin{equation}
d\Omega^2=d\theta^2+\sin^2\theta d\phi^2.
\end{equation}
Hence, after using the coordinate transformation
\begin{equation}
\label{BL_to_EF}
du=dt-\frac{dr}{\sqrt{fg}},
\end{equation} 
the line element in equation (\ref{spherically}) takes the form
\begin{equation}
\label{spherically_EF}
ds^2=-f(r)du^2-2\sqrt{\frac{f(r)}{g(r)}}dudr+h(r)d\theta^2+h(r)\sin^2\theta d\phi^2.
\end{equation}
\\
The last expression is just the line element (\ref{spherically}) expressed in EF coordinates $(u,r,\theta,\phi)$.\\ 

As mentioned above, the contravariant tensor $g^{\mu\nu}$, obtained from the line element in equation (\ref{spherically_EF}), can be expressed in terms of the null tetrad by the following relations \cite{Toshmatov17}
\begin{equation}
\label{tetrad_metric}
g^{\mu\nu}=-l^\mu n^\nu-l^\nu n^\mu+m^\mu\overline{m}^\nu+m^\nu\overline{m}^\mu\ ,
\end{equation}
where
\begin{equation}
\label{definition_tetrad}
\begin{aligned}
l^\mu&=\delta^\mu_r\ ,\\
n^\mu&=\sqrt{\frac{g(r)}{f(r)}}\delta^\mu_\nu-\frac{f(r)}{2}\delta^\mu_\nu\ ,\\
m^\mu&=\frac{1}{\sqrt{2h(r)}}\delta^\mu_\theta+\frac{i}{\sqrt{2h(r)}\sin\theta}\delta^\mu_\phi\ ,\\
\overline{m}^\mu&=\frac{1}{\sqrt{2h(r)}}\delta^\mu_\theta-\frac{i}{\sqrt{2h(r)}\sin\theta}\delta^\mu_\phi.\\
\end{aligned}
\end{equation}
\\
Note that the vectors $l^\mu$ and $n^\mu$ are real, while $m^\mu$ is a complex vector. Furthermore, we use the notation in which $\overline{m}^\mu$ is the complex conjugate of $m^\mu$. These vectors satisfy the orthogonality and isotropic conditions \cite{Newman65b, Toshmatov17}
\begin{equation}
\label{orthogonality}
l^\mu m_\mu=l^\mu\overline{m}_\mu=n^\mu m_\mu=n^\mu\overline{m}_\mu=0,
\end{equation}
and
\begin{equation}
\label{isotropic}
l^\mu l_\mu=n^\nu n_\mu=m^\mu m_\nu=\overline{m}^\mu \overline{m}_\nu=0\ ,
\end{equation}
respectively.\\

Now, from equation (\ref{tetrad_metric}), it is possible to consider a complex transformation, not of the line element (\ref{spherically_EF}), but of the tetrad in equation (\ref{definition_tetrad}). Thus, after using the transformation \cite{Toshmatov17} 
\begin{equation}
\label{complex_transformation}
\begin{array}{cc}
u\rightarrow u-ia\cos\theta,&r\rightarrow r+ia\cos\theta,
\end{array}
\end{equation} 
the tetrad in equation (\ref{definition_tetrad}) takes the form 
\begin{equation}
\label{tetrad_complex}
\begin{aligned}
l^\mu&=\delta^\mu_r\ ,\\
n^\mu&=\sqrt{\frac{G}{F}}\delta^\mu_u-\frac{1}{2}F\delta^\mu_r\ ,\\
m^\mu&=\frac{1}{\sqrt{2\Sigma}}\left[\delta^\mu_\theta+ia(\delta^\mu_u-\delta^\mu_r)\sin\theta+\frac{i}{\sin\theta}\delta^\mu_\phi\right]\ ,\\
\overline{m}^\mu&=\frac{1}{\sqrt{2\Sigma}}\left[\delta^\mu_\theta-ia(\delta^\mu_u-\delta^\mu_r)\sin\theta-\frac{i}{\sin\theta}\delta^\mu_\phi\right]\ .\\
\end{aligned}
\end{equation}
\\
Here it is assumed that the functions $f$, $g$, $h$ have changed to $F$, $G$, and $\Sigma$ respectively (see \cite{Toshmatov17} for details). Therefore, these functions depend on $r$, $a$, and $\theta$. Hence, with the help of equation (\ref{tetrad_metric}), the contravariant components of the metric tensor in the new coordinate system are
\begin{equation}
\begin{aligned}
\label{metric_componentsI}
g^{uu}&=\frac{a^2\sin^2\theta}{\Sigma},\\
g^{ur}&=-\sqrt{\frac{G}{F}}-\frac{a^2\sin^2\theta}{\Sigma},\\
g^{u\phi}&=\frac{a}{\Sigma},\\
g^{rr}&=G+\frac{a^2\sin^2\theta}{\Sigma},\\
g^{r\phi}&=-\frac{a}{\Sigma},\\
g^{\theta\theta}&=\frac{1}{\Sigma},\\
g^{\phi\phi}&=\frac{1}{\Sigma\sin^2\theta},\\
\end{aligned}
\end{equation}
\\
and the covariant components of the metric tensor, which can be calculated using the relation 
\begin{equation}
\label{inverse}
g^{\alpha\nu}g_{\alpha\mu}=\delta^\nu_\mu, 
\end{equation}
are
\begin{equation}
\begin{aligned}
g_{uu}&=-F,\\
g_{ur}&=-\sqrt{\frac{F}{G}},\\
g_{u\phi}&=a\left(F-\sqrt{\frac{F}{G}}\right)\sin^2\theta,\\
g_{r\phi}&=a\sqrt{\frac{F}{G}}\sin^2\theta,\\
g_{\theta\theta}&=\Sigma,\\
g_{\phi\phi}&=\sin^2\theta\left[\Sigma+a^2\left(2\sqrt{\frac{F}{G}}-F\right)\sin^2\theta\right].\\
\end{aligned}
\end{equation}
\\
Then, we come back from the EF coordinates to BL coordinates by using the following transformations~\cite{Azreg14} 
\begin{equation}
\begin{aligned}
du&=dt+\lambda(r)dr,\\
d\phi&=d\tilde{\phi}+\chi(r)dr,\\
\end{aligned}
\end{equation}
where the functions $\lambda(r)$ and $\chi(r)$ can be found using the fact in which all non-diagonal components of the metric tensor, except the coefficient $g_{t\phi}$ ($g_{\phi t}$), are equal to zero. Therefore, we obtain  that~\cite{Azreg14} 
\begin{equation}
\begin{aligned}
\lambda(r)&=-\frac{k(r)+a^2}{g(r)h(r)+a^2},\\
\chi(r)&=-\frac{a}{g(r)h(r)+a^2},\\
\end{aligned}
\end{equation}
with
\begin{equation}
k(r)=\sqrt{\frac{g(r)}{f(r)}}h(r)\ ,
\end{equation}
and
\begin{equation}
\begin{aligned}
F(r,\theta)&=\frac{(gh+a^2\cos^2\theta)\Sigma}{(k^2+a^2\cos^2\theta)^2}\ ,\\\\
G(r,\theta)&=\frac{gh+a^2\cos^2\theta}{\Sigma}.
\end{aligned}
\end{equation}
Finally, the rotating version of the spherically symmetric line element in equation (\ref{spherically}) is
\begin{equation}
\label{rotating_line_element}
\begin{aligned}
ds^2=&-\frac{(gh+a^2\cos^2\theta)\Sigma}{(k+a^2\cos^2\theta)^2}dt^2+\frac{\Sigma}{gh+a^2}dr^2\\
&-2a\sin^2\theta\left[\frac{k-gh}{(k+a^2\cos^2)^2}\right]\Sigma d\phi dt+\Sigma d\theta^2\\
&+\Sigma\sin^2\theta\left[1+a^2\sin^2\theta\frac{2k-gh+a^2\cos^2\theta}{(k+a^2\cos^2\theta)^2}\right]d\phi^2;
\end{aligned}
\end{equation}  
where we have removed the tilde on $\phi$.
\section{Non-linear magnetic charged black hole solution \label{sectionIII}}
The non-linear magnetic charged black hole surrounded by quintessence was obtained in reference \cite{Nam18} by considering Einstein gravity coupled to a non-linear electromagnetic field. It is described by the following equations
\begin{equation}
\label{Einstein_gravity_QE}
G^\nu_{\;\;\mu}=2\left[\frac{\partial\mathcal{L}(F)}{\partial F}F_{\mu\rho}F^{\nu\rho}-\frac{1}{4}\delta^\nu_\mu\mathcal{L}(F)+T^\nu_{\;\;\mu}(quintessence)\right],
\end{equation}
with
\begin{equation}
\begin{aligned}
\nabla_\mu\left(\frac{\partial\mathcal{L}(F)}{\partial F}F^{\mu\nu}\right)&=0,\\
\nabla_\mu{}^\ast F^{\nu\mu}&=0,
\end{aligned}
\end{equation}
and where $\mathcal{L}(F)$ is a function of the invariant $F_{\mu\nu}F^{\mu\nu}/4\equiv F $, and $F_{\mu\nu}$ is the electromagnetic tensor defined by
\begin{equation}
F_{\mu\nu}=\partial_\mu A_\nu-\partial_\nu A_\mu.
\end{equation}
Explicitly, the non-linear electrodynamic term is given by~\cite{Nam18} 
\begin{equation}
\label{L(F)}
\mathcal{L}(F)=\frac{3M}{|Q|^3}\frac{(2Q^2F)^{\frac{3}{2}}}{[1+(2Q^2F)^{\frac{3}{4}}]^2}.
\end{equation}
The solution, in BL coordinates, is given by the line element~\cite{Nam18} 
\begin{equation}
\label{Charged_QE}
ds^2=-f(r)dt^2+f(r)^{-1}dr^2+r^2d\theta^2+r^2\sin^2\theta d\phi^2\ ,
\end{equation}
with\footnote{It is important to point out the following: according to reference~\cite{Kiselev03}, $f(r)$ is defined as
\begin{equation*}
f(r)=1-\frac{2Mr^2}{r^3+Q^3}+\frac{C}{r^{3\omega_q+1}};
\end{equation*}
where $C<0$ in the case of quintessence. Nevertheless, since we use positive values of $C$, we express $f(r)$ as in Eq.~(\ref{f_QE}).}
\begin{equation}
\label{f_QE}
f(r)=1-\frac{2Mr^2}{r^3+Q^3}-\frac{C}{r^{3\omega_q+1}}=1-\frac{2\rho(r)}{r},
\end{equation}
where $2\rho(r)$ is expressed as in Eq.~(\ref{definitionsI}). Here $C$ is the positive normalization factor (from now on the quintessence parameter) and $\omega_q$ is the quintessential state parameter; which is constrained to the interval~\cite{Nam18}
\begin{equation}
-1<\omega_q<-\frac{1}{3}.
\end{equation}
Using the mathematica package RGTC~\footnote{http://www.inp.demokritos.gr/7Esbonano/RGTC/} we found, from the line element (\ref{Charged_QE}), the non-zero components of Einstein's tensor
\begin{equation}
\label{Non-rotating Einstein tensor}
\begin{aligned}
G^t_{\;\;t}=G^r_{\;\;r}=-\frac{2 \rho'}{r^2}&=\frac{3\omega_qC}{r^{3 (1+\omega_q)}}-\frac{6 M Q^3}{\left(Q^3+r^3\right)^2}\\
G^\theta_{\;\;\theta}=G^\phi_{\;\;\phi}=-\frac{\rho''}{r}&=-\frac{6 M Q^6}{\left(Q^3+r^3\right)^3}+\frac{12M r^3Q^3}{\left(Q^3+r^3\right)^3}\\
&-\frac{3}{2}   (3 \omega_q+1) \frac{C \omega_q}{r^{3 (\omega_q +1)}}\\
\end{aligned}
\end{equation}

Note that for $C=0$ the line element (\ref{Charged_QE}) reduces to the Hayward-Like black hole \cite{Hayward06}. For $Q=0$ the metric reduces to the Schwarzschild solution surrounded by quintessence. In this particular case, the energy-momentum tensor reduces to that considered by Kiselev~\cite{Kiselev03}
\begin{equation}
\label{EnergyTensor Kiselev}
\begin{aligned}
T^t_{\;t}=T^r_{\;r}&=\frac{3\omega_qC}{2r^{3(1+\omega_q)}}=-\rho_q\\\\
T^\theta_{\;\theta}=T^\phi_{\;\phi}&=\frac{(3\omega_q+1)}{2}\rho_q.
\end{aligned}
\end{equation}
Finally, for $C=0$, $Q=0$  Eq.~(\ref{Charged_QE}) reduces to the Schwarzschild black hole.\\

\section{Rotating and Non-linear magnetic charged black hole solution  \label{sectionIV}}
The rotating solution is obtained from Eq.~(\ref{rotating_line_element}). In our case, 
\begin{equation}
\begin{aligned}
g(r)&=f(r)=1-\frac{2Mr^2}{r^3+Q^3}-\frac{C}{r^{3\omega_q+1}}\ ,\\
h(r)&=k(r)=r^2.
\end{aligned}
\end{equation}
Thus, the line element (\ref{rotating_line_element}) takes the form
\begin{equation}
\label{Rotating_charged_QE}
\begin{aligned}
ds^2=&-\left[1-\frac{2\rho r}{\Sigma}\right]dt^2+\frac{\Sigma}{\Delta}dr^2-\frac{4a\rho r\sin^2\theta}{\Sigma}dt d\phi\\
&+\Sigma d\theta^2+\sin^2\theta\left[r^2+a^2+\frac{2a^2\rho r\sin^2\theta}{\Sigma}\right]d\phi^2\ ,
\end{aligned}
\end{equation} 
where 
\begin{equation}
\label{definitionsI}
\begin{aligned}
\Delta&=r^2-2\rho r+a^2\ ,\\
\Sigma&=r^2+a^2\cos^2\theta\ ,\\
2\rho&=\frac{2Mr^3}{Q^3+r^3}+Cr^{-3\omega_q}\ .
\end{aligned}
\end{equation}

\subsection{Equations of state\label{Energy tensor}}
Using the Mathematica package RGTC we found, from the line element (\ref{Rotating_charged_QE}), the non-zero components of the Einstein tensor $G_{\mu\nu}$ are given by 
\begin{equation}
\label{Rotating Einstein tensor}
\begin{aligned}
	G_{tt}&=\frac{2(r^4+a^2r^2-2r^3\rho-a^4\sin^2\theta\cos^2\theta)\rho'}{\Sigma^3}\\
	&-\frac{a^2r\sin^2\theta\rho''}{\Sigma^2},\\\\
	G_{t\phi}&=\frac{2a\sin^2\theta[(a^2+r^2)(a^2\cos^2\theta-r^2)+2r^3\rho]\rho'}{\Sigma^3}\\
	&+\frac{ar\sin^2\theta (a^2+r^2)\rho''}{\Sigma^2},\\\\
	G_{rr}&=-\frac{2r^2\rho'}{\Sigma\Delta},\\\\
	G_{\theta\theta}&=-\frac{2a^2\cos^2\theta\rho'}{\Sigma}-r\rho'',\\\\
	G_{\phi\phi}&=-\frac{a^2\sin^2\theta(a^2+r^2)(a^2+(a^2+2r^2)\cos2\theta)\rho'}{\Sigma^3}\\
	&-\frac{4a^2r^3\sin^4\theta\rho\rho'}{\Sigma^3}-\frac{r\sin^2\theta(a^2+r^2)^2\rho''}{\Sigma^2},
\end{aligned}
\end{equation}
where the prime ${}'$ represents a derivative with respect to $r$. 

In order to obtain the equations of state, we use the standard orthonormal basis~\cite{Misner73}
\begin{equation}
\label{tetrad}
\begin{aligned}
e^\mu_{(t)}&=\frac{1}{\sqrt{\Delta  \Sigma }}\left(a^2+r^2,0,0,a\right),\\
e^\mu_{(r)}&=\frac{\sqrt{\Delta}}{\sqrt{\Sigma}}\left(0,1,0,0\right),\\
e^\mu_{(\theta)}&=\frac{1}{\sqrt{\Sigma}}\left(0,0,1,0\right),\\
e^\mu_{(\phi)}&=-\frac{1}{\sqrt{\Sigma \sin^2\theta}}\left(a\sin^2\theta,0,0,1\right).\\
\end{aligned}
\end{equation}
Here we use the notation in which tetrad indices are enclosed in parentheses in order to distinguish from tensor indices (which are not enclosed)~\cite{Chandrasekhar98}. Hence, after using Eq.~(\ref{Einstein_gravity_QE}), we obtain the relations between the components of the stress-energy tensor and the Einstein tensor related to the orthonormal basis in the form
\begin{equation}
\label{stress-energy tensor}
\begin{aligned}
T_{(t)(t)}&=\frac{1}{2}e^\mu_{(t)}e^\nu_{(t)}G_{\mu\nu}=\epsilon\\
T_{(r)(r)}&=\frac{1}{2}e^\mu_{(r)}e^\nu_{(r)}G_{\mu\nu}=p_r\\
T_{(\theta)(\theta)}&=\frac{1}{2}e^\mu_{(\theta)}e^\nu_{(\theta)}G_{\mu\nu}=p_\theta\\
T_{(\phi)(\phi)}&=\frac{1}{2}e^\mu_{(\phi)}e^\nu_{(\phi)}G_{\mu\nu}=p_\phi,
\end{aligned}
\end{equation}
from which we obtain 
\begin{equation}
\label{equations of state}
\begin{aligned}
\epsilon=-p_r&=\frac{r^2\rho'}{\Sigma^2}\\
p_\theta=p_\phi&=\epsilon-\frac{2\rho'+r\rho''}{2\Sigma}.
\end{aligned}
\end{equation}
As expected, the components of the energy-momentum tensor in Eq.~(\ref{equations of state}) contains two contributions: a non-linear magnetic-charged part and the quintessence. This is due to the fact that $\rho$ is defined as in Eq.~(\ref{definitionsI}). Now, in the next subsection, we study the energy conditions for a rotating and non-linear magnetic-charged black hole surrounded by quintessence.

\subsection{Energy conditions}
It is known that all stress-tensors representing what is believed to be physically reasonable matter are diagonalizable~\cite{Wald84}. In this sense, using such  a diagonalizaition, the \textit{weak}, \textit{strong} and \textit{dominant} energy conditions are, respectively, 
\begin{equation*}
\begin{array}{ccc}
\epsilon\geq0&\text{and}&\epsilon+p_i\geq 0, \\
\end{array}
\end{equation*}

\begin{equation*}
\begin{array}{ccc}
\epsilon+\sum p_i\geq 0&\text{and}&\epsilon+p_i\geq 0
\end{array}
\end{equation*}
and,
\begin{equation}
\label{energy conditions}
\epsilon\geq |p_i|
\end{equation}
where $i=r, \theta, \phi$.
\begin{figure*}[t]
	a.\includegraphics[scale=0.6]{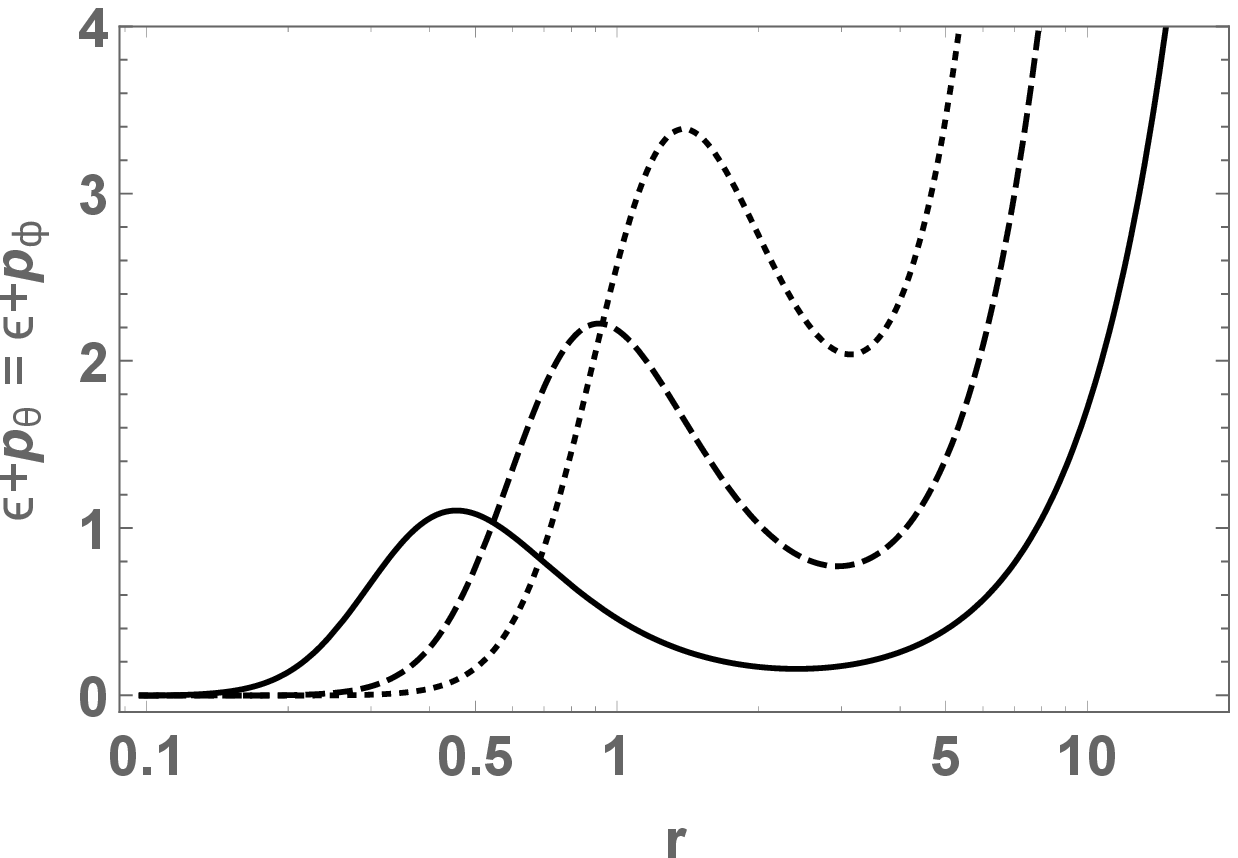}
	\hspace{1cm}
    \includegraphics[scale=0.64]{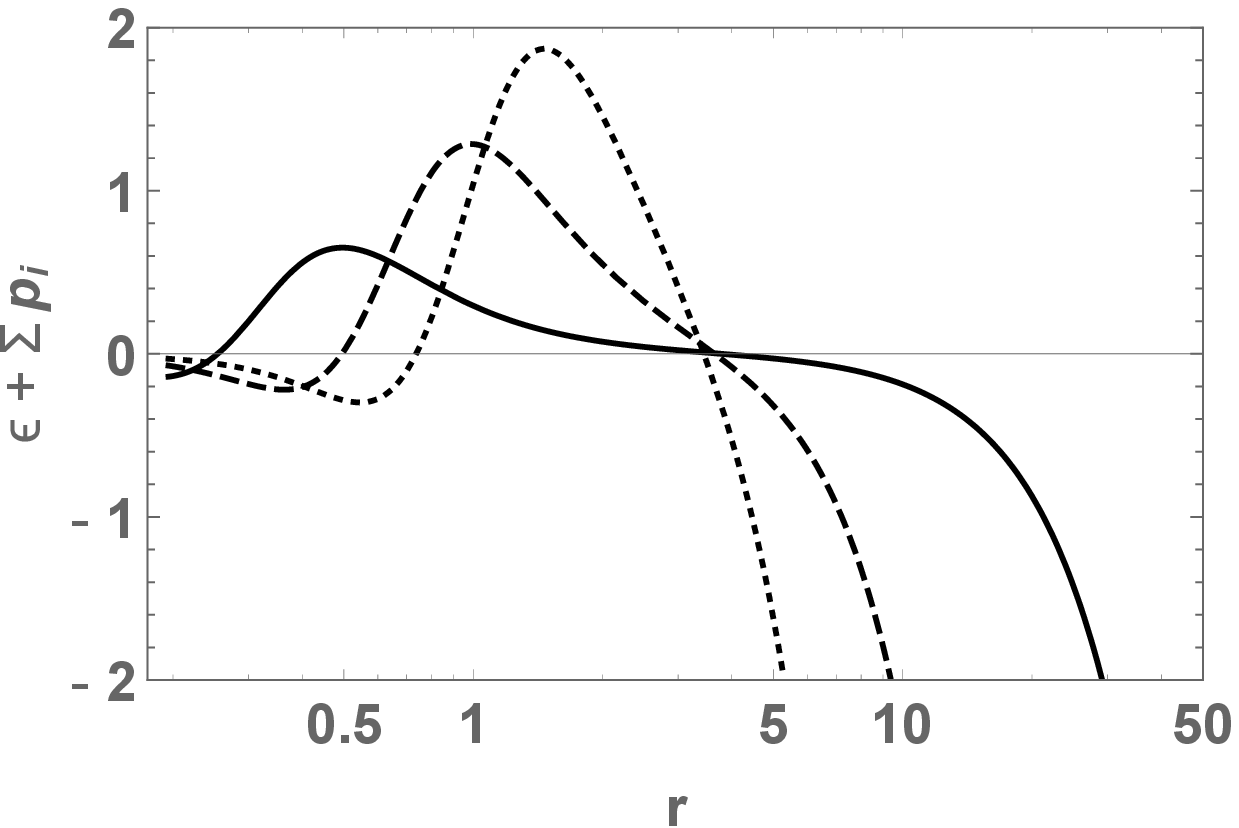}b.\\
	c.\includegraphics[scale=0.89]{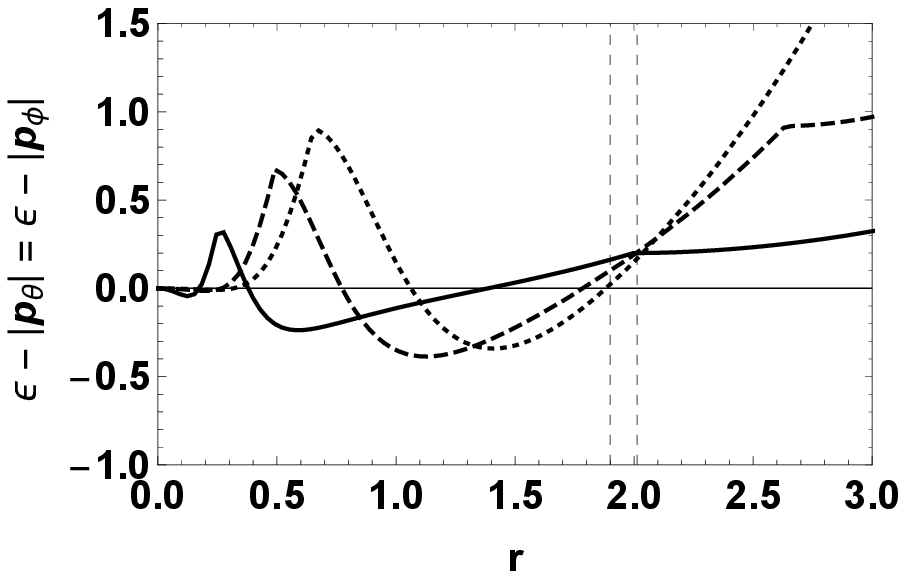}
	\hspace{0.4cm}
	\includegraphics[scale=0.66]{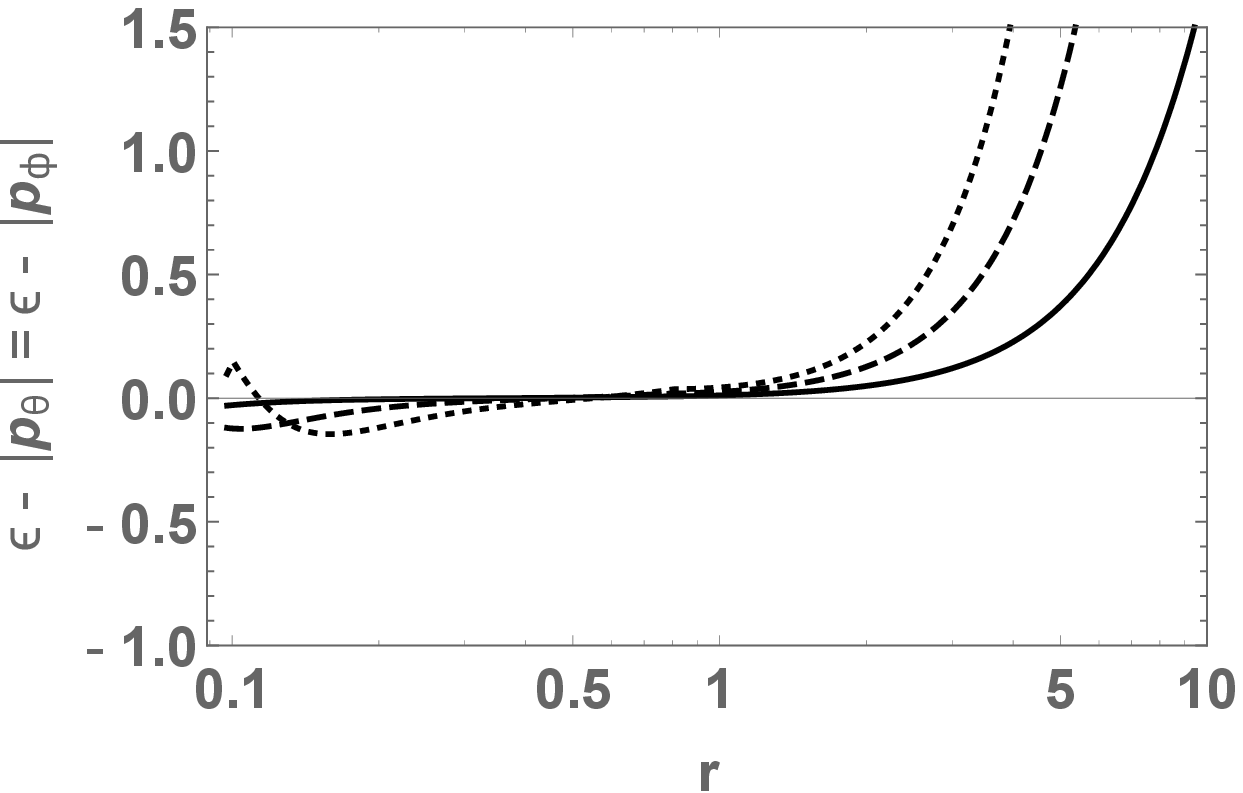}d.
	\caption{Energy condition as function of $r$ for different values of $Q$, $a$, $\omega_q$ and $C$. In the figures a and b, we consider: solid line $Q=0.3$, $a=0.1$, $\omega_q=-0.4$ and $C=0.01$; dashed line $Q=0.6$, $a=0.2$, $\omega_q=-0.5$ and $C=0.02$; and dotted line $Q=0.9$, $a=0.3$, $\omega_q=-0.6$ and $C=0.03$ . In figure c, we consider the following values from Table~\ref{tab1} ($\omega_q=-2/3$): $Q=0.3$ and $C=0.01$ (solid line), $Q=0.605$ and $C=0.02$ (dashed line), and $Q=0.81$ and $C=0.03$(dotted line). The two vertical lines are located at $r=1.9$ and $r=2.013$. In figure d we considered $Q=0.03$ (solid line), $Q=0.06$ (dashed line) and $Q=0.09$  dotted line. \label{fig0}}
\end{figure*}
From Eq.~(\ref{equations of state}) the weak energy condition reduces to 
\begin{equation}
\label{weak}
\begin{aligned}
\epsilon=\frac{3MQ^3 r^4}{\Sigma^2(Q^3+r^3)^2}-\frac{3}{2}\frac{C \omega_q r^{1-3\omega_q}}{\Sigma^2}&\geq0\\
\end{aligned}
\end{equation}
and,
\begin{equation}
\label{weakI}
\begin{aligned}
\epsilon+p_r&=0\\
\epsilon+p_\theta&=\epsilon+p_\phi\geq 0.
\end{aligned}
\end{equation}
According to Eq.~(\ref{weak}), since $-1<\omega_q<-1/3$ and $C>0$, the energy density $\epsilon$ is always positive if $Q>0$. Note that Eq.~(\ref{weak}) does not depend on the spin parameter $a$. Similarly, $\epsilon+p_\theta=\epsilon+p_\phi\geq0$, is also satisfied if $Q>0$ (see Fig.\ref{fig0}a). Therefore, the line element obtained in Eq.~(\ref{Rotating_charged_QE}) satisfies the weak energy condition. 

In the case of the strong energy condition, the relation $\epsilon+p_i$ is satisfied if $Q>0$ (see Eq.~(\ref{weakI}) and Fig.~\ref{fig0}a). Nevertheless, when the expression $\epsilon+p_r+p_\theta+p_\phi$ is considered, the strong energy condition it is not satisfied for $r>3.4$ (see Fig.~\ref{fig0}b). Furthermore, as Fig.~\ref{fig0}b also shows, for $r<3.4$ it is possible to see regions where the strong energy condition is satisfied and regions where it is not.

The dominant energy condition (DEC) is plotted in Fig~\ref{fig0}c. From the figure, we see regions where the DEC is satisfied and regions where it is not. The condition is always satisfied for $r>1.37906$ (when $C=0.01$ and $Q=0.3$), $r>1.77211$ (when $C=0.02$ and $Q=0.605$) and $r>1.87948$ (when $C=0.03$ and $Q=0.81$).  If we then compare these values of $r$ with their respective outer horizons $r_+$ (see Tab.~\ref{tab1}), we conclude that they are inside $r_+$. For example, when $C=0.01$ and $Q=0.3$, the outer horizon is $r_+=2.013$. In this case, as mentioned before, the DEC is already satisfied for $r>1.37906$. Therefore, the DEC is always satisfied for $r>r_+$. A similar situation occurs with the other two cases. This result is in agreement with the claim that the region outside the outer horizon is the only region observationally meaningful. Finally, if we consider $Q<<1$, these regions become smaller and the dominant energy condition tends to be satisfied (see Fig.\ref{fig0}d).

\subsection{Electromagnetic vector potential}

The gauge field of the spherically symmetric regular black hole solution with the magnetic charge is $A_\mu=Q\cos\theta \delta^\phi_\mu$~\cite{Fan16}. However, in the rotating spacetimes, due to axial symmetry, the non-zero components for the electromagnetic tensor are  $F_{tr}$, $F_{t\theta}$, $F_{t\phi}$ and $F_{\theta\phi}$ ~\cite{Dymnikova15} and the vector potential takes the form~\cite{Erbin15a}
\begin{equation}
\label{vector potential}
A_\mu=-\frac{Q a \cos\theta}{\Sigma}\delta^t_\mu+\frac{Q(r^2+a^2)\cos\theta}{\Sigma}\delta^\phi_\mu.
\end{equation}

\begin{table*}
	\label{tableI}
	\centering
	\begin{tabular}{|c|c|c|c|c|c|c|c|c|c|c|c|c|}
		\hline
		\multirow{2}{*}{}&
		\multicolumn{3}{|c|}{$C=0.01$} &
		\multicolumn{3}{|c|}{$C=0.02$} &
		\multicolumn{3}{|c|}{$C=0.03$} &
		\multicolumn{3}{|c|}{$C=0.04$}\\
		\hline
		Q& $r_-$ & $r_+$ & $r_q$ & $r_-$ & $r_+$ & $r_q$ & $r_-$ & $r_+$ & $r_q$ &$r_-$ & $r_+$ & $r_q$  \\
		\hline
		0.100&0.080&2.034&97.95842168 &0.080 &2.079 &47.91310621 &  0.080 & 2.129 & 31.196697& 0.080 & 2.184 & 22.808295\\
		\hline
		0.205&0.159&2.0259&97.95842168  &0.159  &2.071  &47.91310635  & 0.159 &2.121 &31.196698 & 0.159 & 2.176 & 22.808297\\
		\hline
		0.3&0.234&2.013&97.95842172  &0.325  & 2.038 &47.91310673 &  0.234 & 2.108& 31.196699& 0.234 & 2.163 & 22.808301  \\
		\hline
		0.4&0.325 &1.992 &97.95842180 &0.399 &2.017 & 47.91310746 & 0.324 & 2.088&31.196702&0.316 &2.145 &22.808308\\
		\hline
		0.49&0.399 &1.971 &97.95842195 & 0.501 &1.982 &47.91310867 & 0.398 & 2.067 &31.196706&0.393 &2.124 &22.808321\\
		\hline
		0.605&0.502 &1.935 &97.95842214 &0.596 & 1.944&47.91311048 & 0.499 & 2.033 &31.196713& 0.498& 2.090 & 22.808339\\
		\hline
		0.7&0.598 &1.895 &97.95842242 &0.717 &1.886 &47.91311300 & 0.593 & 1.996 &31.19672341& 0.591& 2.054 &22.808365\\
		\hline
		0.81& 0.722 &1.836 &97.95842279 & 0.717& 1.886& 47.91311635 & 0.713 & 1.941 &31.196736&0.708 &2.001 &22.808400\\
		\hline
		0.9&0.839 &1.770 &97.95842328 &0.830 &1.825 &47.91312065 & 0.823& 1.884 &31.196752& 0.815 & 1.947 &22.808444\\
		\hline
		1&1 &1.666 &97.95842384 &0.982 &1.730 & 47.91312573& 0.966 &  1.797 &31.1967730& 0.951&1.868 &22.808500\\
		\hline
	\end{tabular}
	\caption{Values for the inner ($r_-$), outer ($r_+$), and quintessential ($r_q$) horizons as a function of $Q$ for different values of $C$. We set $a=0.1$ and $M=1$. \label{tab1}}
\end{table*}

\section{Rotating black hole with the parameter $\omega_q=-2/3$\label{sectionV}}  
In this section, we focus our attention to the case in which $\omega_q=-2/3$. Hence, after substitution, we have that
\begin{equation}
\label{definitions}
\begin{aligned}
2\rho&=\frac{2Mr^3}{r^3+Q^3}+Cr^2.
\end{aligned}
\end{equation}   
Therefore, when $Q=0$, the line element in Eq.(\ref{Rotating_charged_QE}) reduces to the Kerr solution surrounded by quintessence \cite{Toshmatov17}. When $a=0$, we obtain the non-linear charger magnetic black hole obtained in reference \cite{Nam18}. Finally, for $C=0$, $Q=0$ and $a=0$ Eq.~(\ref{Rotating_charged_QE}) reduces to the Schwarzschild black hole.

Now, we study the behavior of the event horizon, the ergosphere, and the ZAMO.

\subsection{Event horizons}
To compute the event horizons we use the condition $g^{rr}=0$. Therefore, for the line element  (\ref{Rotating_charged_QE}), we obtain 
\begin{equation}
g^{rr}=\frac{\Delta}{\Sigma}=\frac{a^2-C r^3-\frac{2 M r^4}{Q^3+r^3}+r^2}{a^2 \cos ^2(\theta )+r^2},
\end{equation}
from which 
\begin{equation}
\label{Delta}
\Delta=a^2-C r^3-\frac{2 M r^4}{Q^3+r^3}+r^2=0.
\end{equation}

\begin{figure}[h]
\includegraphics[scale=0.38]{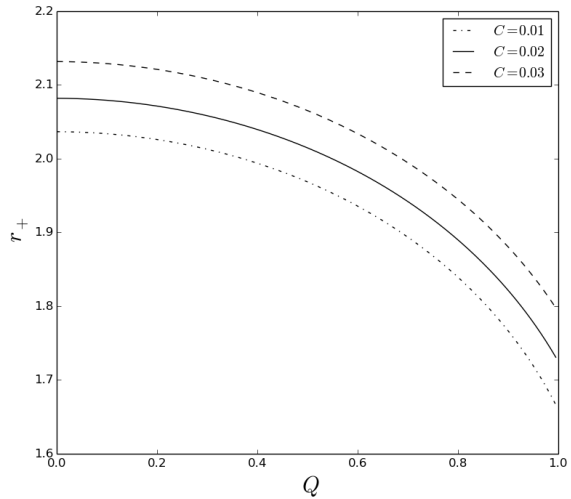}
\caption{$r_+$ vs. $Q$ for different values of $C$. We set $a=0.1$\label{fig1}}
\end{figure}
\begin{figure*}[t]
a.\includegraphics[scale=0.4]{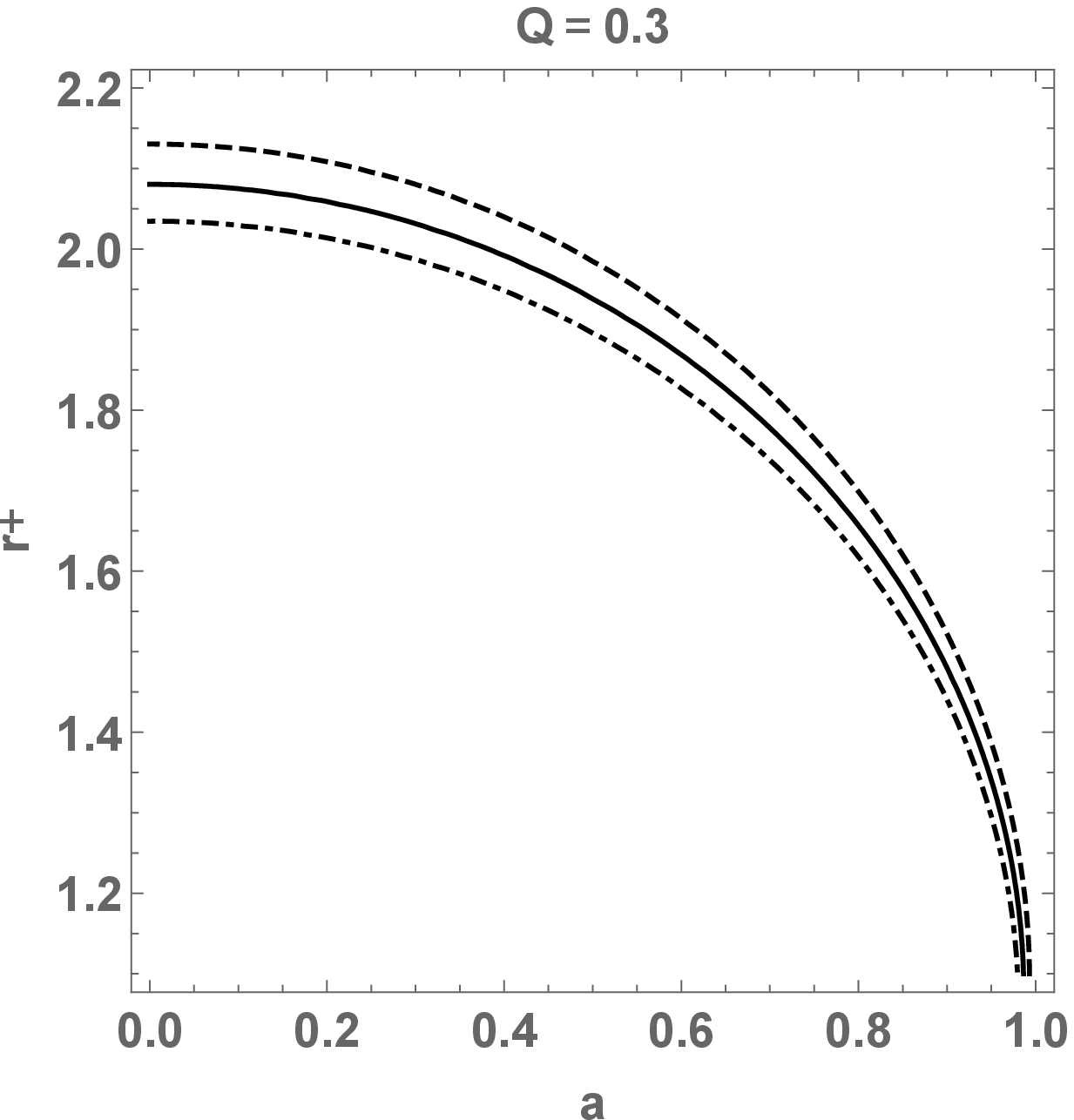}
\hspace{0.5cm}
b.\includegraphics[scale=0.4]{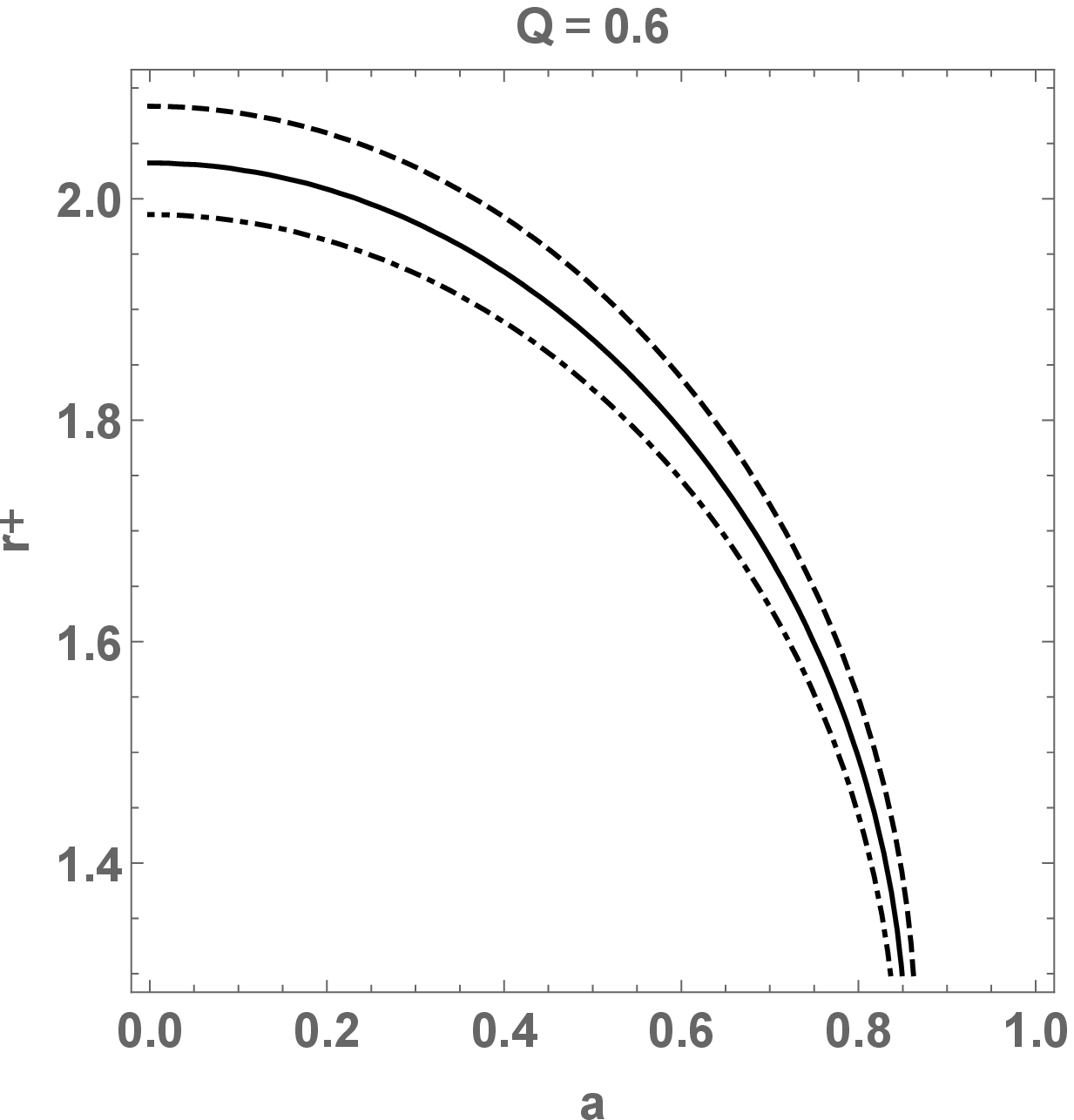}
\hspace{0.5cm}
c.\includegraphics[scale=0.4]{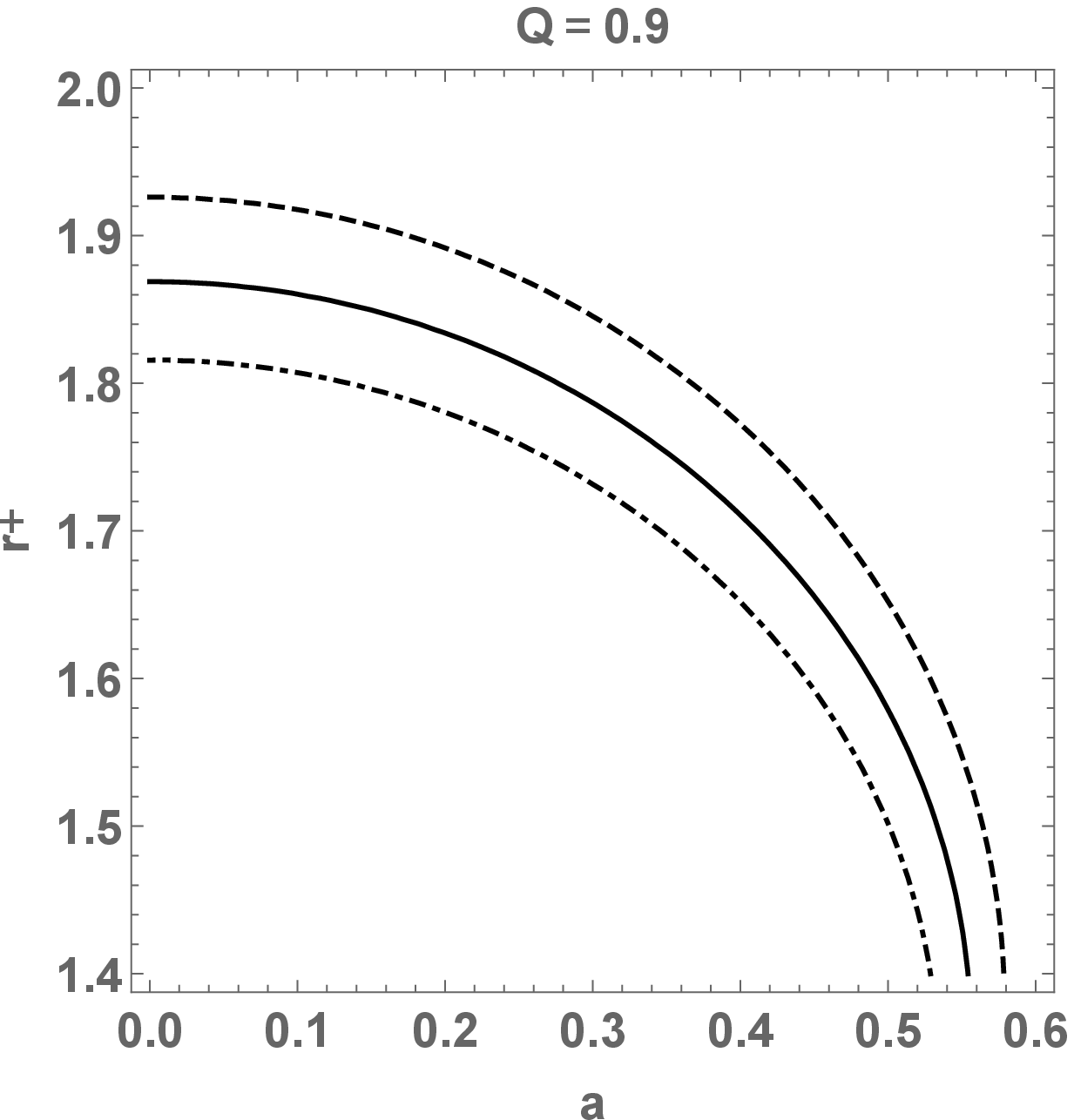}
\caption{Plot of the outer horizon $r_+$ vs. the spin parameter $a$ for different values of $Q$ and $C=0.01$ (dot-dashed line), $C=0.02$ (continuous line), and $C=0.03$ (dashed line). \label{fig1B}}
\end{figure*}

Hence, equation (\ref{Delta}) can be expressed as (with $M=1$)
\begin{equation}
\label{poly_horizon}
Cr^6-r^5+2r^4-(a^2-CQ^3)r^3-Q^3r^2-a^2Q^3=0.
\end{equation}

\begin{figure*}[t]
a.\includegraphics[scale=0.278]{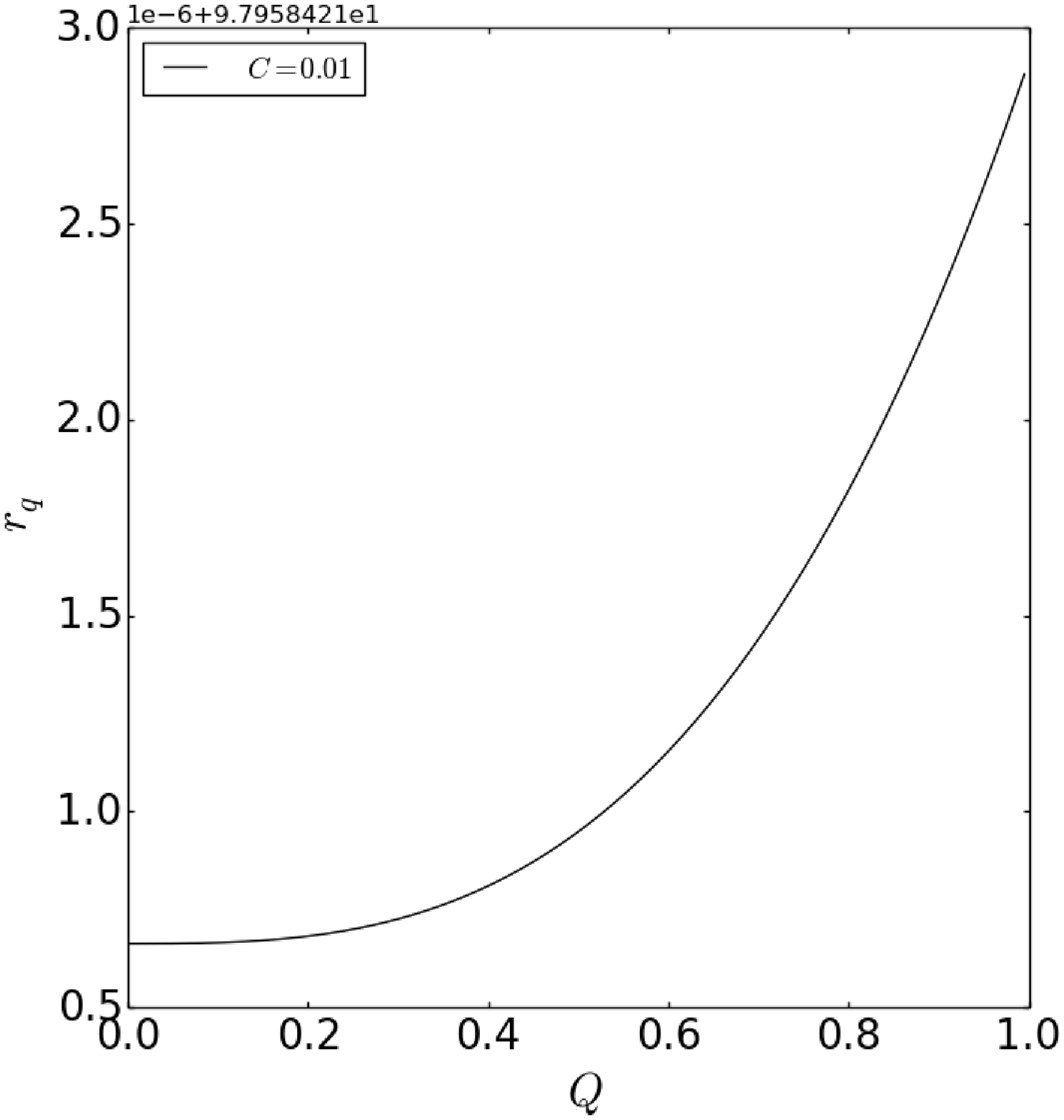}
b.\includegraphics[scale=0.278]{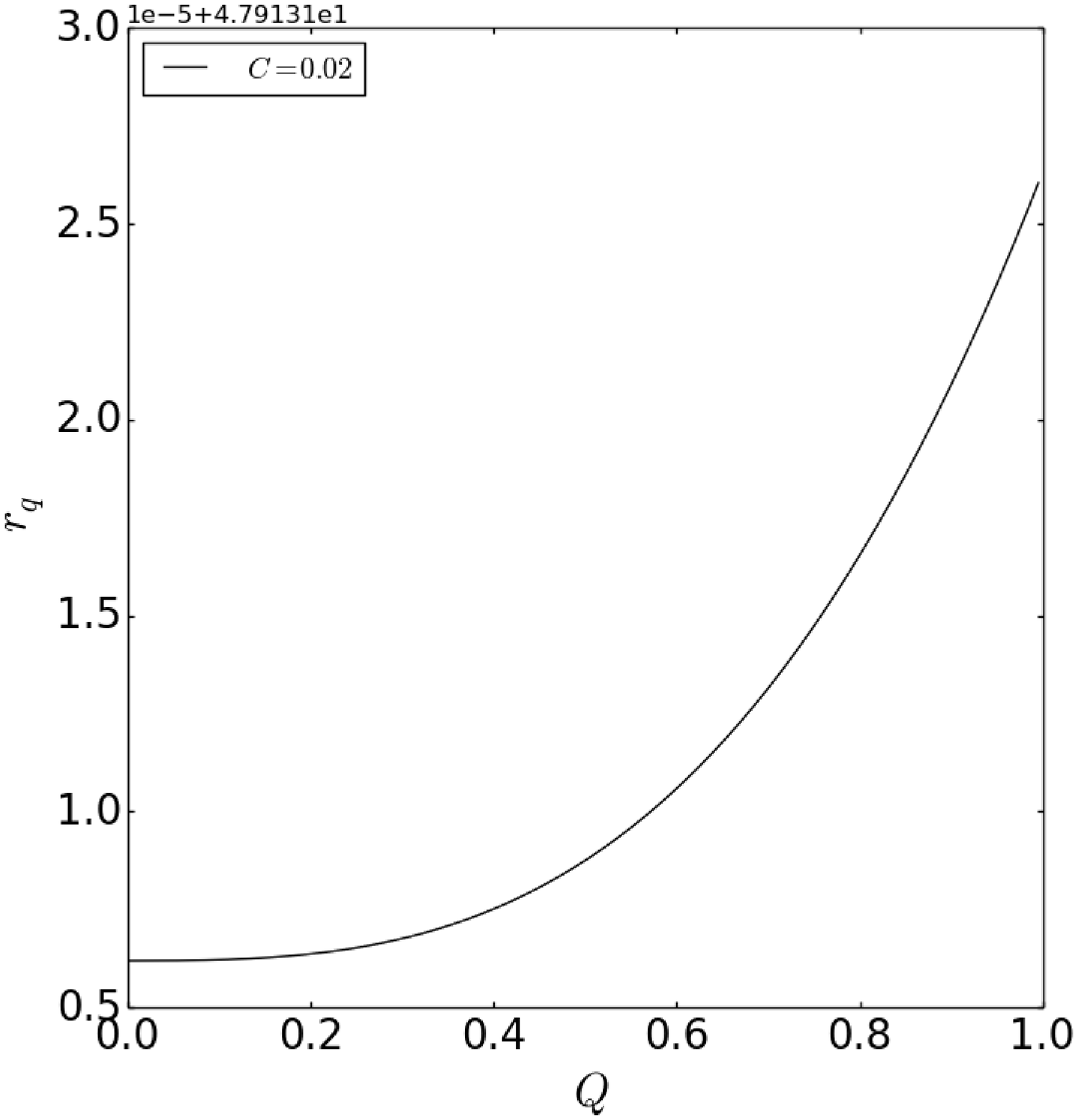}
c.\includegraphics[scale=0.278]{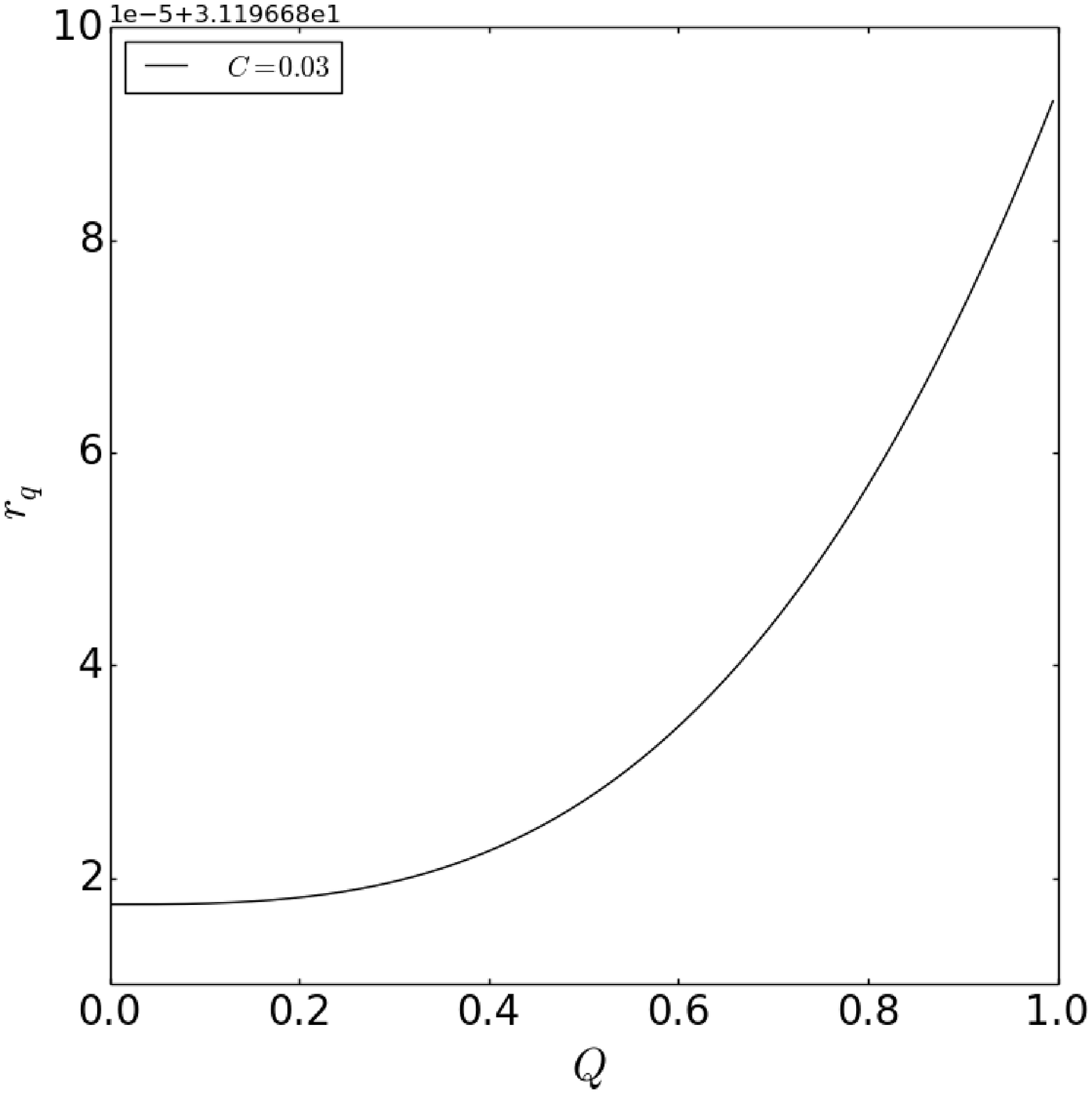}
\caption{$r_q$ vs. $Q$ for different values of $C$. Note the scale used at the top of each plot. In figure $a$, for example, to the value $2.0$ in the vertical axis corresponds to $r_q=97.95841+2\times 10^{-6}$. Similarly for figures $b$ and $c$. We set $a=0.1$ in all the plots.\label{fig2}}
\end{figure*}
\noindent  This equation cannot be solved analytically but numerically. Therefore, in order to find the roots of equation (\ref{poly_horizon}), we set $a=0.1$ and considered different values of the quintessential parameter $C$. Then, for different values of $Q$ in the interval $0\leq Q\leq1$, we compute the roots. We found three real roots: $r_-$, $r_+$ and $r_q$, which represent the inner and outer black hole horizons, and the quintessential cosmological horizon respectively.       

Table~\ref{tab1} shows the values of the inner, outer, and quintessential horizon for different values of $Q$ and $C$. From Table~\ref{tab1}, we see that $r_-$ and $r_q$ increase while $r_+$ decreases as $Q$ increases. On the other hand, for fixed values of $Q$, the quintessential horizon $r_q$ decreases while $r_+$ increases as $C$ increases.

Fig.~\ref{fig1} shows the outer horizon $r_+$ as a function of $Q$. In contrast with the inner horizon, the outer horizon $r_+$ decreases when the charge $Q$ increases. Furthermore, the greater the values of $C$, the greater the values of the outer radius. In Fig.~\ref{fig1B}, we plotted the outer radius $r_+$ as a function of the spin parameter $a$ for different values of $Q$ and $C$. From the figure we can obtain some constrains for $r_+$. For example, in Fig.~\ref{fig1B} panel $a$, the outer radius exists between $r_+=0.1$ and $r_+=2.2$. For $a>1$, $r_+$ does not exist. Furthermore, if $Q$ increases, the existence of $r_+$ is constrained by $a>0.87$ and $a>0.58$ when $Q=0.6$ and $Q=0.9$, respectively (see panels b and c in Fig.~\ref{fig1B}). 

On the other hand, Fig.~\ref{fig2} shows the quintessential cosmological horizon as a function of the charge $Q$ for different values of $C$. The behavior is quite similar between figures: $r_q$ increases as the charge $Q$ increases. Note that the changes in the quintessential horizon are of order $10^{-6}$ ($C=0.01$) and $10^{-5}$ ($C=0.02$ and $C=0.03$) (see Table~\ref{tab1} for details). The value of $r_q$ is bigger for small values of $C$. For example, for $C=0.01$ the value of the quintessential radius is of order $r_q=97.9584$ while for $C=0.03$ the value is $r_q=31.1966$. As can be seen clearly from Fig.~\ref{fig2} $a$ and $c$. 

\begin{figure*}[t]
a.\includegraphics[scale=0.4]{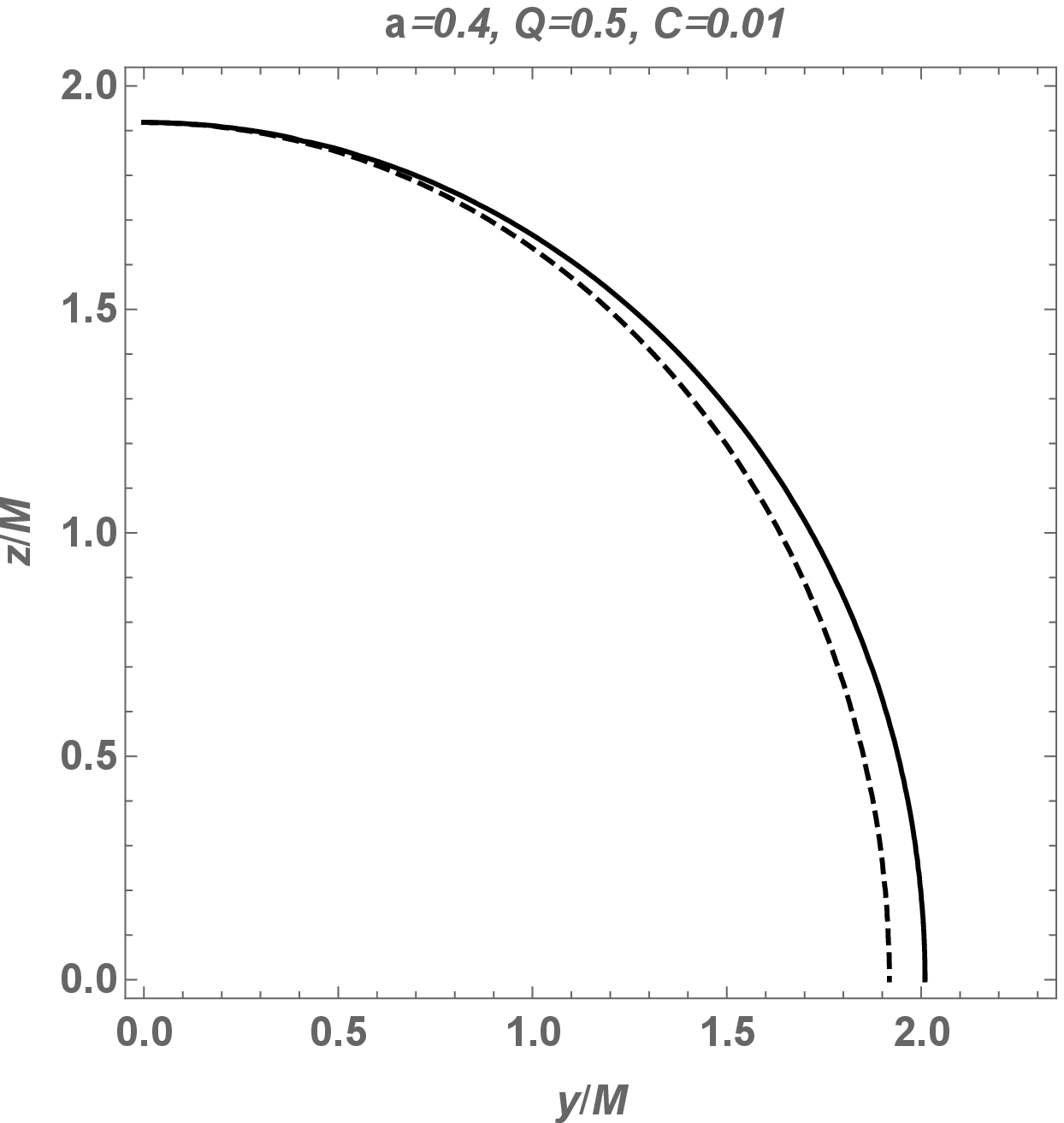}
\hspace{0.5cm}
b.\includegraphics[scale=0.4]{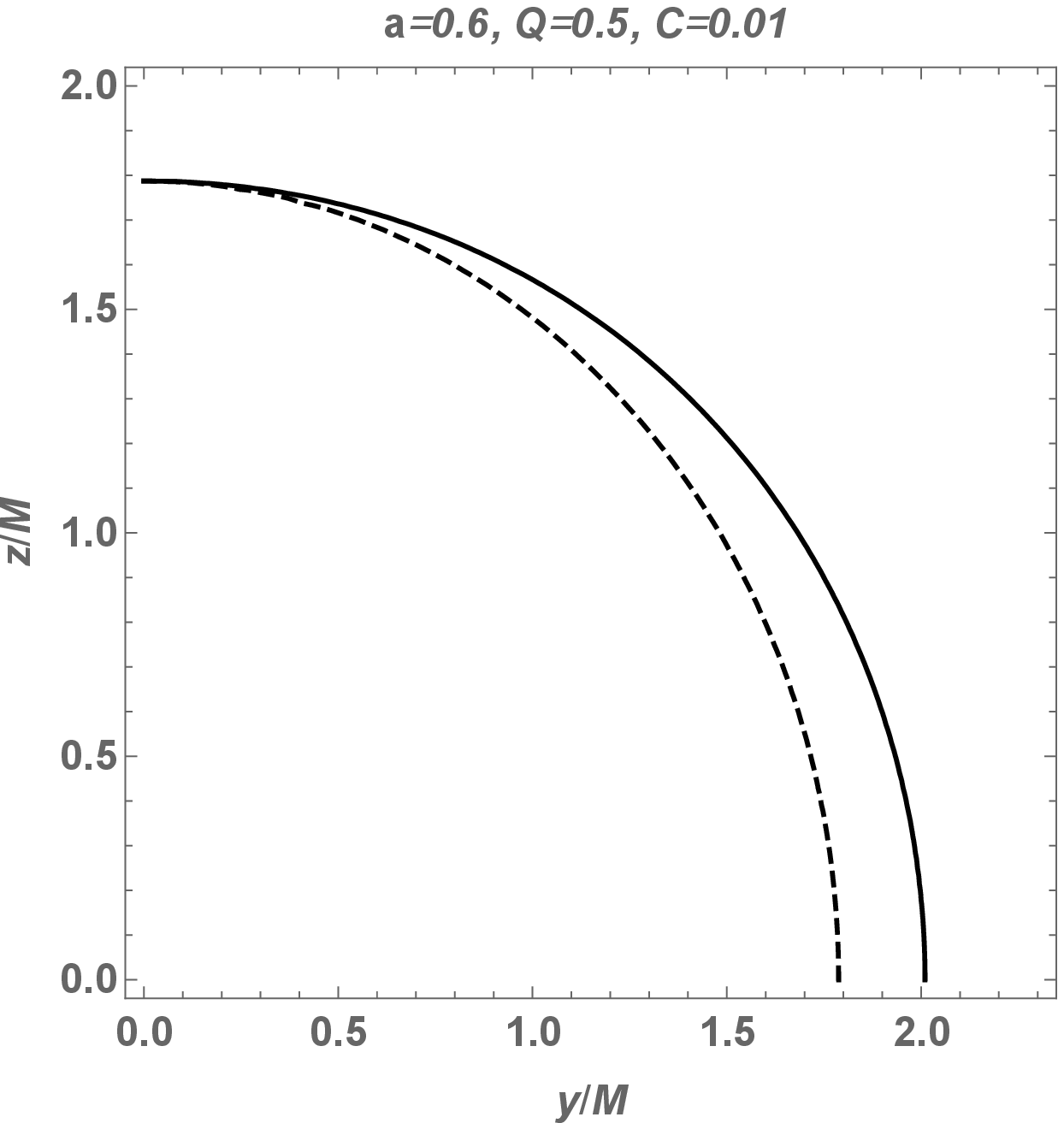}
\hspace{0.5cm}
c.\includegraphics[scale=0.4]{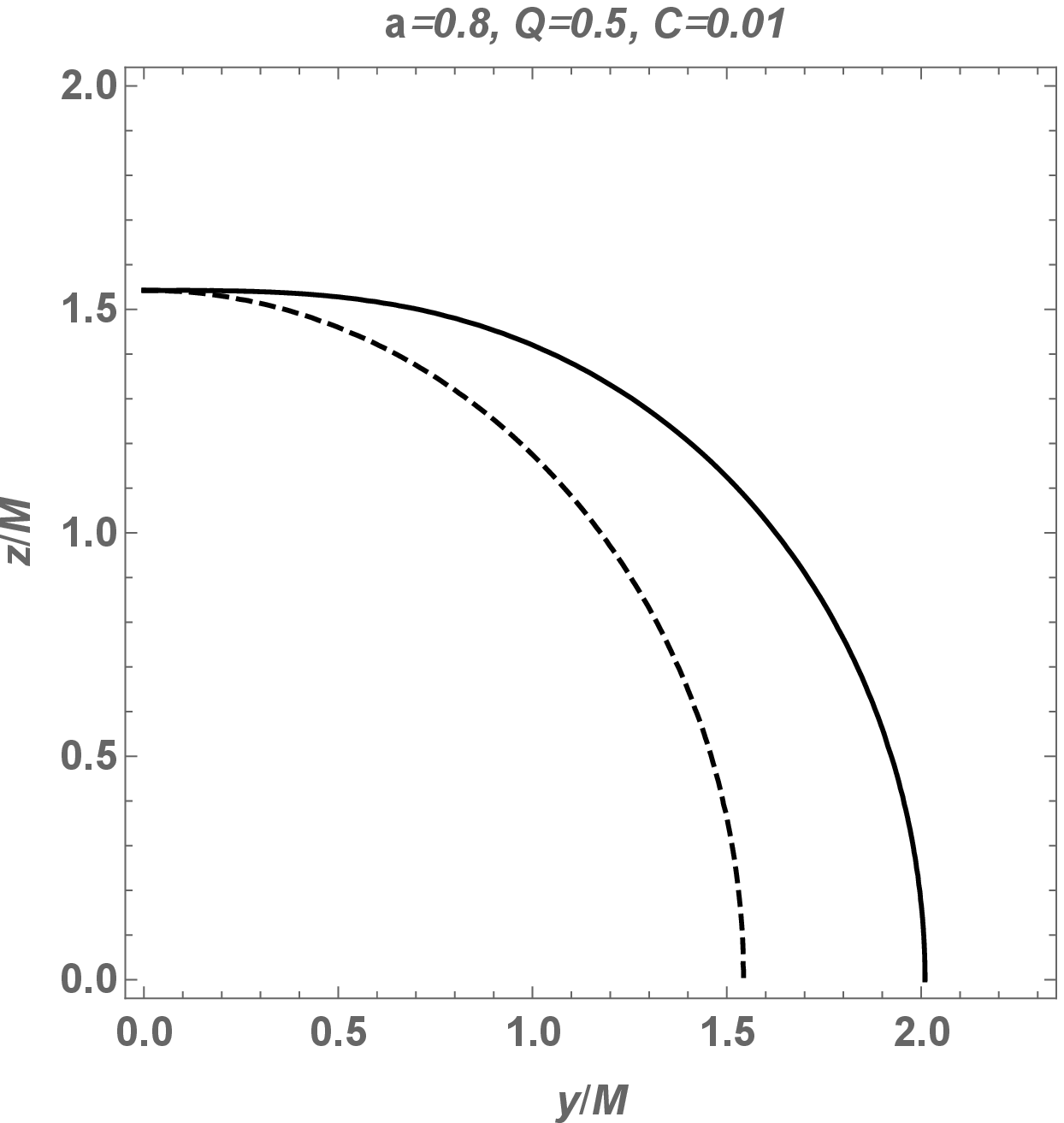}\\
d.\includegraphics[scale=0.4]{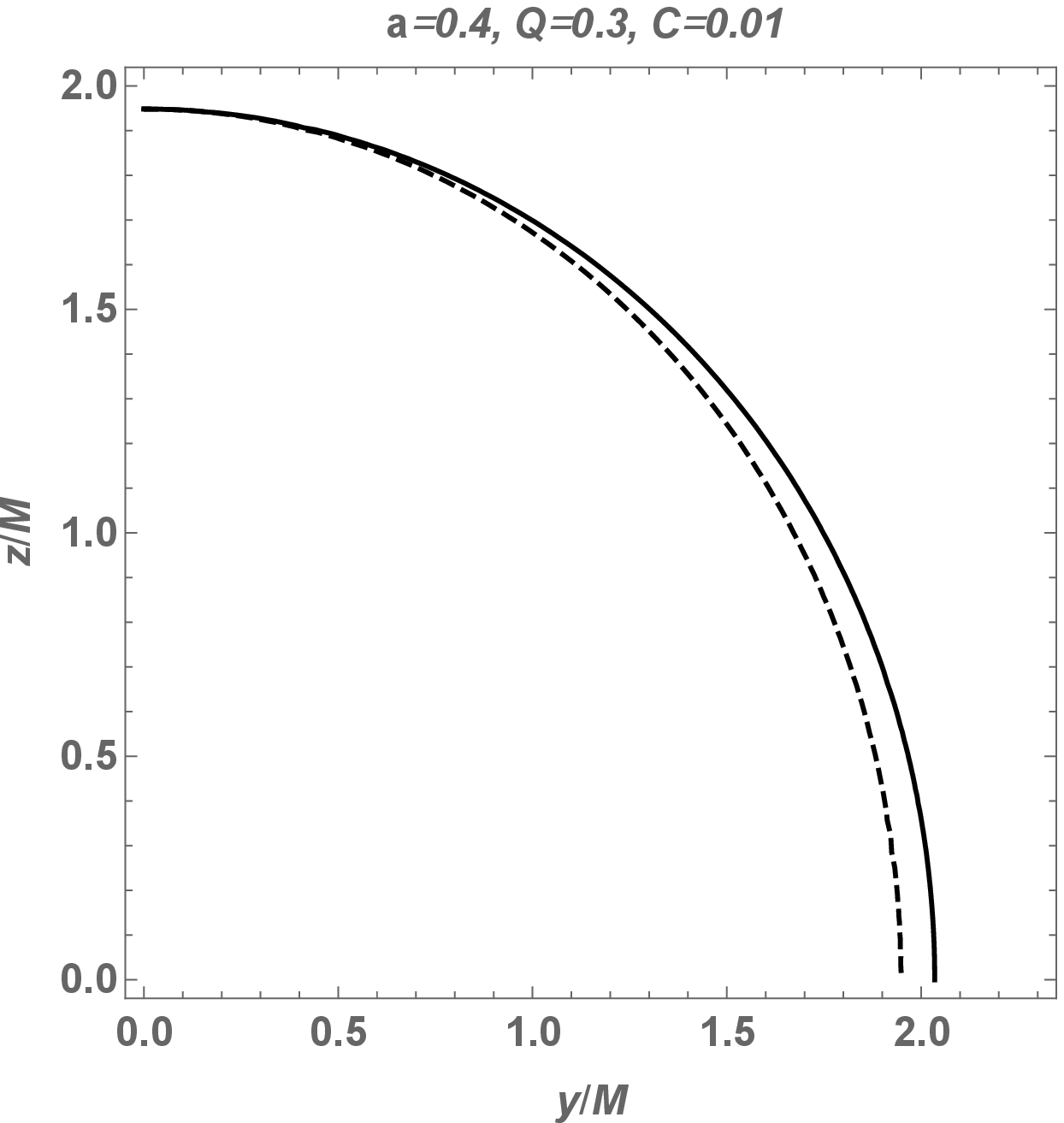}
\hspace{0.5cm}
e.\includegraphics[scale=0.4]{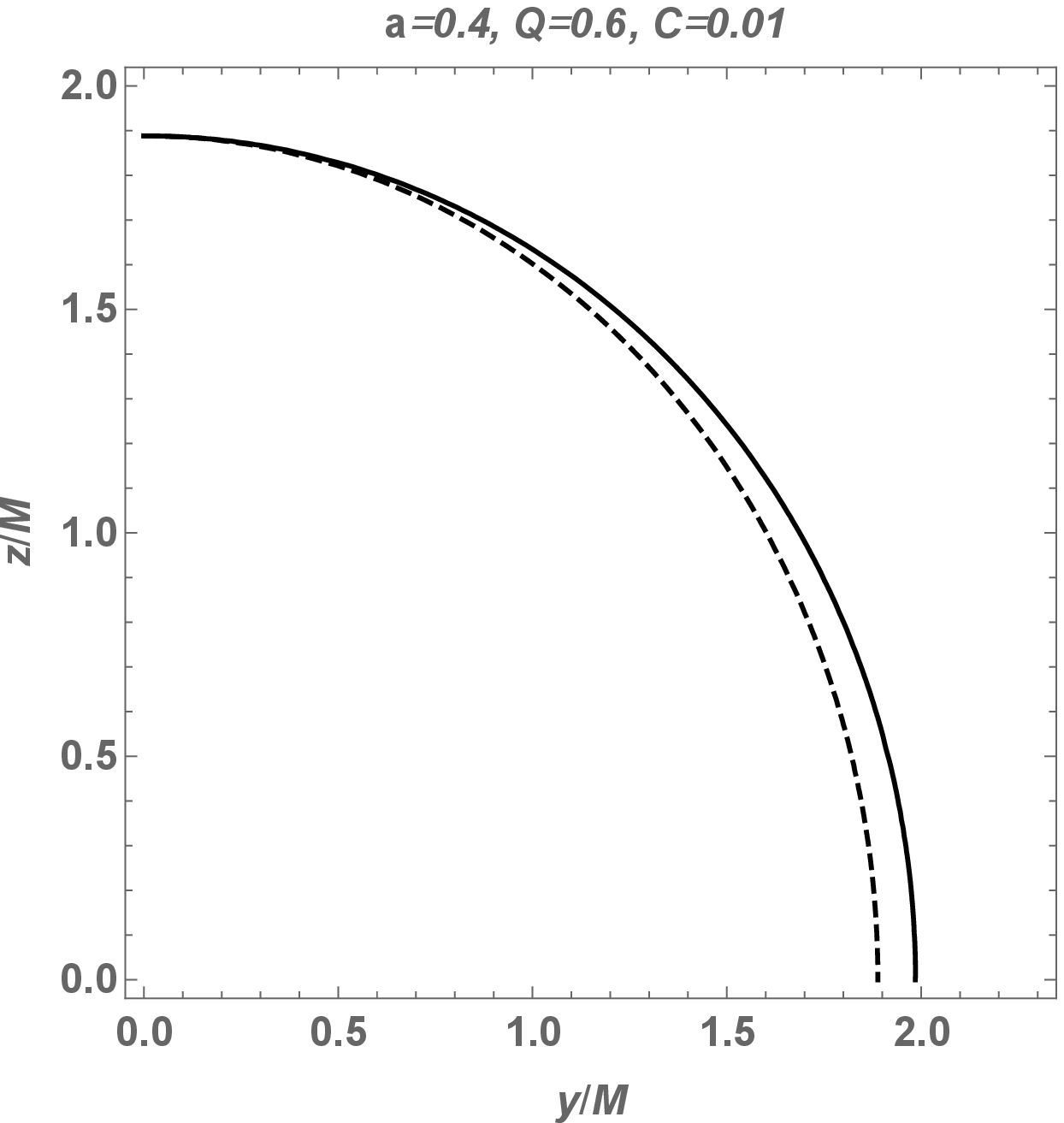}
\hspace{0.5cm}
f.\includegraphics[scale=0.4]{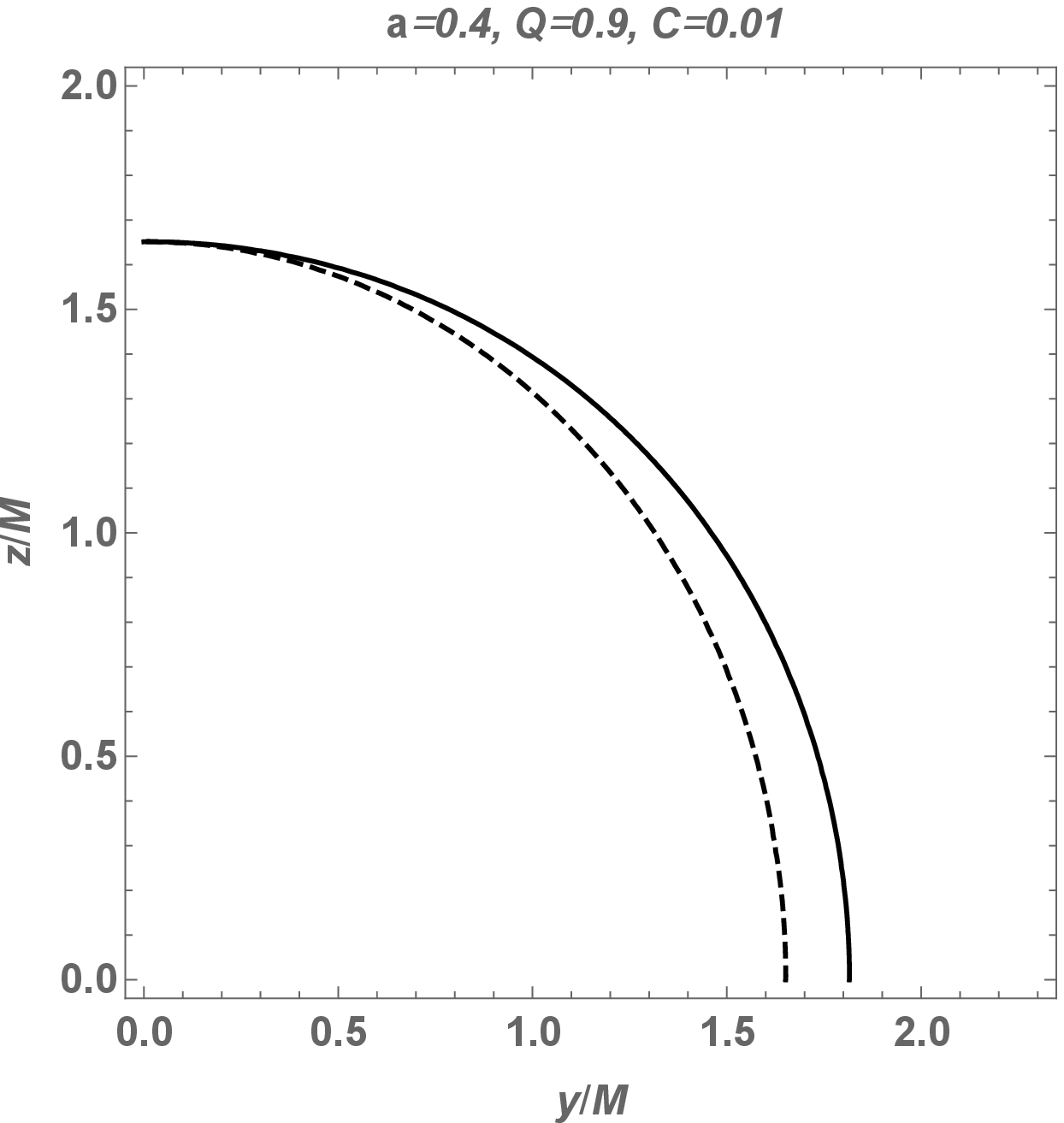}
\caption{Shape and size of the horizon (dashed line) and the ergosphere (continuous line) of the rotating and non-linear charged black hole.\label{fig3}}
\end{figure*}

\subsection{Ergosphere}

The ergosphere is a region located between the event horizon and the static limit surface, which is defined by equation \cite{Toshmatov17}
\begin{equation}
\label{ergosphere_definition}
g_{tt}(r_{st},\theta)=\left(1-\frac{2\rho r_{st}}{\Sigma}\right)=0.
\end{equation} 
Using the second and third equations in (\ref{definitionsI}), we obtain
\begin{equation}
\label{ergosphere_equation}
Cr^6_{st}-r^5_{st}+2r^4_{st}-(a\cos^2\theta-CQ^3)r^3_{st}-a^2Q^3\cos^2\theta=0.
\end{equation}
In the equatorial plane ($\theta=\pi/2$), equation (\ref{ergosphere_equation}) takes the form
\begin{equation}
\label{ergosphere_equatorial}
Cr^6_{st}-r^5_{st}+2r^4_{st}+CQ^3r^3_{st}=0.
\end{equation} 
Note that, as in the case of Kerr black hole, at the equatorial plane $r_{st}$ does not depend on the rotation parameter $a$ (see Fig.~\ref{fig3} a, b and c). After factorization, equation (\ref{ergosphere_equatorial}) can be reduced to
\begin{equation}
Cr^3_{st}-r^2_{st}+2r_{st}+CQ^3=0.
\end{equation} 
In the polar caps ($\theta=0,\pi$), equation (\ref{ergosphere_equation}) reduces to
\begin{equation}
\label{ergosphere_polar_caps}
Cr^6_{st}-r^5_{st}+2r^4-(a-CQ^3)r^3_{st}-Q^3r^2_{st}-a^2Q^3=0.
\end{equation} 
Note that the equations governing the event horizon and the ergosphere, equations (\ref{poly_horizon}) and (\ref{ergosphere_polar_caps}) respectively, coincide at the polar cap, in agreement with the result obtained in \cite{Toshmatov17}.

In Fig.~\ref{fig3} we plot the shape and size of the horizon and the ergosphere. When the spin parameter $a$ is increased, while keeping constant the charge $Q$ and the quintessence parameter $C$, the radius of the event horizon decreases and the area of the ergo-region increases. Similar behavior is observed when the charge $Q$ is increased while keeping constant the spin and quintessence parameters. Nevertheless, the increment of the ergo-region is stronger in the first case: when the spin parameter $a$ is increased. On the other hand, from Figs.~\ref{fig3} d, e and f, we conclude that the ergosphere radius does depend on the charge $Q$ in the equatorial plane.      

\begin{figure*}[t]
a.\includegraphics[scale=0.38]{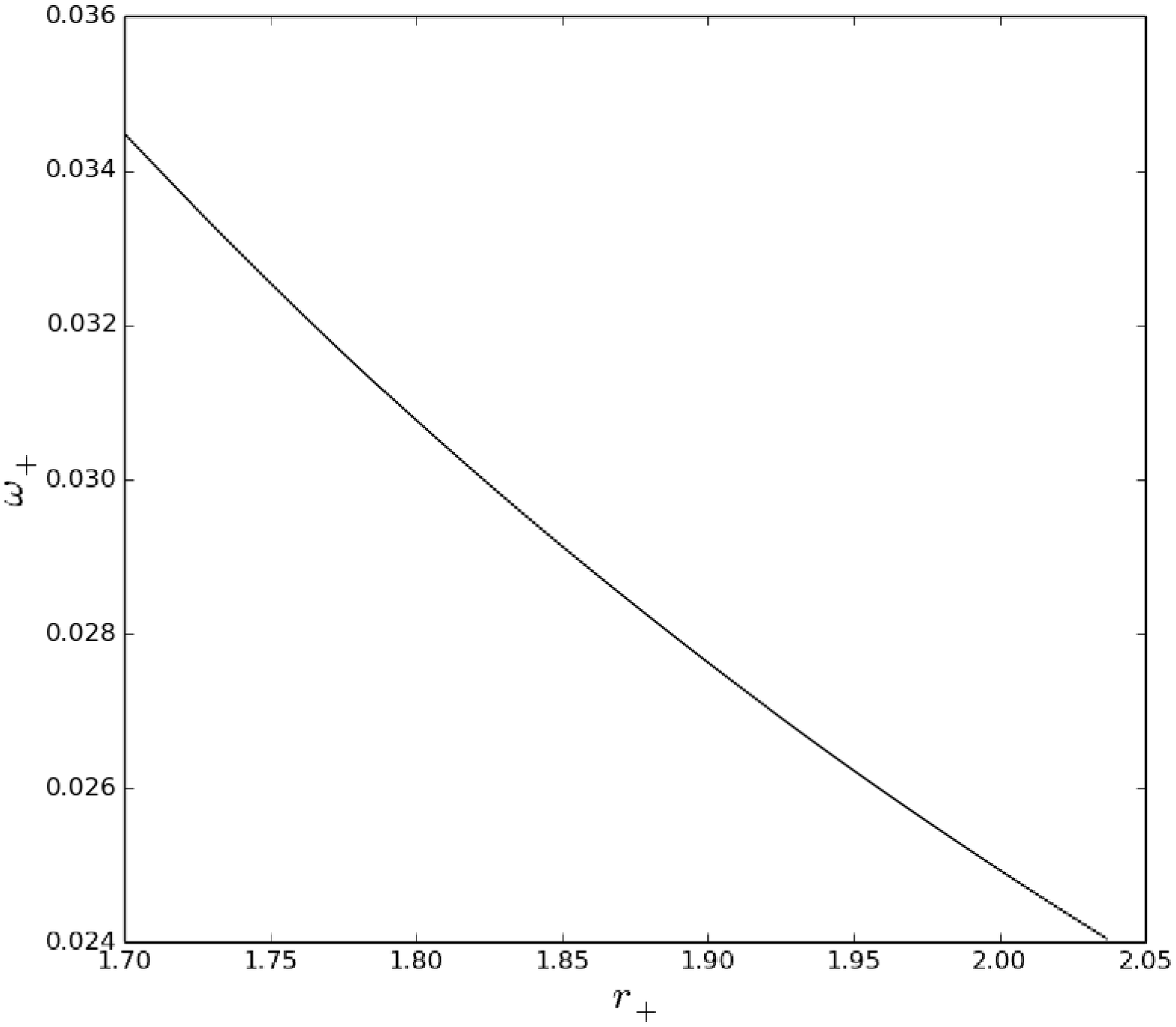}
b.\includegraphics[scale=0.38]{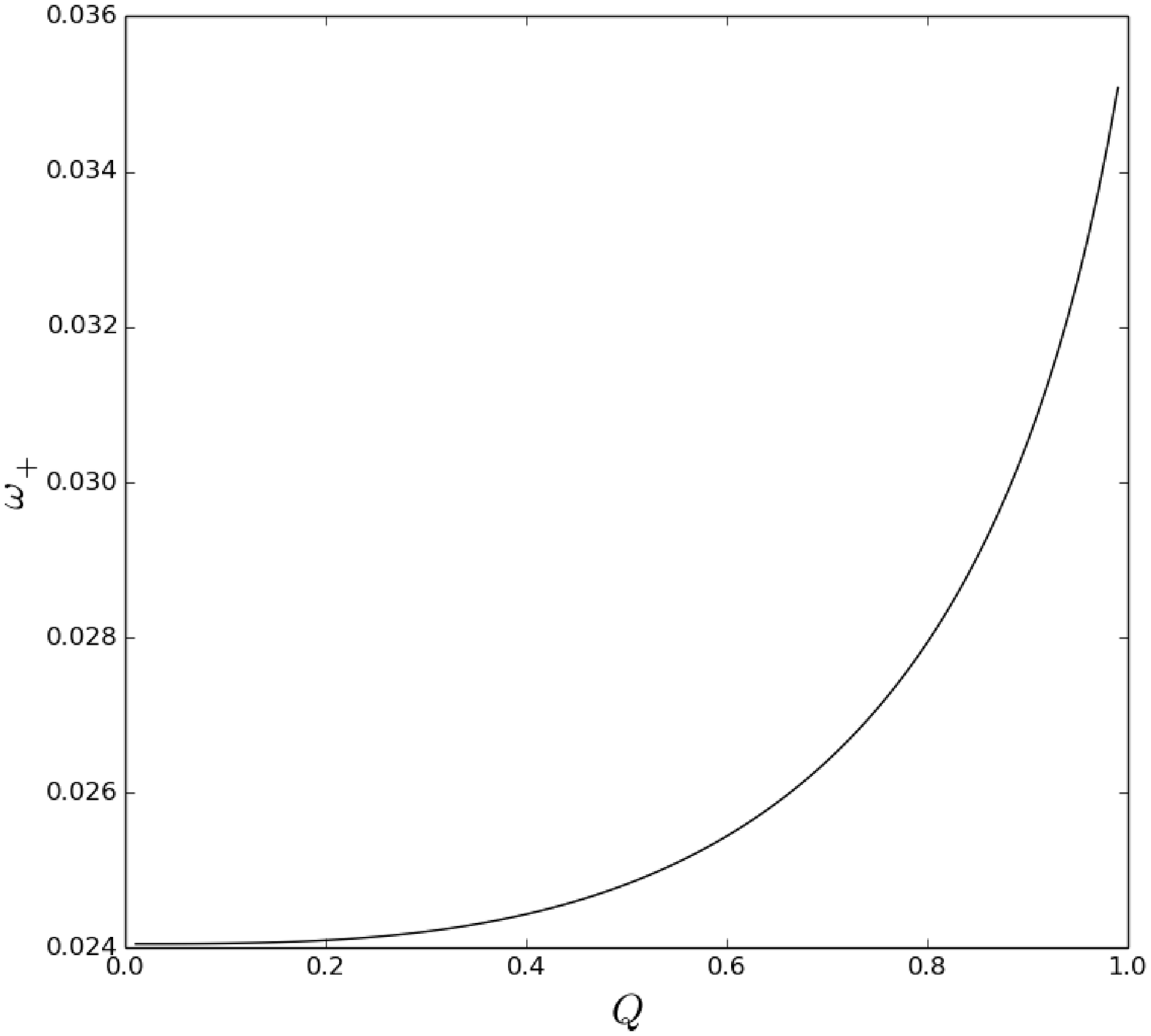}b.
\caption{a) $\omega_+$ vs. $r_+$. b) $\omega_+$ vs. $Q$. In both plots we solve equation (\ref{poly_horizon}) setting $a=0.1$, and $C=0.01$ \label{fig4}}
\end{figure*}

\subsection{ZAMO}
The geodesic motion of a particle on a space-time with metric $g_{\mu\nu}$ is governed by the Lagrangian \cite{Bambi17e}
\begin{equation}
\label{lagrangian}
\mathcal{L}=\frac{1}{2}g_{\mu\nu}\dot{x}^\mu\dot{x}^\nu,
\end{equation}
where the dot denotes the total derivative with respect to an affine parameter.\\

In equation (\ref{Rotating_charged_QE}), the components of the metric  does not depend on the coordinates $t$ and $\phi$. Therefore, two constants of motion are obtained: the specific energy  $E$, and the axial component of the specific angular momentum $L_z$. Hence, after using the Euler-Lagrange equations, we obtain that \cite{Bambi17e}
\begin{equation}
\label{Euler-Lagrange-equations}
\begin{aligned}
\frac{\partial\mathcal{L}}{\partial \dot{t}}=g_{tt}\dot{t}+g_{t\phi}\dot{\phi}&=-E\\
\frac{\partial\mathcal{L}}{\partial\dot{\phi}}=g_{t\phi}\dot{t}+g_{\phi\phi}\dot{\phi}&=L_z
\end{aligned}
\end{equation}  
In contrast with Newtonian gravity, where the angular momentum of a spinning object has no gravitational effect, in general relativity, the angular momentum alters the geometry of the space-time. As a consequence, it is possible for a test-particle to have a non-vanishing angular velocity even if its specific angular momentum $L_z$ vanishes \cite{Bambi17e}. This effect, known as dragging, can be obtained from the second equation in (\ref{Euler-Lagrange-equations}) after setting $L_z=0$.

The angular velocity of a zero angular momentum observer (ZAMO) is defined by \cite{Carroll03}
\begin{equation}
\label{ZAMO}
\omega=-\frac{g_{\phi t}}{g_{\phi\phi}}.
\end{equation}   
Thus, using the line element (\ref{Rotating_charged_QE}), we obtain that 
\begin{equation}
\label{ZAMO_QE}
\begin{aligned}
\omega&=\frac{2a\rho r\sin^2\theta}{\Sigma\sin^2\theta\left(r^2+a^2+\frac{2\rho ra^2\sin^2\theta}{\Sigma}\right)}\\
&=\frac{2a\rho r}{(r^2+a^2)^2-a^2\Delta\sin^2\theta}.
\end{aligned}
\end{equation}
In the last step, we used the relation $2\rho r=r^2+a^2-\Delta$ in the denominator. At the horizon $r_h$, where $\Delta=0$, the ZAMO can be obtained by 
\begin{equation}
\label{ZAMO_horizon}
\omega_h=\omega(r_h)=\frac{a}{r^2_h+a^2}.
\end{equation}
Note that $r_h$ can be the inner, outer, or quintessential horizon.

 In Figs.~\ref{fig4}.a and b it is plotted $\omega_+$ as a function of $r_+$ and $Q$, respectively. From the figures, we see that $\omega_+$ increases as the charge increases, while it decreases as $r_+$ increases. Note that the change in $\omega_+$ is small in both cases.    
\section{Equatorial circular orbits\label{sectionVI}}

In this section, we focus our attention on equatorial circular orbits ($\dot{r}=0$ and $\theta=\pi/2$). These orbits are the most  relevant for astrophysical observations because they are expected to describe the motion of the gas in possible geometrically thin accretion disks around black holes, where the effect of the gas pressure is negligible in comparison with that of the gravitational field. Equatorial orbits are vertically stable and, if the inner part of the accretion disk is initially not on the equatorial plane perpendicular to the black hole spin, it adjusts to the equatorial plane due to the Bardeen-Petterson effect~\cite{Bardeen:1975zz}.

To begin with, we consider the Lagrangian obtained from Eq.~(\ref{lagrangian})
\begin{equation}
2\mathcal{L}=g_{tt}\dot{t}^2+g_{rr}\dot{r}^2+g_{\theta\theta}\dot{\theta}^2+2g_{t\phi}\dot{t}\dot{\phi}
+g_{\phi\phi}\dot{\phi}^2.
\end{equation}
If we consider the case $\theta=\pi/2$ (the equatorial plane), the Lagrangian takes the form
\begin{equation}
\label{Lagrangian}
2\mathcal{L}=g_{tt}\dot{t}^2+g_{rr}\dot{r}^2+2g_{t\phi}\dot{t}\dot{\phi}
+g_{\phi\phi}\dot{\phi}^2.
\end{equation} 
By doing so, we are assuming that the motion of test particles is constrained to a single plane where the coordinate $\theta$ is a constant. In this sense, $\dot{\theta}=0$ and the Lagrangian reduces to Eq.~(\ref{Lagrangian}). As mentioned before, the spacetime described in Eq.~(\ref{Rotating_charged_QE}) has two conserved quantities: the specific energy $E$, and the specific angular momentum $L_z$. Consequently, the momenta $p_t$, $p_\phi$, $p_r$, and $p_\theta$ are given by \cite{Oteev16,Bambi17e}
\begin{equation}
\label{momenta}
\begin{aligned}
p_t&=\frac{\partial \mathcal{L}}{\partial \dot{t}}=g_{tt}\dot{t}+
g_{t\phi}\dot{\phi}=-E\\
p_\phi&=\frac{\partial \mathcal{L}}{\partial \dot{\phi}}=g_{t\phi}\dot{t}+g_{\phi\phi}\dot{\phi}=L_z\\
p_r&=\frac{\partial \mathcal{L}}{\partial \dot{r}}=g_{rr}\dot{r}\\
p_\theta&=\frac{\partial \mathcal{L}}{\partial \dot{\theta}}=0.
\end{aligned}
\end{equation}

Hence, after using Eq.~(\ref{momenta}), the Lagrangian defined in Eq.~(\ref{Lagrangian}) can be expressed as \cite{Oteev16}
\begin{equation}
\label{hamiltonian}
2\mathcal{L}=-E\dot{t}+g_{rr}\dot{r}^2+L_z\dot{\phi}=\kappa,
\end{equation} 
where $\kappa$ is equal to $-1$, $0$, or $1$ for time-like, light-like, and space-like geodesics, respectively. Then, if we consider the first and second equations in Eq.~(\ref{momenta}), we have that 
\begin{equation}
\label{dotphi}
\dot{\phi}=-\frac{E+g_{tt}\dot{t}}{g_{t\phi}}.
\end{equation} 
Hence, the second equation in Eq.~(\ref{momenta}) takes the form
\begin{equation}
\label{dott}
\dot{t}=\frac{g_{\phi\phi}E+L_zg_{t\phi}}{g^2_{t\phi}-g_{\phi\phi}g_{tt}}, 
\end{equation}
and Eq.~(\ref{dotphi}) reduces to 
\begin{equation}
\label{phidot1}
\dot{\phi}=-\frac{g_{t\phi}E+g_{tt}L_z}{g^2_{t\phi}-g_{\phi\phi}g_{tt}}.
\end{equation} 

\begin{figure*}[t]
	\begin{center}
		a.\includegraphics[scale=0.6]{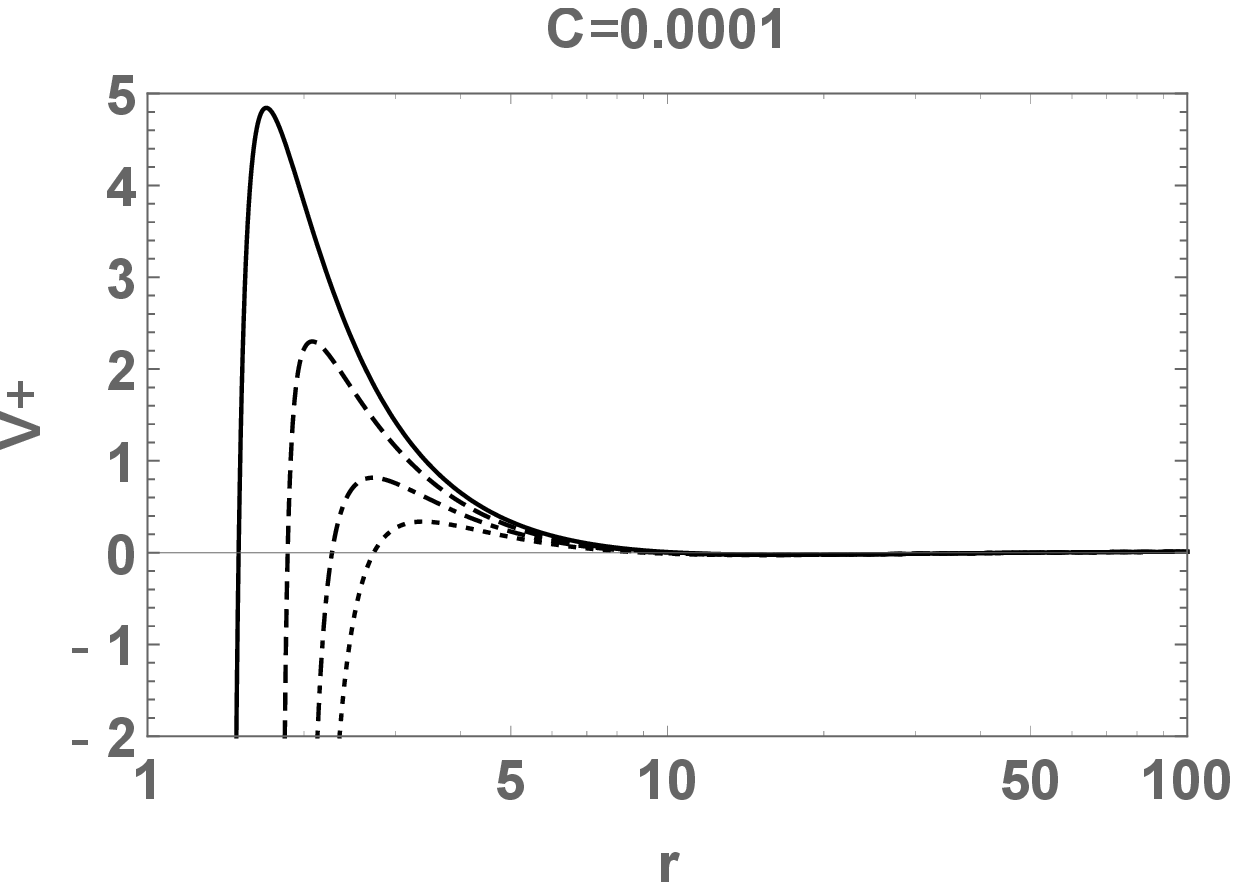}
		\includegraphics[scale=0.6]{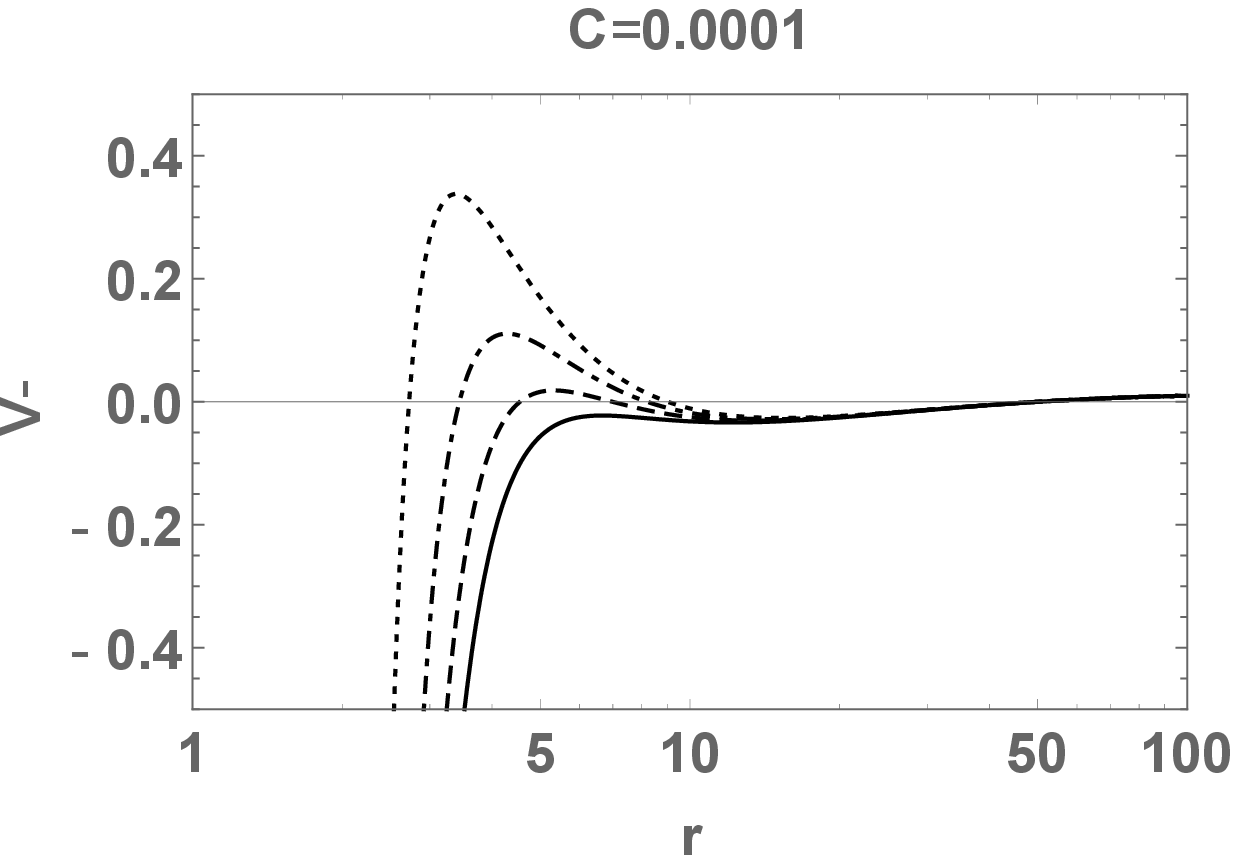}b.
		\caption{Plot of $V_+$ and $V_-$ vs. $r$ for different values of $C$ when $a=0$ (dotted line), $a=0.3$ (Dot-Dashed line), $a=0.6$ (Dashed line), and $a=0.9$ (Continuous line). In the figures we considered $Q=0.3$, $\omega_q=-2/3$. We set the values of $E=0.981$ and $L_z=\pm4.4$.\label{fig5}}
	\end{center}
\end{figure*}
\begin{figure*}[t]
\begin{center}
a.\includegraphics[scale=0.43]{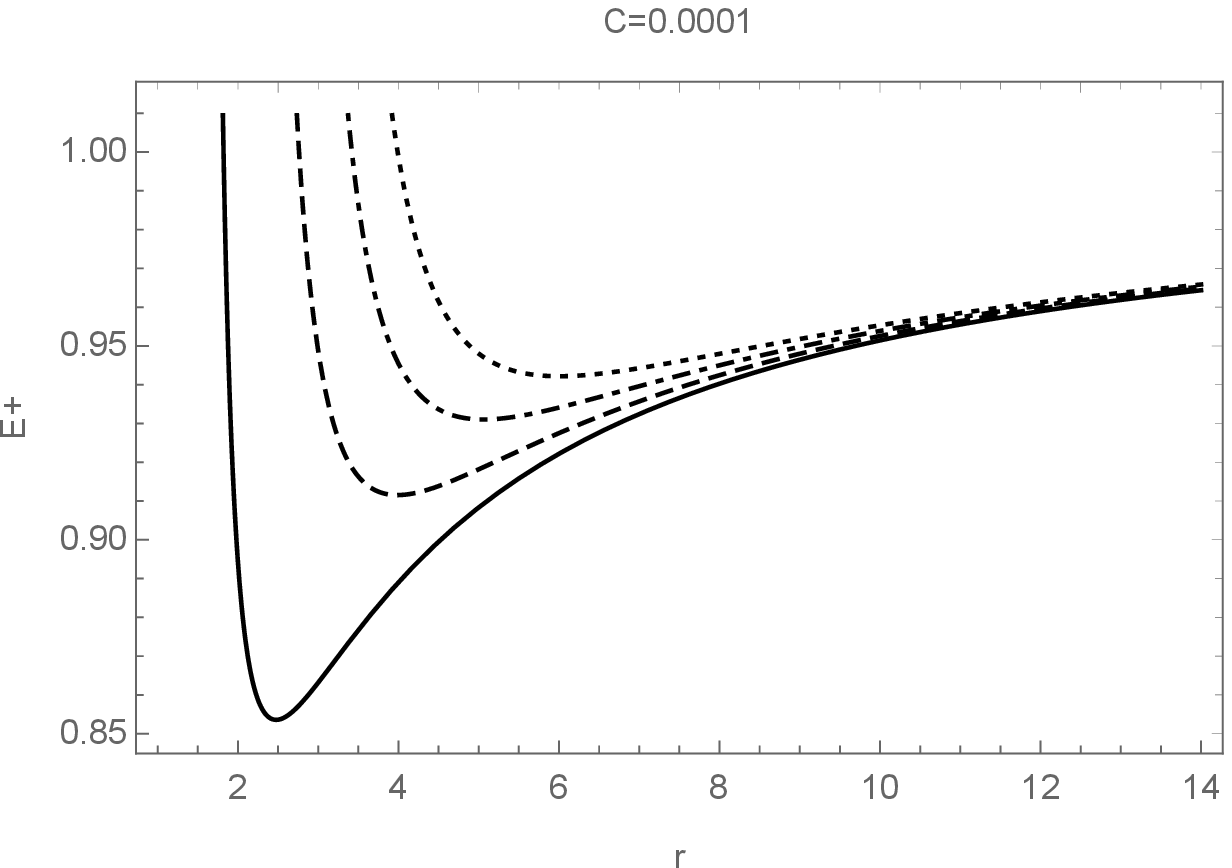}
\hspace{0.3cm}
b.\includegraphics[scale=0.43]{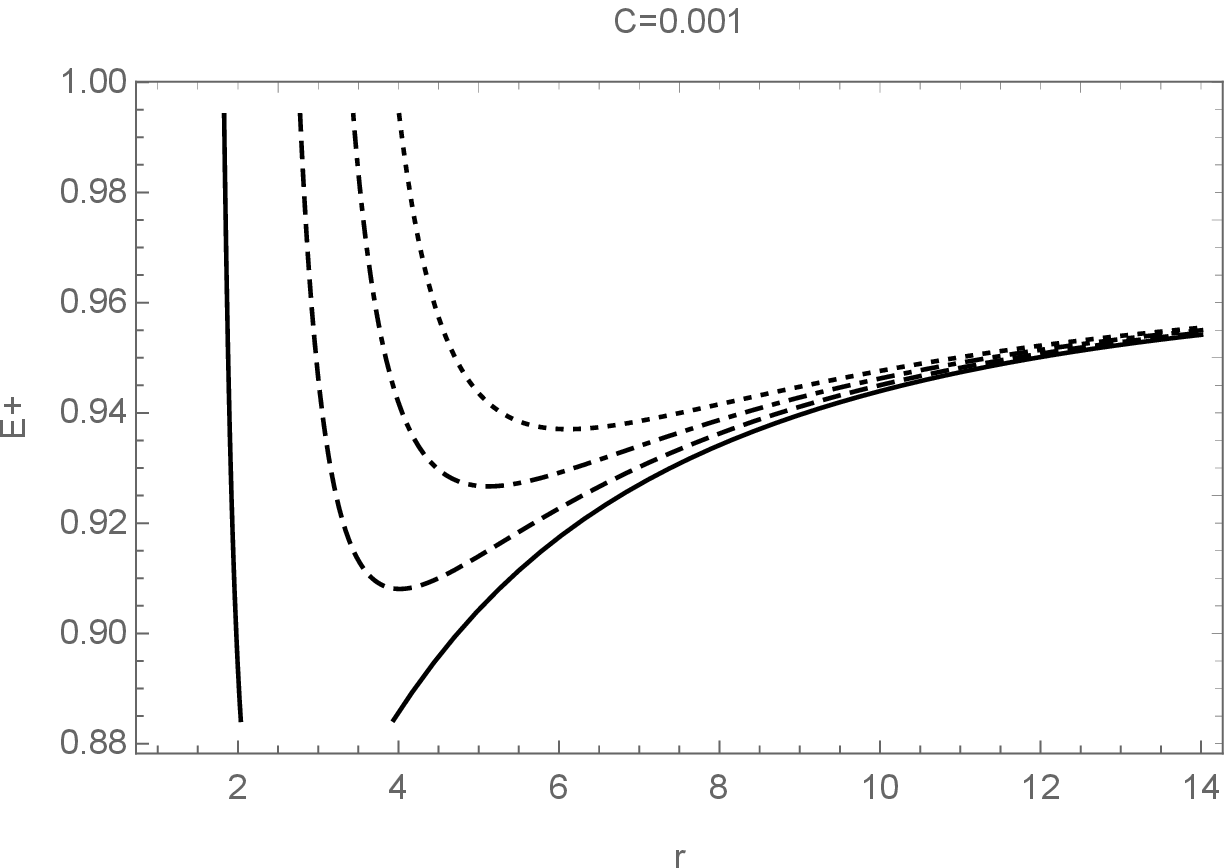}
\hspace{0.3cm}
c.\includegraphics[scale=0.43]{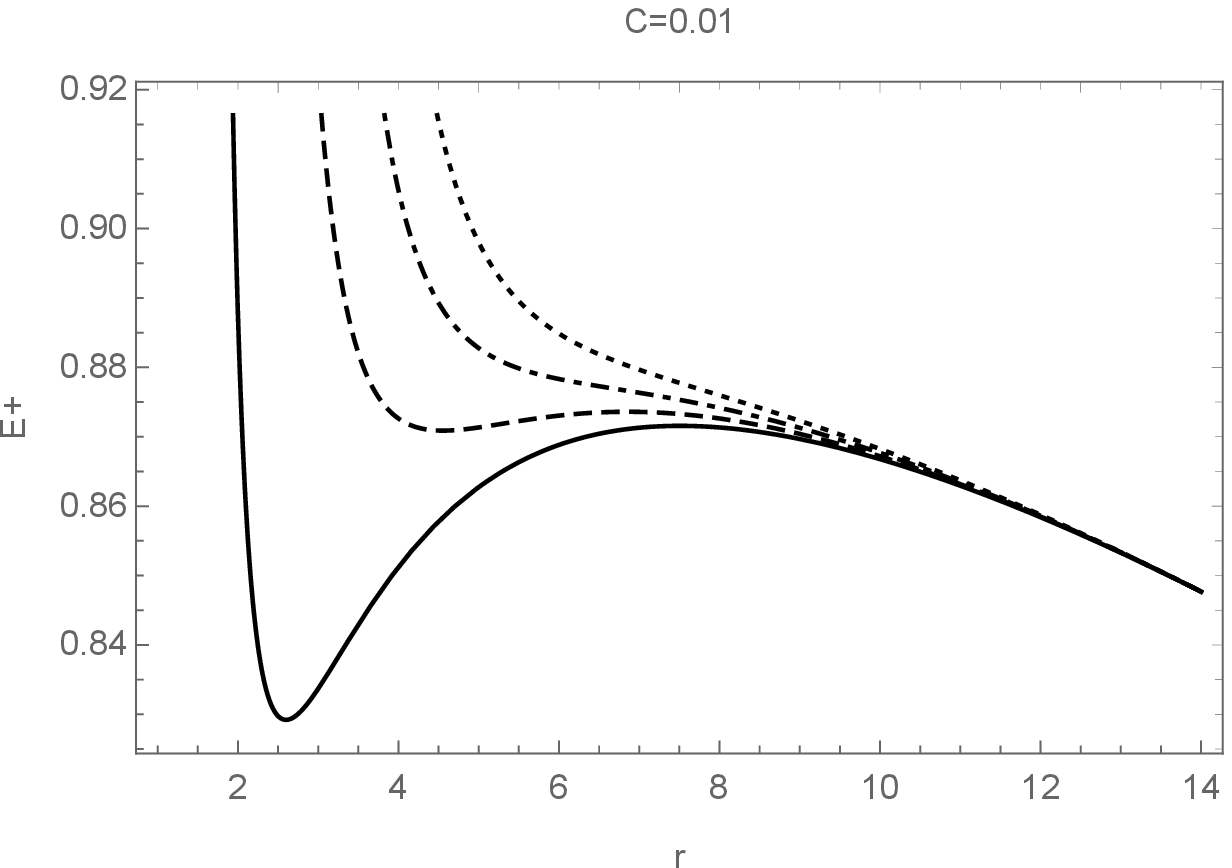}\\
d.\includegraphics[scale=0.43]{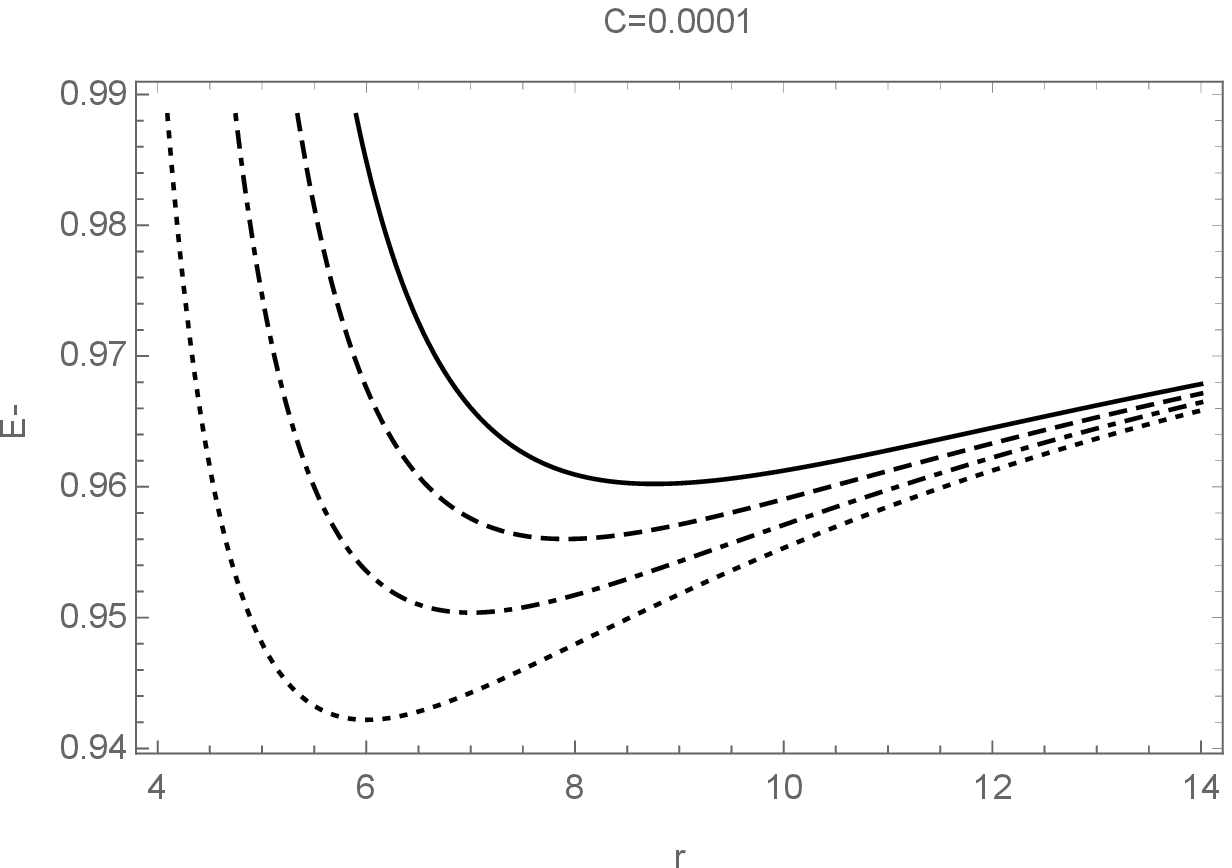}
\hspace{0.3cm}
e.\includegraphics[scale=0.43]{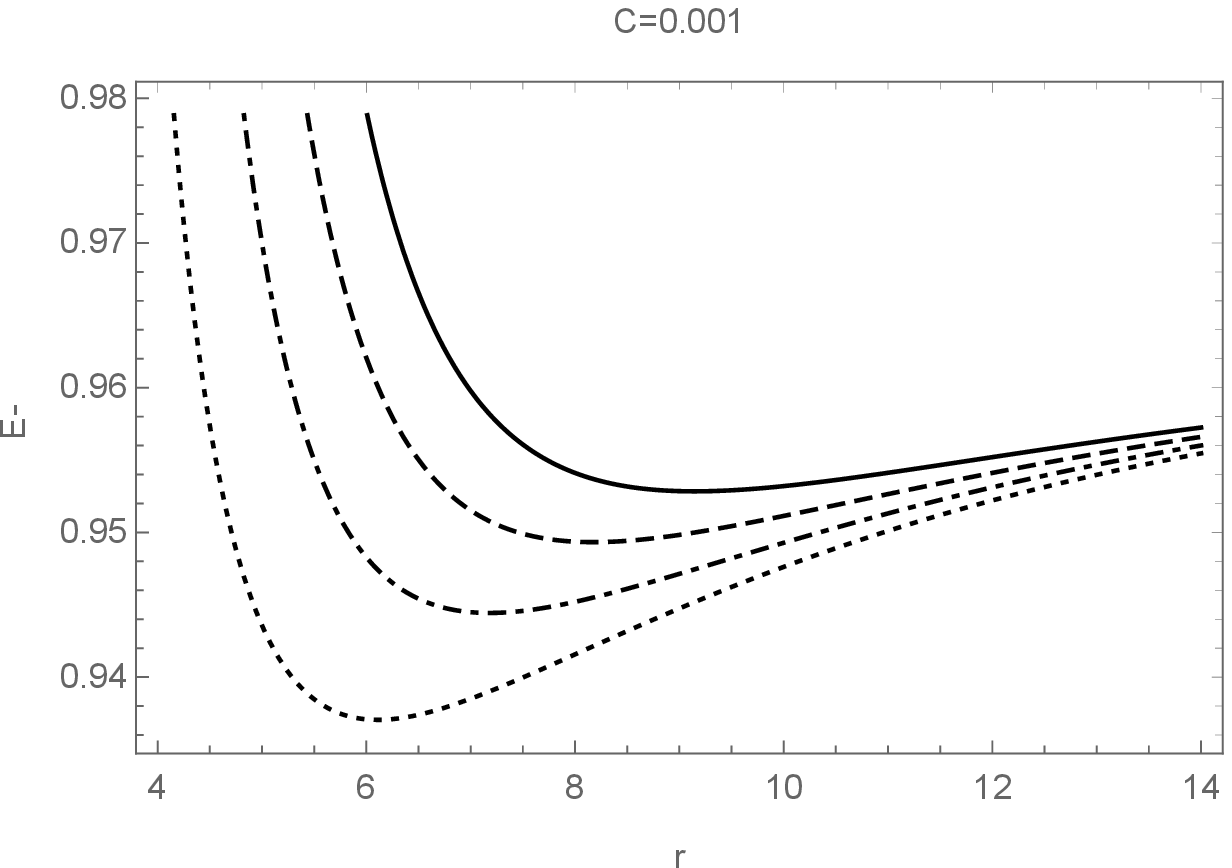}
\hspace{0.3cm}
f.\includegraphics[scale=0.43]{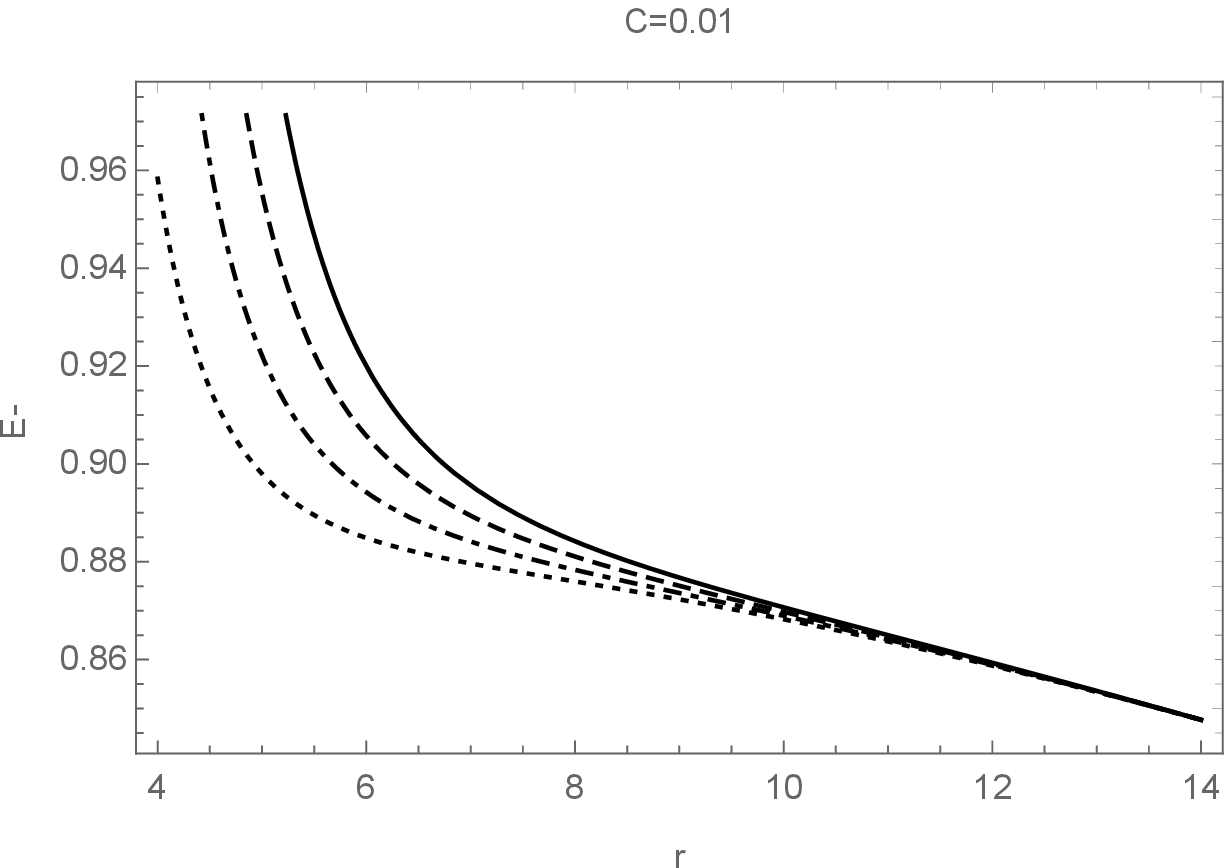}		
\caption{Plot of $E_+$ and  $E_-$ vs. $r$ for different values of $C$ when $a=0$ (dotted line), $a=0.3$ (Dot-Dashed line), $a=0.6$ (Dashed line), and $a=0.9$ (Continuous line). In all the figures we considered $Q=0.3$, $\omega_q=-2/3$.\label{fig6}}
\end{center}
\end{figure*}
Replacing Eqs.~(\ref{dott}) and (\ref{phidot1}) into Eq.~(\ref{hamiltonian}), and considering particles moving along time-like geodesics ($\kappa=-1$), we obtain \cite{Bambi17e}
\begin{equation}
\label{dotr}
g_{rr}\dot{r}^2+V_{\text{eff}}(r)=0
\end{equation}
According to equation (\ref{dotr}), the effective potential is given by
\begin{equation}
\label{effective_potential}
V_{\text{eff}}(r)=1-\frac{g_{\phi\phi}E^2+2g_{t\phi}EL_z+g_{tt}L^2_z}{g^2_{t\phi}-g_{tt}g_{\phi\phi}}.
\end{equation} 
In the equatorial plane, circular orbits are located at the zeros and turning points of the effective potential. This means that $\dot{r}=0$ from which $V_{\text{eff}}=0$, and $\ddot{r}=0$ which requires that $\partial_r V_{\text{eff}}=0$. Using these conditions, we can obtain the values of the specific energy $E$ and angular momentum $L_z$ required to have circular orbits. Nevertheless, we will follow the analysis done in Ref.~\cite{Bambi17e} to obtain these values.

The geodesic equation is given by
\begin{equation}
\label{geodesic_equation}
\frac{d^2x^\mu}{d\tau^2}+\Gamma^\mu_{\alpha\beta}\frac{dx^\alpha}{d\tau}
\frac{dx^\beta}{d\tau}=0,
\end{equation}  
where $\tau$ is an affine parameter and can be written in the form  
\begin{equation}
\label{geodesic_equation_2}
\frac{d}{d\tau}\left(g_{\nu\alpha}\dot{x}^\alpha\right)=\frac{1}{2}(\partial_\nu g_{\alpha\beta})\dot{x}^\alpha\dot{x}^\beta.
\end{equation} 
Since we are interested in equatorial circular orbits ($\dot{r}=\ddot{r}=0$), the last expression reduces to 
\begin{equation}
0=\partial_r g_{tt}\dot{t}^2+2\partial_r g_{\phi t}\dot{\phi}\dot{t}+\partial_r g_{\phi\phi}\dot{\phi}^2.
\end{equation}
Now, dividing by $\dot{t}^2$ we obtain 
\begin{equation}
\label{quadratic}
0=\partial_r g_{tt}+2\partial_r g_{\phi t}\Omega+\partial_r g_{\phi\phi}\Omega^2,
\end{equation} 
where $\Omega=\dot{\phi}/\dot{t}$ is the angular velocity. Equation (\ref{quadratic}) is a quadratic equation with solution 
\begin{equation}
\label{Omega}
\Omega_\pm=\frac{-\partial_r g_{\phi t}\pm\sqrt{(\partial_r g_{\phi t})^2-\partial_r g_{\phi\phi}\partial_r g_{tt}}}{\partial_r g_{\phi\phi}}.
\end{equation}
According to equation (\ref{Omega}), we have two possible orbits in the equatorial plane: co-rotating orbits ($\Omega_+$), where the angular momentum is parallel to the spin of the central object; and the counter-rotating orbits ($\Omega_-$), with the angular momentum anti-parallel to the spin of the central object. Moreover, Eq.~(\ref{Omega}) also shows that the absolute values of the angular velocities for co-rotating and counter-rotating circular orbits at the same radius are different from each other.

The specific energy and angular momentum in equation (\ref{momenta}) can be expressed as 
\begin{equation}
\label{EL}
\begin{aligned}
(g_{tt}+g_{t\phi}\Omega)\dot{t}&=-E\ ,\\
(g_{t\phi}+g_{\phi\phi}\Omega)\dot{t}&=L_z\ .
\end{aligned}
\end{equation}   
Now, form $g_{\mu\nu}\dot{x}^\mu\dot{x}^\nu=-1$ with $\dot{r}=\dot{\theta}=0$, we have that 
\begin{equation}
g_{tt}\dot{t}^2+g_{t\phi}\dot{t}\dot{\phi}+g_{\phi\phi}\dot{\phi}^2=-1,
\end{equation}
from which, after using $\Omega=\dot{\phi}/\dot{t}$, we obtain that 
\begin{equation}
\dot{t}=\frac{1}{\sqrt{-g_{tt}-2g_{t\phi}\Omega-g_{\phi\phi}\Omega^2}}.
\end{equation}
Therefore, in the equatorial plane, the specific energy and angular momentum in equation (\ref{EL}) reduces to \cite{Bambi17e}
\begin{equation}
\label{E_and_L_I}
\begin{aligned}
E&=-\frac{g_{tt}+g_{t\phi}\Omega_\pm}{\sqrt{-g_{tt}-2g_{t\phi}\Omega_\pm-g_{\phi\phi}\Omega^2_\pm}}\ ,\\\\
L_z&=\frac{g_{t\phi}+g_{\phi\phi}\Omega_\pm}{\sqrt{-g_{tt}-2g_{t\phi}\Omega_\pm-g_{\phi\phi}\Omega^2_\pm}}.
\end{aligned}
\end{equation}
Where we use equation (\ref{Omega}). Thus, after using the metric components, the specific energy and the angular momentum reduce to \cite{Toshmatov17}
\begin{equation}
\label{E_and_L}
\begin{aligned}
E^2_{\pm}&=\frac{8\Delta(a^2-\Delta)^2+2r\Delta\Delta '(a^2-\Delta)-a^2r^2\Delta'^2}{r^2[16\Delta(\Delta-a^2)+r\Delta'(r\Delta'-8\Delta)]}\\\\
&\pm\frac{2\sqrt{2}a\Delta\sqrt{(2a^2-2\Delta+r\Delta')^3}}{r^2[16\Delta(\Delta-a^2)+r\Delta'(r\Delta'-8\Delta)]}\,
\end{aligned}
\end{equation}

\begin{equation}
\begin{aligned}
L^2_{z\pm}&=\frac{8a^2\Delta^3-r^2(r^2+a^2)^2\Delta'^2}{r^2[16\Delta(\Delta-a^2)+r\Delta'(r\Delta'-8\Delta)]}\\\\
&-\frac{2\Delta^2[8a^2(r^2+a^2)+4r^4+a^2r\Delta']}{r^2[16\Delta(\Delta-a^2)+r\Delta'(r\Delta'-8\Delta)]}\\\\
&+\frac{2(r^2+a^2)\Delta[4a^2(r^2+a^2)+r(3r^2+a^2)\Delta']}{r^2[16\Delta(\Delta-a^2)+r\Delta'(r\Delta'-8\Delta)]}\\\\
&\pm\frac{(r^2+a^2)(2(r^2+a^2)+r\Delta')-2(2r^2+a^2)\Delta}{r^2[16\Delta(\Delta-a^2)+r\Delta'(r\Delta'-8\Delta)]}\\\\
&\times 2\sqrt{2}a\Delta\sqrt{(2a^2-2\Delta+r\Delta')^3}\ .
\end{aligned}
\end{equation}
Here, ($'$) means the derivative with respect to $r$. The sings $+$ and $-$ stand for the co-rotating and counter-rotating particle orbits, respectively.

\begin{figure*}[t]
\begin{center}
a.\includegraphics[scale=0.43]{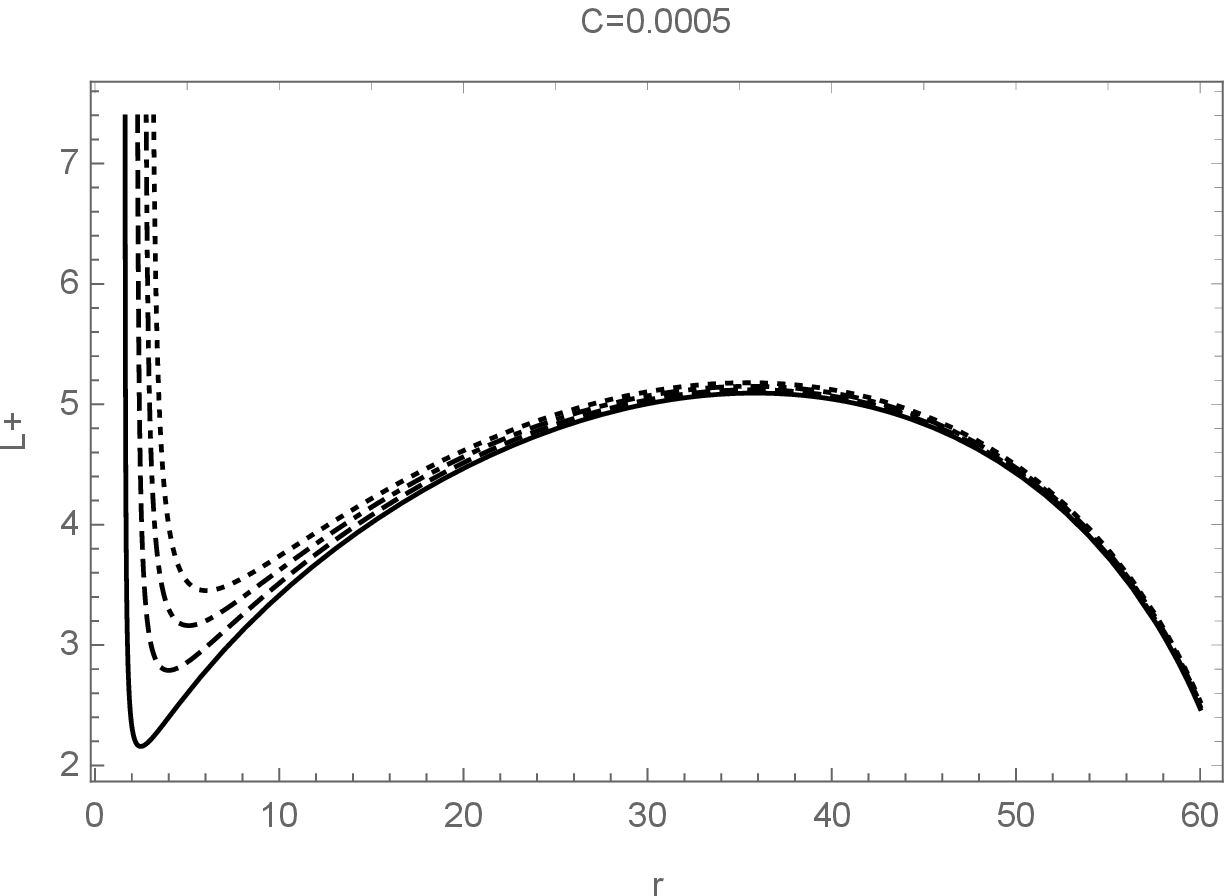}
\hspace{0.3cm}
b.\includegraphics[scale=0.43]{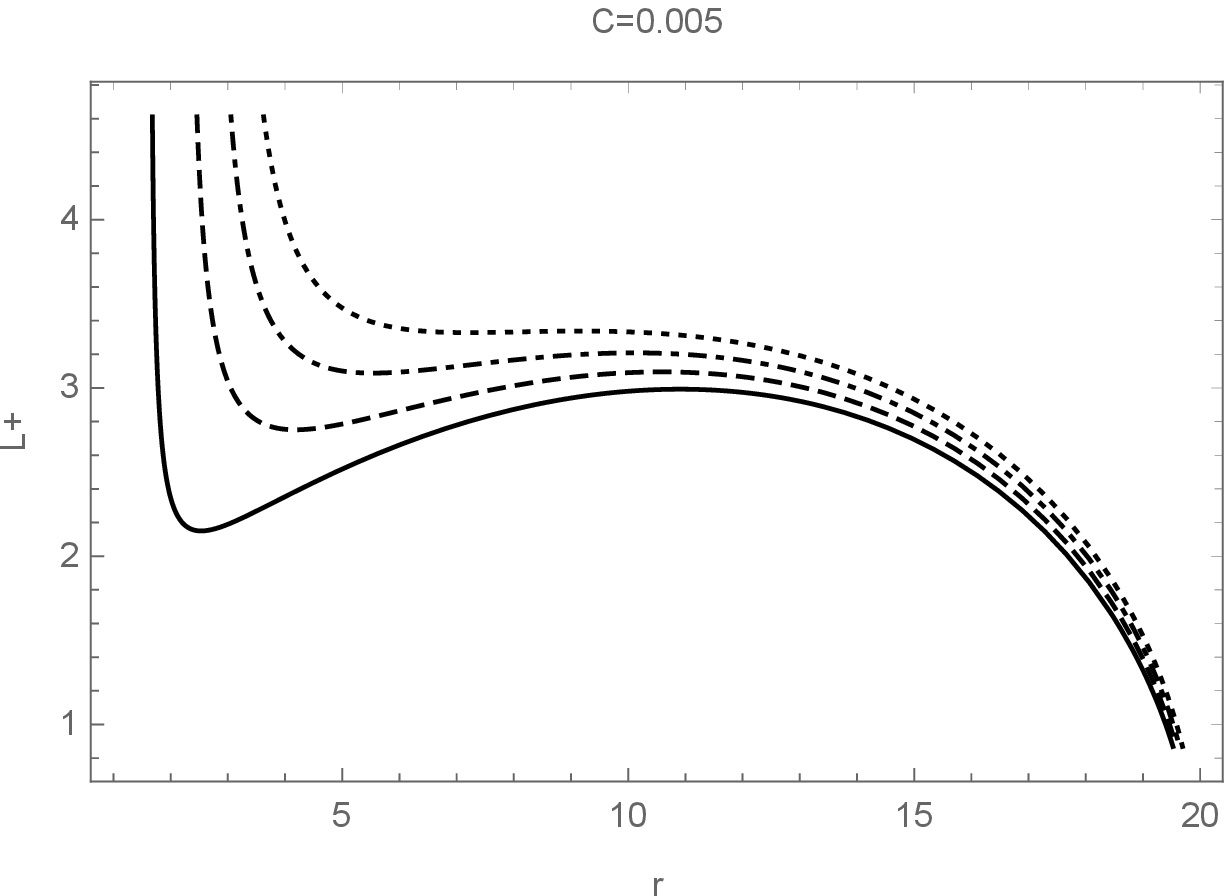}
\hspace{0.3cm}
c.\includegraphics[scale=0.43]{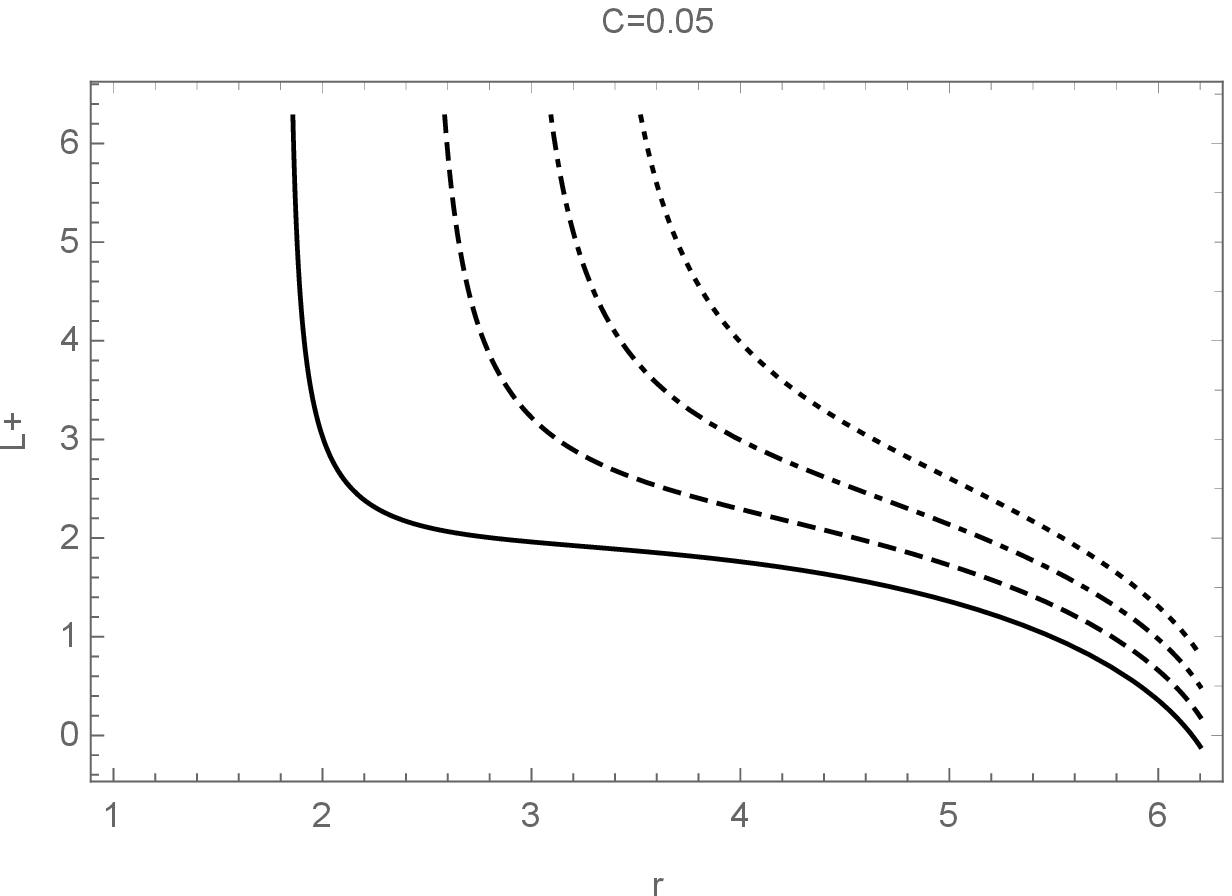}\\
d.\includegraphics[scale=0.43]{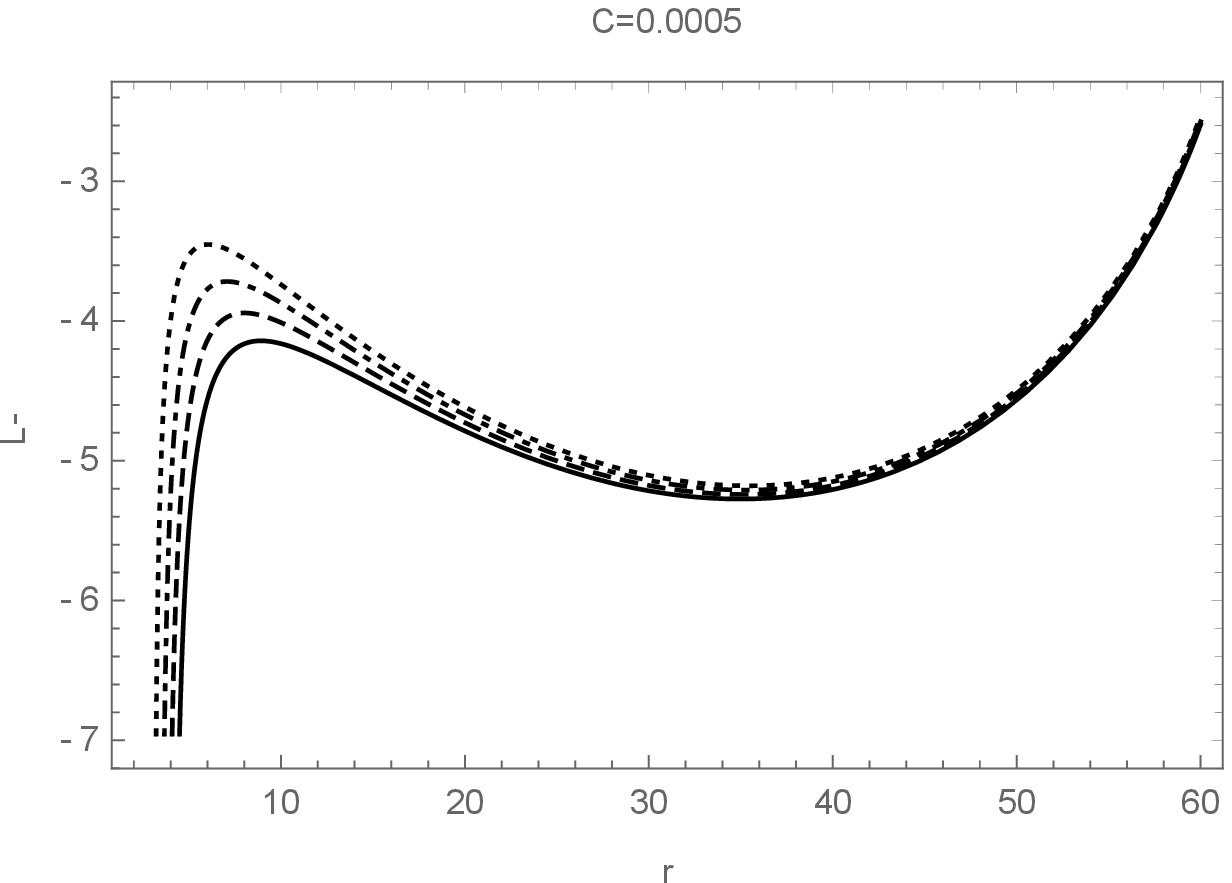}
\hspace{0.3cm}
e.\includegraphics[scale=0.43]{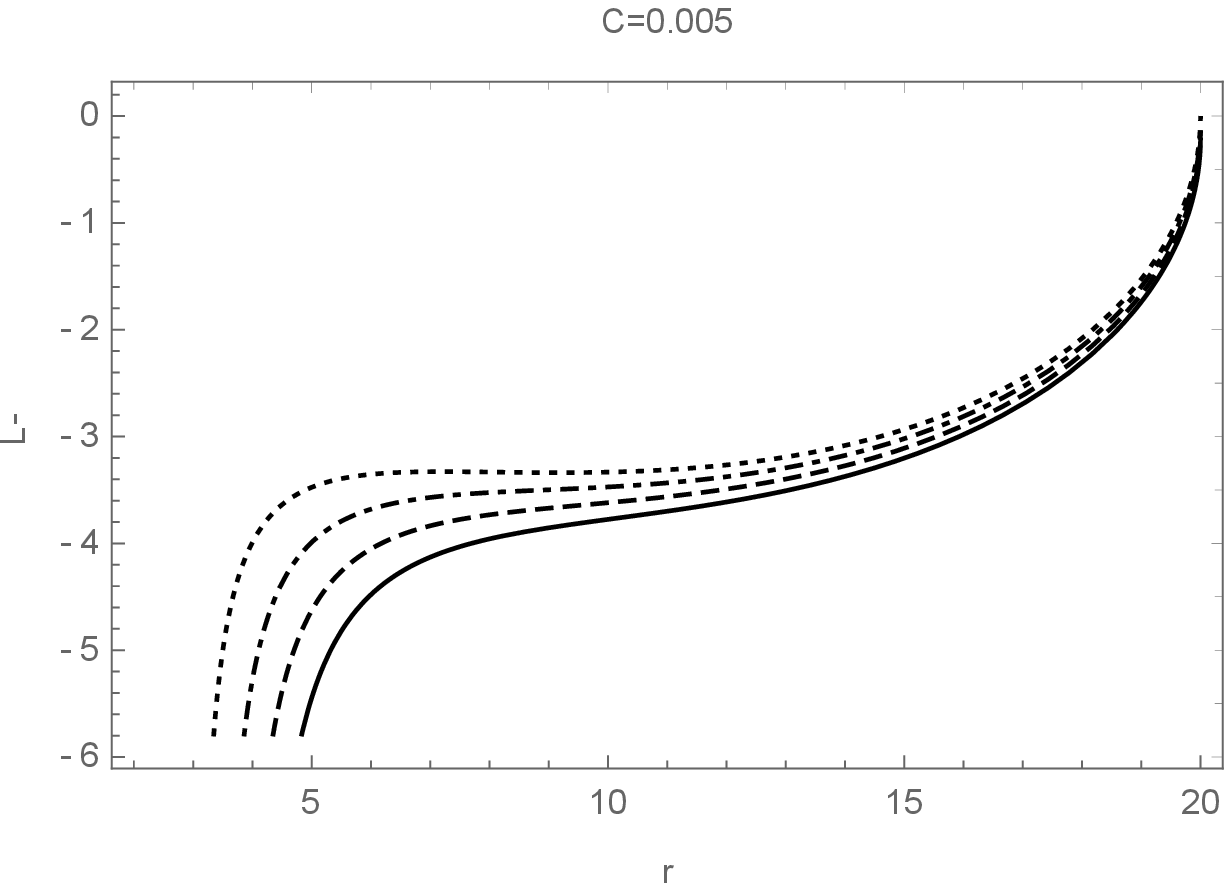}
\hspace{0.3cm}
f.\includegraphics[scale=0.43]{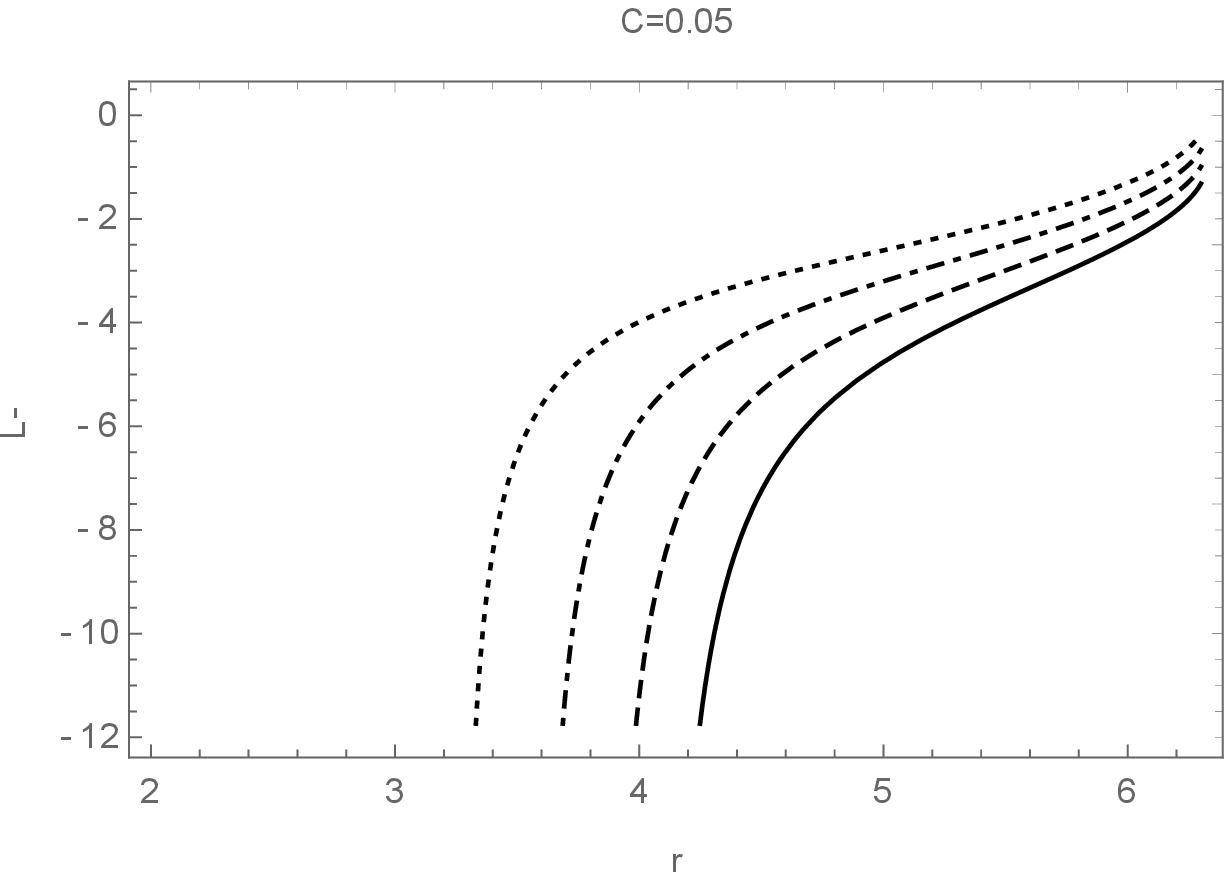}		
\caption{Plot of $L_+$ and  $L_-$ vs. $r$ for different values of $C$ when $a=0$ (dotted line), $a=0.3$ (Dot-Dashed line), $a=0.6$ (Dashed line), and $a=0.9$ (Continuous line). In all the figures we considered $M=1$, $Q=0.3$, $\omega_q=-2/3$.\label{fig7}}
\end{center}
\end{figure*}
 
In Figs.~\ref{fig5}-\ref{fig7} we plotted, respectively, the effective potential, and the specific energy and angular momentum of the co-rotating and counter-rotating circular orbits as a function of the radius for different values of the spin parameter $a$ and the quintessence parameter $C$. From Fig.~\ref{fig6} co-rotating test particles (those with $E_+$) are able to move along circular orbits that are closer to the black hole having small energy when the spin parameter is increased. The reason for such a behavior is related to the dragging effect which increases as the spin parameter increases. On the other hand, counter-rotating particles are allowed to move far from the horizon with greater specific energy than the co-rotating particles. \\

\subsection{Static radius giving limit on the existence of circular geodesics}
According to equation (\ref{E_and_L}), the specific energy $E_{\pm}$ will take real values if the condition \cite{Toshmatov17}
\begin{equation}
\label{condition}
2a^2-2\Delta-r\Delta'\geq0
\end{equation} 
is satisfied. Hence, equation (\ref{condition}) constrains the radius of the circular orbits to those values satisfying a relation of the form $r\leq r_C$, where $r_C$ is the \textit{static radius} at which the test particle can be located at an unstable-equilibrium position. It was shown in~\cite{Toshmatov17} that at this point, the gravitational attraction of the black hole is just balanced by the cosmic repulsion due to the quintessential field .\\

\begin{table*}
\label{tableII}
\centering
\begin{tabular}{|c|c|c|c|c|c|c|}
\hline
Q& $C=0.001$ & $C=0.002$ & $C=0.003$ & $C=0.01$&$C=0.02$&$C=0.03$\\
\hline
& $r_C$ & $r_C$ & $r_C$ & $r_C$&$r_C$&$r_C$\\
\hline
0.1&44.7214&31.6228&25.8199&14.1421&9.99998&8.16494 \\
\hline
0.2&44.7214&31.6228&25.8199&14.1421&9.99984& 8.16473\\
\hline
0.3&44.7213&31.6227&25.8198&14.1419&9.99946&8.16416 \\
\hline
0.4&44.7213&31.6226&25.8197&14.1415&9.99872&8.16304 \\
\hline
0.5&44.72212&31.6225&25.8195&14.1409&9.9975&8.16121 \\
\hline
0.6&44.7211&31.6223&25.8192&14.1387&9.99568&8.15847 \\ 
\hline
0.7&44.7210&31.6221&25.8189&14.137&9.99313&8.15464 \\
\hline
0.8&44.7208&31.6218&25.8184&14.1348&9.98973&8.14953 \\
\hline
0.9&44.7206&31.6213&25.8177&14.1348&9.98536&8.14294 \\
\hline
1&44.7204&31.6208&25.8169&14.1321&9.97989&8.13467 \\
\hline
\end{tabular}
\caption{\label{tab2}Values for the \textit{static radius} ($r_c$) for different values of $Q$ and $C$. We set $M=1$.}
\end{table*}

\begin{table*}
\label{tableIII}
\centering
\begin{tabular}{|c|c|c|c|c|c|c|c|c|c|c|c|c|}
\hline
\multirow{2}{*}{}&
\multicolumn{3}{|c|}{$C=0.001$} &
\multicolumn{3}{|c|}{$C=0.002$} &
\multicolumn{3}{|c|}{$C=0.003$} &
\multicolumn{3}{|c|}{$C=0.004$}\\
\hline
a& $r_{\gamma_1}$ & $r_{\gamma_2}$ & $r_{\gamma_3}$ & $r_{\gamma_1}$ & $r_{\gamma_2}$ & $r_{\gamma_3}$ & $r_{\gamma_1}$ & $r_{\gamma_2}$ & $r_{\gamma_3}$ & $r_{\gamma_1}$ & $r_{\gamma_2}$ & $r_{\gamma_3}$  \\
\hline
0.1&0.004457 &2.886697 &3.117851 &0.004457 &2.891229 &3.122380 &0.004457 &2.895790 &3.126936 &0.004457 &2.900380 &3.131518\\
\hline
0.2&0.017992 &2.763666 &3.227286 &0.017992 &2.768169 &3.231783 &0.017992 &2.772701 &3.236306 &0.017992 &2.777262 &3.240853\\
\hline
0.3&0.041119 &2.634448 &3.333277 &0.041119 &2.638900 &3.337723 &0.041119 &2.643382 &3.342191 &0.041119 &2.647893 &3.346683  \\
\hline
0.4&0.074793 &2.497713 &3.436197 &0.074793 &2.502093 &3.440572 &0.074792 &2.506502 &3.444968 &0.074792 &2.510942 &3.449384\\
\hline
0.5&0.120613 &2.351554 &3.536355 &0.120611 &2.355841 &3.540640 &0.120609 &2.360157 &3.544943 &0.120608 &2.364503&3.549265\\
\hline
0.6&0.181230 &2.193060 &3.634010 &0.181224 &2.197235 &3.638186 &0.181218 &2.201439 &3.642378 &0.181212 &2.205671 &3.646585\\
\hline
0.7&0.261301 &2.017356 &3.729381 &0.261282 &2.021405 &3.733430 &0.261263 &2.025482 &3.737491 &0.261244 &2.029587 &3.741565\\
\hline
0.8&0.370088 &1.814991 &3.822657 &0.370026 &1.818922 &3.826561 &0.369964 &1.822879 &3.830473 &0.369902 &1.826863 &3.834395\\
\hline
0.9&0.531638 &1.561754 &3.914002 &0.531400 &1.565677 &3.917741 &0.531162 &1.569624 &3.921487 &0.530925 &1.573595 &3.925237\\
\hline
1  &0.965119 &1.038336 &4.003565 &0.951561 &1.055371 &4.007114 &0.941481 &1.068951 &4.010674  &0.933191 &1.080763 &4.014236\\
\hline
\end{tabular}
\caption{\label{tab3}Values of $r_{\gamma_1}$, $r_{\gamma_2}$, and $r_{\gamma_3}$ for different values of $a$ and $C$. We consider the limit $Q<<1$.}
\end{table*}

\begin{figure*}[t]
	\begin{center}
		a.\includegraphics[scale=0.43]{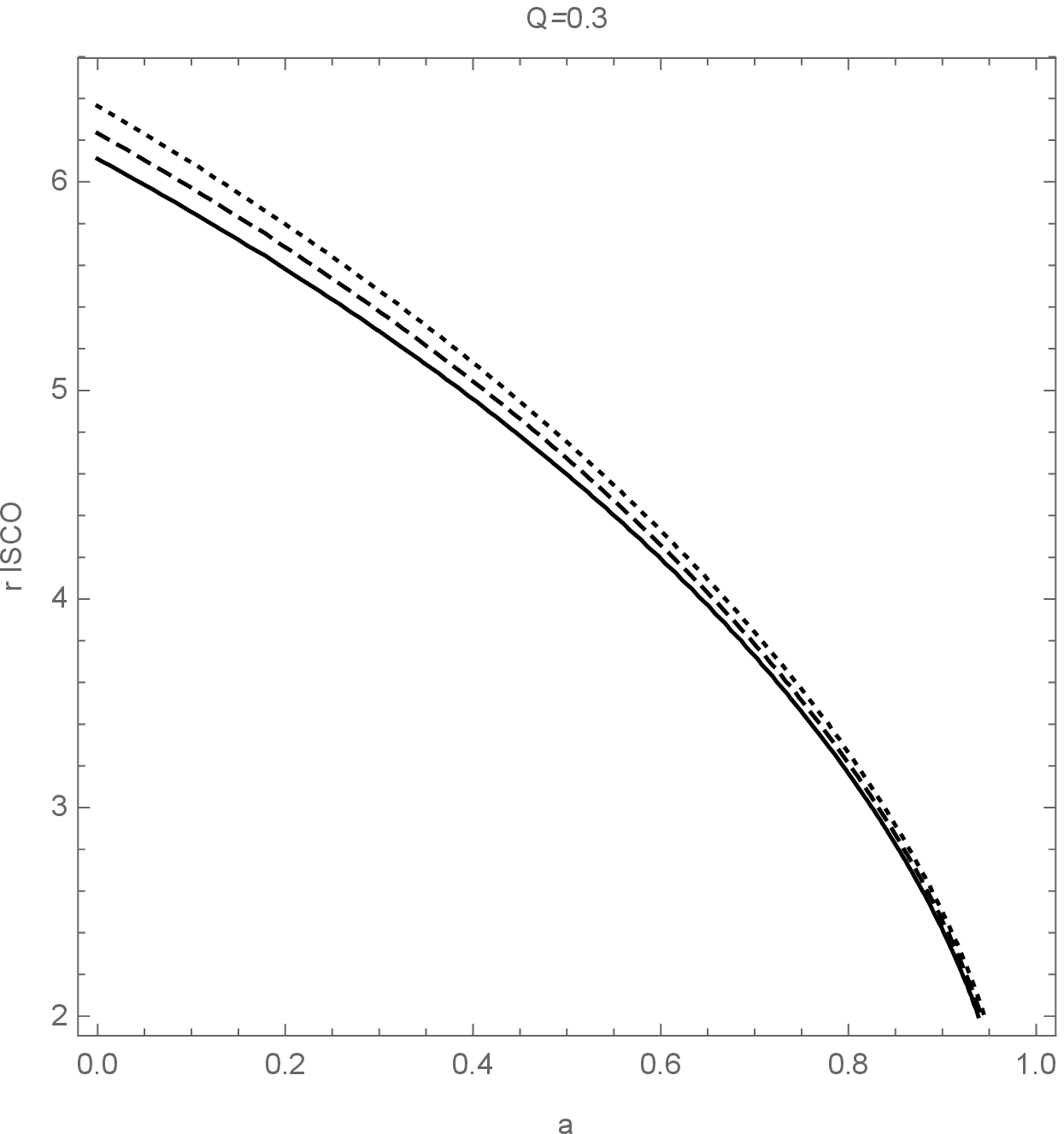}
		\hspace{0.2cm}
		b.\includegraphics[scale=0.43]{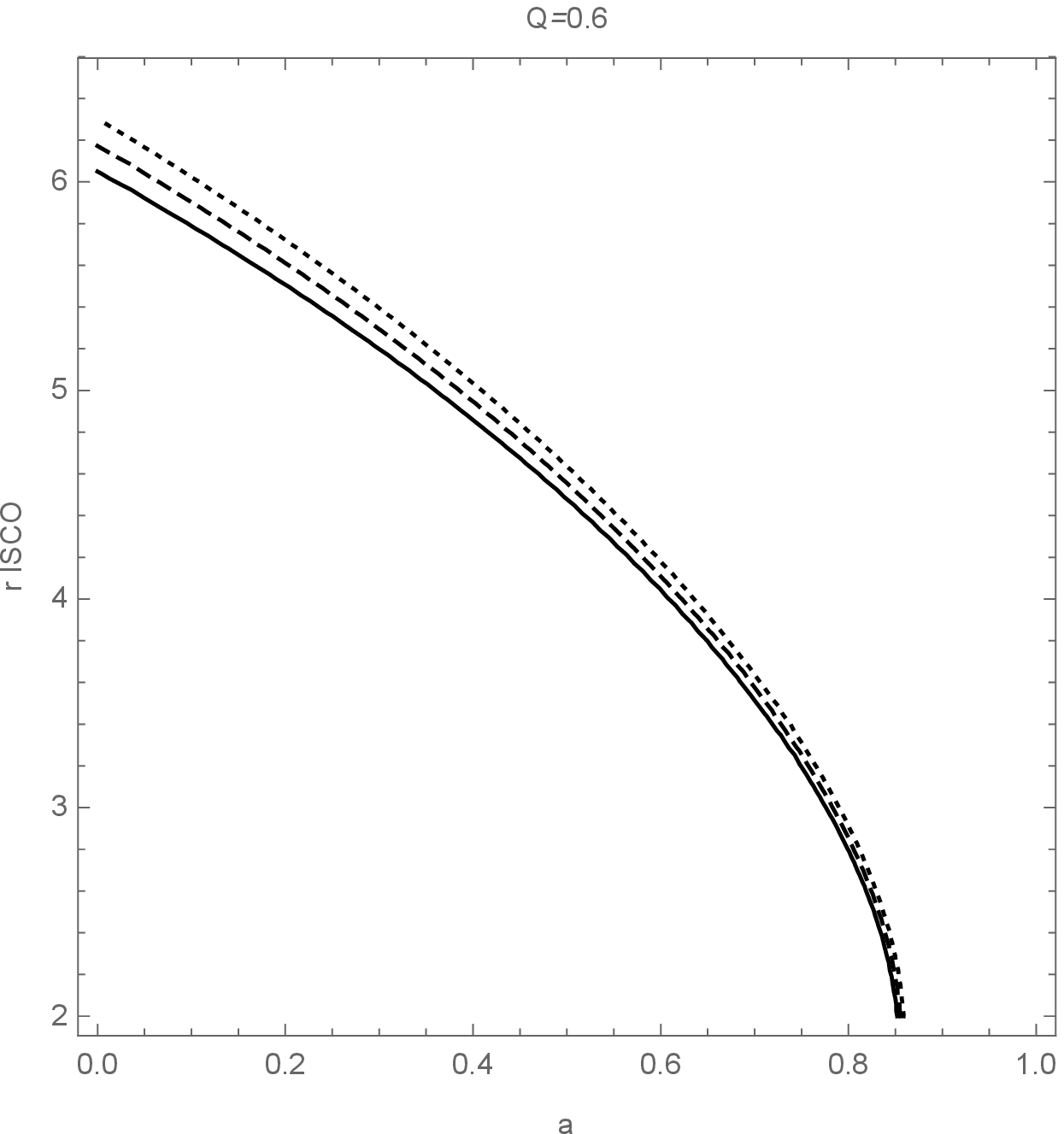}
		\hspace{0.2cm}
		c.\includegraphics[scale=0.43]{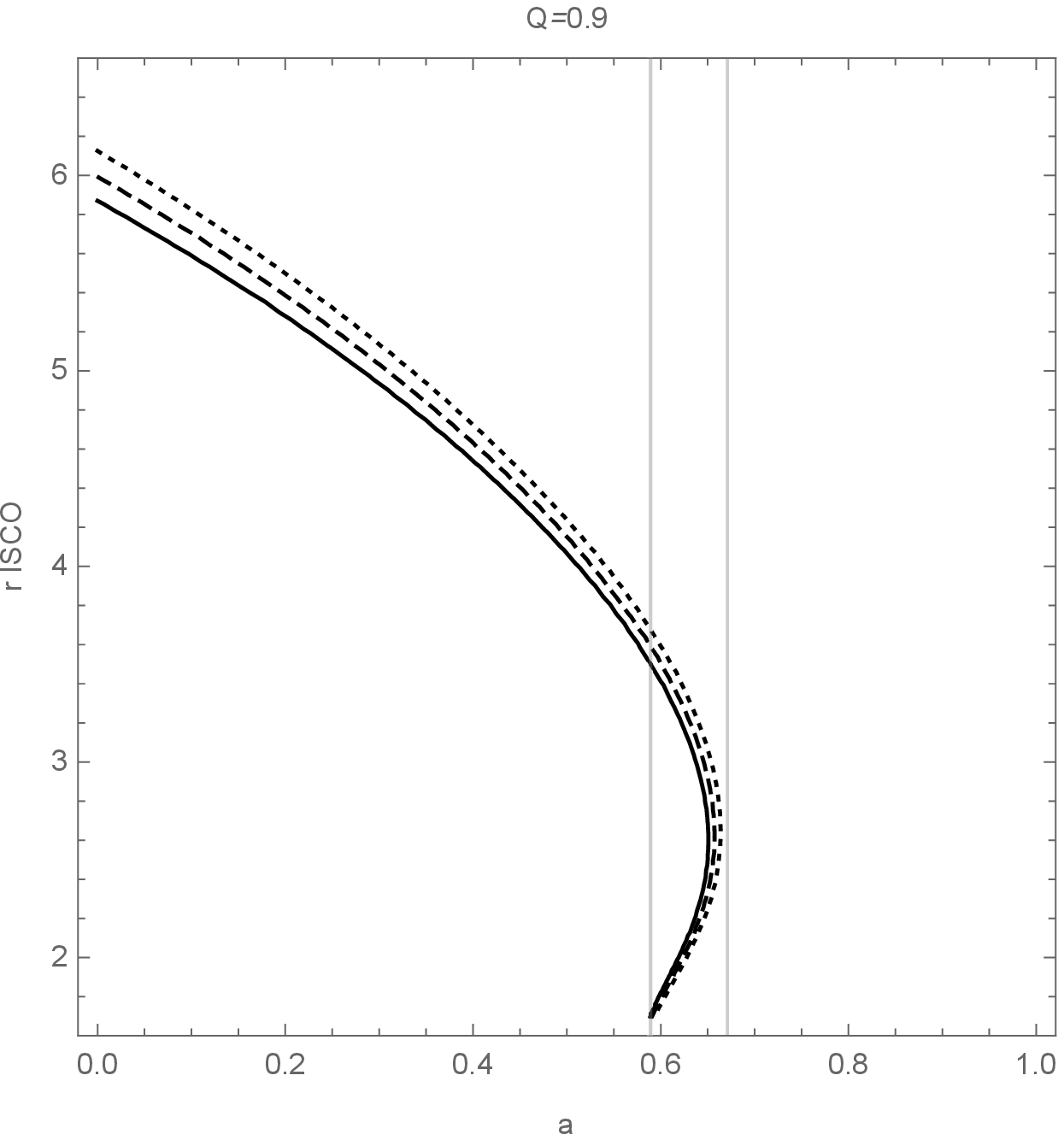}
		\caption{Plot of $r_{ISCO}$ vs. $a$ for different values of of $Q$ when $C=0.001$ (Continuous line), $C=0.002$ (Dashed line), $C=0.003$ (Dotted line). In all the figures we considered $M=1$, and $\omega_q=-2/3$.\label{fig8}}
	\end{center}
\end{figure*}
In the case of the line element (\ref{Rotating_charged_QE}) the condition in equation (\ref{condition}) reduces to 
\begin{equation}
\label{condition_QE}
Cr^6-2r^4+2CQ^3r^3+4Q^3r+CQ^6\leq 0.
\end{equation}
Note that equation (\ref{condition_QE}) reduces to the condition  
\begin{equation}
r\leq r_C\equiv\sqrt{\frac{2}{C}}\ ,
\end{equation} 
when $Q=0$ (see equation (82) in reference \cite{Toshmatov17}). Moreover, it is also important to point out that the condition in equation (\ref{condition_QE}) is independent of $a$.

In Table~\ref{tab2} we show the static radius $r_C$ for different values of $C$ and $Q$. From the table, it is possible to see that, for constant values of the charge $Q$, the static radius decreases as the quintessence parameter $C$ increases. On the other hand, for constant values of the quintessence parameter $C$, the static radius increases when the charge $Q$ decreases but not significantly. For example, when $C=0.0005$ the value of $r_C$ changes from $63.2456$ to $63.2451$, when the charge $Q$ goes from $0.1$ to $1$, respectively.

\subsection{Photon circular geodesics}
According to equation (\ref{E_and_L_I}), $E$ and $L_z$ diverge when
\begin{equation}
g_{tt}+2g_{t\phi}\Omega_\pm+g_{\phi\phi}\Omega^2_\pm=0. 
\end{equation} 
It is known that this happens at the radius of the photon orbit $r_{\gamma}$. It is usual to find circular orbits for $r > r_\gamma$ but not for $r < r_\gamma$. In such a case, the radius of the photon orbit is the equatorial circular orbit with the smallest radius for both massless and massive particles. In this sense, massive particles can reach the photon orbit in the limit of infinite specific energy $E$. However, in some particular space-times, such as quintessence space-time, circular orbits with $r< r_\gamma$ may exist, and it is also possible the presence of two or more photon orbits \cite{Bambi17e}.

From equation (\ref{E_and_L}), we find that $r_\gamma$ is given by the condition

\begin{equation}
\label{photon_orbit}
16\Delta(\Delta-a^2)+r\Delta'(r\Delta'-8\Delta)=0,
\end{equation}  
or
\begin{equation}
\label{photon_orbit_a}
\begin{aligned}
0&=8a^2CQ^{12}+(32a^2Q^9+4Q^{12})r-4CQ^{12}r^2\\
&+(32a^2CQ^9+C^2Q^{12})r^3+(48a^2Q^6+16Q^9)r^4\\
&-16CQ^9r^5+(48a^2CQ^6-24Q^6+4C^2Q^9)r^6\\
&+(24Q^6+12CQ^6)r^7-24CQ^6r^8\\
&+(32a^2CQ^3-48Q^3+6C^2Q^6)r^9\\
&+(16Q^3-16a^2+24CQ^3)r^{10}\\
&+(36-16CQ^3)r^{11}+(8a^2C-24+4C^2Q^3)r^{12}\\
&+(4+12C)r^{13}-4Cr^{14}+C^2r^{15},
\end{aligned}
\end{equation}
In order to study the behavior of $r_\gamma$, we consider the case in which $Q\ll 1$. Therefore, equation (\ref{photon_orbit_a}) reduces to 
\begin{equation}
\label{photon_orbit_b}
C^2r^5-4Cr^4+(4+12C)r^3+(8a^2C-24)r^2+36r-16a^2=0\ .
\end{equation}

In Table~\ref{tab3} we show the behavior of $r_{\gamma}$ for different values of the spin and quintessence parameters in the limit $Q<<1$. According to the table, when the spin $a$ is constant and the quintessence parameter $C$ increases, the photon radius $r_{\gamma_2}$ increases while $r_{\gamma_3}$ decreases. On the other hand, for fixed values of the quintessence parameter $C$, $r_{\gamma_2}$ decreases and $r_{\gamma_3}$ increases when the spin parameter increases.    

\subsection{Innermost Stable Circular Orbits (ISCO)}
The radius of the innermost stable circular orbit (ISCO) $r_{ISCO}$ is defined by \cite{Bambi17e}
\begin{equation}
\label{definition_ISCO}
\frac{\partial^2 V_{\text{eff}}}{\partial r^2}=0.
\end{equation}

In Fig.~\ref{fig8} we plot $r_{ISCO}$ as a function of the spin parameter $a$ for different values of $Q$ when the quintessence parameter takes the values $C=0.001, 0.002$ and $0.003$. According to the figure, $r_{ISCO}$ decreases as the spin $a$ parameter increases. Note that the difference is bigger for small values of the spin parameter $a$. A similar effect is obtained when $Q$ is increased. Moreover, the $r_{ISCO}$ increases as the value of the quintessence parameter $C$ increases. On the other hand, as Fig.\ref{fig8}c shows, there are two regions (outside the event horizon $r_+$) where the innermost stable circular orbits may exist. This feature, according to the figure, starts when the value of the spin parameter $a\approx0.589$ and ends at $a\approx0.671$. For values of $a$ greater than $0.671$ the existence of $r_{ISCO}$ is not possible.

\section{Conclusion}
In this work we have obtained the rotating solution of a non-linear magnetic-charged black hole surrounded by quintessence using a modified version of the NJA discussed in reference \cite{Toshmatov17,Azreg14}. In section \ref{sectionV}, using $\omega_q=-2/3$, we studied the behavior of the event horizon, the ergosphere, and the ZAMO for different values of the charge $Q$, the spin parameter $a$ and the quintessence parameter $C$. In the case of the event horizons, we solved equation (\ref{poly_horizon}) numerically and found three solutions: the inner, outer, and quintessence horizons. According to the results shown in table I, for constant values of the quintessence parameter $C$ and $a=0.1$, the outer horizon $r_+$ decreases as the charge $Q$ increases, while for constant values of $Q$ and $a=0.1$, the outer radius increases as $C$ increases. 

On the other hand, according to Fig.~\ref{fig1B}, we found that the existence of the outer horizon $r_+$ is constrained by the values of the charge $Q$ and the spin parameter $a$. In Fig.~\ref{fig1B} panel a, we see that $r_+$ exists for $0\leq a\leq1$; however, this interval reduces if the charge $Q$ increases ($0\leq a\leq0.58$ when $Q=0.9$). Therefore, $r_+$ strongly depends on the charge $Q$ and the spin parameter $a$. In the same way, we found that shape and size of the horizon and the ergosphere depend on the charge $Q$, the spin parameter $a$, and the quintessence parameter $C$. According to Fig.~\ref{fig3}, when the spin parameter $a$ is increased, the radius of the event horizon decreases and the area of the ergo-region increases. A similar behavior was observed when the charge $Q$ was increased. In addition, by comparing Figs.~\ref{fig3}a, b ,and c with Figs.~\ref{fig3}d, e, and f, we conclude that the increment of the ergo-region was greater when we increased the spin parameter $a$. 

On the other hand, we found that the ergosphere radius does depend on the charge $Q$ at the equatorial plane (see Figs.~\ref{fig3} d, e and f) but does not depend on the spin parameter (see Figs.\ref{fig3}a, b and c) when the spin parameter and the charge are kept constant. This result agrees with that obtained in~\cite{Toshmatov17}.

In section \ref{sectionVI}, we studied the effective potential, and the specific energy and angular momentum of the co-rotating and counter-rotating circular orbits. We found that co-rotating test particles (those with $E _+$) are able to move along circular orbits that are closer to the black hole having small energy when the spin parameter is increased (see Figs.~\ref{fig6}a, b, and c). This is due to the dependence of $r_+$ on the charge and the spin parameter as we have shown in Fig.~\ref{fig3}.a, b, and c. On the other hand, we found that counter-rotating particles are allowed to move far from the horizon with greater specific energy than the co-rotating particles.

In this work we also found that the photon circular orbits do not depend strongly on the charge $Q$. The contribution in equation (\ref{photon_orbit_a}) due to the charge is small even if we consider the interval $0\leq Q<1$. Therefore, equation (\ref{photon_orbit_b}) is in agreement with the limit imposed by the static radius $r_C$. This can be seen easily when comparing Tables~\ref{tab2} and \ref{tab3} for $C=0.003$.

Finally, we found that the innermost stable circular orbits are constrained to the values of the charge $Q$. For example, for $Q=0.3$, Fig.~\ref{fig8} shows that $r_{ISCO}$ exists only for values of $a$ less than $0.95$. When $Q=0.9$, we see that $r_{ISCO}$ exists for values of the spin parameter less than $0.7$. Moreover, the $r_{ISCO}$ increases as the value of the quintessence parameter $C$ increases keeping the constraint (approximately) with the spin parameter $a$.   \\

\begin{acknowledgments}
The authors thank Hrishikesh Chakrabarty, Askar Abdikamalov, Pritom Mondal, and Dimitry Ayzenberg for helpful suggestions. This work was supported by the Innovation Program of the Shanghai Municipal Education Commission, Grant No.~2019-01-07-00-07-E00035, and the National Natural Science Foundation of China (NSFC), Grant No.~11973019. C.A.B.G. also acknowledges support from the China Scholarship Council (CSC), grant No.~2017GXZ019022.  This research
is supported by Grants No. VA-FA-F-2-008, No. MRB-AN-2019-29
and No. YFA-Ftech-2018-8 of the Uzbekistan Ministry for Innovative
Development, and by the Abdus Salam International Centre for Theoretical
Physics through GrantNo.OEA-NT-01. This research is partially
supported by an Erasmus+ exchange grant between SU and NUUz. A.A. thanks Silesian University in Opava and the Nazarbayev University for hospitality. 
\end{acknowledgments}

\appendix


\bibliography{gravreferences}

\begin{thebibliography}{51}%
\makeatletter
\providecommand \@ifxundefined [1]{%
 \@ifx{#1\undefined}
}%
\providecommand \@ifnum [1]{%
 \ifnum #1\expandafter \@firstoftwo
 \else \expandafter \@secondoftwo
 \fi
}%
\providecommand \@ifx [1]{%
 \ifx #1\expandafter \@firstoftwo
 \else \expandafter \@secondoftwo
 \fi
}%
\providecommand \natexlab [1]{#1}%
\providecommand \enquote  [1]{``#1''}%
\providecommand \bibnamefont  [1]{#1}%
\providecommand \bibfnamefont [1]{#1}%
\providecommand \citenamefont [1]{#1}%
\providecommand \href@noop [0]{\@secondoftwo}%
\providecommand \href [0]{\begingroup \@sanitize@url \@href}%
\providecommand \@href[1]{\@@startlink{#1}\@@href}%
\providecommand \@@href[1]{\endgroup#1\@@endlink}%
\providecommand \@sanitize@url [0]{\catcode `\\12\catcode `\$12\catcode
  `\&12\catcode `\#12\catcode `\^12\catcode `\_12\catcode `\%12\relax}%
\providecommand \@@startlink[1]{}%
\providecommand \@@endlink[0]{}%
\providecommand \url  [0]{\begingroup\@sanitize@url \@url }%
\providecommand \@url [1]{\endgroup\@href {#1}{\urlprefix }}%
\providecommand \urlprefix  [0]{URL }%
\providecommand \Eprint [0]{\href }%
\providecommand \doibase [0]{http://dx.doi.org/}%
\providecommand \selectlanguage [0]{\@gobble}%
\providecommand \bibinfo  [0]{\@secondoftwo}%
\providecommand \bibfield  [0]{\@secondoftwo}%
\providecommand \translation [1]{[#1]}%
\providecommand \BibitemOpen [0]{}%
\providecommand \bibitemStop [0]{}%
\providecommand \bibitemNoStop [0]{.\EOS\space}%
\providecommand \EOS [0]{\spacefactor3000\relax}%
\providecommand \BibitemShut  [1]{\csname bibitem#1\endcsname}%
\let\auto@bib@innerbib\@empty
\bibitem [{\citenamefont {Bambi}\ and\ \citenamefont
  {Dolgov}(2016)}]{Bambi16a}%
  \BibitemOpen
  \bibfield  {author} {\bibinfo {author} {\bibfnamefont {C.}~\bibnamefont
  {Bambi}}\ and\ \bibinfo {author} {\bibfnamefont {A.~D.}\ \bibnamefont
  {Dolgov}},\ }\href@noop {} {\emph {\bibinfo {title} {Introduction to Particle
  Cosmology: The Standard Model of Cosmology and its Open Problems}}}\
  (\bibinfo  {publisher} {Springer-Verlag},\ \bibinfo {address} {Berlin
  Heidelberg},\ \bibinfo {year} {2016})\BibitemShut {NoStop}%
\bibitem [{\citenamefont {{Dodelson}}(2003)}]{Dodelson03}%
  \BibitemOpen
  \bibfield  {author} {\bibinfo {author} {\bibfnamefont {S.}~\bibnamefont
  {{Dodelson}}},\ }\href@noop {} {\emph {\bibinfo {title} {Modern Cosmology}}}\
  (\bibinfo  {publisher} {Academic Press},\ \bibinfo {year} {2003})\BibitemShut
  {NoStop}%
\bibitem [{\citenamefont {{Planck Collaboration}}\ \emph
  {et~al.}(2014)\citenamefont {{Planck Collaboration}}, \citenamefont {{Ade}},
  \citenamefont {{Aghanim}}, \citenamefont {{Armitage-Caplan}}, \citenamefont
  {{Arnaud}}, \citenamefont {{Ashdown}}, \citenamefont {{Atrio-Barandela}},
  \citenamefont {{Aumont}}, \citenamefont {{Baccigalupi}}, \citenamefont
  {{Banday}},\ and\ \citenamefont {et~al.}}]{Planck14}%
  \BibitemOpen
  \bibfield  {author} {\bibinfo {author} {\bibnamefont {{Planck
  Collaboration}}}, \bibinfo {author} {\bibfnamefont {P.~A.~R.}\ \bibnamefont
  {{Ade}}}, \bibinfo {author} {\bibfnamefont {N.}~\bibnamefont {{Aghanim}}},
  \bibinfo {author} {\bibfnamefont {C.}~\bibnamefont {{Armitage-Caplan}}},
  \bibinfo {author} {\bibfnamefont {M.}~\bibnamefont {{Arnaud}}}, \bibinfo
  {author} {\bibfnamefont {M.}~\bibnamefont {{Ashdown}}}, \bibinfo {author}
  {\bibfnamefont {F.}~\bibnamefont {{Atrio-Barandela}}}, \bibinfo {author}
  {\bibfnamefont {J.}~\bibnamefont {{Aumont}}}, \bibinfo {author}
  {\bibfnamefont {C.}~\bibnamefont {{Baccigalupi}}}, \bibinfo {author}
  {\bibfnamefont {A.~J.}\ \bibnamefont {{Banday}}}, \ and\ \bibinfo {author}
  {\bibnamefont {et~al.}},\ }\href {\doibase 10.1051/0004-6361/201321591}
  {\bibfield  {journal} {\bibinfo  {journal} {Astronomy and Astrophysics}\
  }\textbf {\bibinfo {volume} {571}},\ \bibinfo {eid} {A16} (\bibinfo {year}
  {2014})},\ \Eprint {http://arxiv.org/abs/1303.5076} {arXiv:1303.5076}
  \BibitemShut {NoStop}%
\bibitem [{\citenamefont {{Kazanas}}(1980)}]{Kazanas80}%
  \BibitemOpen
  \bibfield  {author} {\bibinfo {author} {\bibfnamefont {D.}~\bibnamefont
  {{Kazanas}}},\ }\href {\doibase 10.1086/183361} {\bibfield  {journal}
  {\bibinfo  {journal} {Astrophys. J. Lett.}\ }\textbf {\bibinfo {volume}
  {241}},\ \bibinfo {pages} {L59} (\bibinfo {year} {1980})}\BibitemShut
  {NoStop}%
\bibitem [{\citenamefont {{Starobinsky}}(1980)}]{Starobinsky80}%
  \BibitemOpen
  \bibfield  {author} {\bibinfo {author} {\bibfnamefont {A.~A.}\ \bibnamefont
  {{Starobinsky}}},\ }\href {\doibase 10.1016/0370-2693(80)90670-X} {\bibfield
  {journal} {\bibinfo  {journal} {Physics Letters B}\ }\textbf {\bibinfo
  {volume} {91}},\ \bibinfo {pages} {99} (\bibinfo {year} {1980})}\BibitemShut
  {NoStop}%
\bibitem [{\citenamefont {{Guth}}(1981)}]{Guth81}%
  \BibitemOpen
  \bibfield  {author} {\bibinfo {author} {\bibfnamefont {A.~H.}\ \bibnamefont
  {{Guth}}},\ }\href {\doibase 10.1103/PhysRevD.23.347} {\bibfield  {journal}
  {\bibinfo  {journal} {Phys. Rev. D}\ }\textbf {\bibinfo {volume} {23}},\
  \bibinfo {pages} {347} (\bibinfo {year} {1981})}\BibitemShut {NoStop}%
\bibitem [{\citenamefont {{Guth}}\ and\ \citenamefont {{Pi}}(1982)}]{Guth82}%
  \BibitemOpen
  \bibfield  {author} {\bibinfo {author} {\bibfnamefont {A.~H.}\ \bibnamefont
  {{Guth}}}\ and\ \bibinfo {author} {\bibfnamefont {S.-Y.}\ \bibnamefont
  {{Pi}}},\ }\href {\doibase 10.1103/PhysRevLett.49.1110} {\bibfield  {journal}
  {\bibinfo  {journal} {Physical Review Letters}\ }\textbf {\bibinfo {volume}
  {49}},\ \bibinfo {pages} {1110} (\bibinfo {year} {1982})}\BibitemShut
  {NoStop}%
\bibitem [{\citenamefont {{Linde}}(1982)}]{Linde82}%
  \BibitemOpen
  \bibfield  {author} {\bibinfo {author} {\bibfnamefont {A.~D.}\ \bibnamefont
  {{Linde}}},\ }\href {\doibase 10.1016/0370-2693(82)91219-9} {\bibfield
  {journal} {\bibinfo  {journal} {Physics Letters B}\ }\textbf {\bibinfo
  {volume} {108}},\ \bibinfo {pages} {389} (\bibinfo {year}
  {1982})}\BibitemShut {NoStop}%
\bibitem [{\citenamefont {{Albrecht}}\ and\ \citenamefont
  {{Steinhardt}}(1982)}]{Albrecht82}%
  \BibitemOpen
  \bibfield  {author} {\bibinfo {author} {\bibfnamefont {A.}~\bibnamefont
  {{Albrecht}}}\ and\ \bibinfo {author} {\bibfnamefont {P.~J.}\ \bibnamefont
  {{Steinhardt}}},\ }\href {\doibase 10.1103/PhysRevLett.48.1220} {\bibfield
  {journal} {\bibinfo  {journal} {Physical Review Letters}\ }\textbf {\bibinfo
  {volume} {48}},\ \bibinfo {pages} {1220} (\bibinfo {year}
  {1982})}\BibitemShut {NoStop}%
\bibitem [{\citenamefont {{Riess}}\ \emph {et~al.}(1998)\citenamefont
  {{Riess}}, \citenamefont {{Filippenko}}, \citenamefont {{Challis}},
  \citenamefont {{Clocchiatti}}, \citenamefont {{Diercks}}, \citenamefont
  {{Garnavich}}, \citenamefont {{Gilliland}}, \citenamefont {{Hogan}},
  \citenamefont {{Jha}}, \citenamefont {{Kirshner}}, \citenamefont
  {{Leibundgut}}, \citenamefont {{Phillips}}, \citenamefont {{Reiss}},
  \citenamefont {{Schmidt}}, \citenamefont {{Schommer}}, \citenamefont
  {{Smith}}, \citenamefont {{Spyromilio}}, \citenamefont {{Stubbs}},
  \citenamefont {{Suntzeff}},\ and\ \citenamefont {{Tonry}}}]{Riess98}%
  \BibitemOpen
  \bibfield  {author} {\bibinfo {author} {\bibfnamefont {A.~G.}\ \bibnamefont
  {{Riess}}}, \bibinfo {author} {\bibfnamefont {A.~V.}\ \bibnamefont
  {{Filippenko}}}, \bibinfo {author} {\bibfnamefont {P.}~\bibnamefont
  {{Challis}}}, \bibinfo {author} {\bibfnamefont {A.}~\bibnamefont
  {{Clocchiatti}}}, \bibinfo {author} {\bibfnamefont {A.}~\bibnamefont
  {{Diercks}}}, \bibinfo {author} {\bibfnamefont {P.~M.}\ \bibnamefont
  {{Garnavich}}}, \bibinfo {author} {\bibfnamefont {R.~L.}\ \bibnamefont
  {{Gilliland}}}, \bibinfo {author} {\bibfnamefont {C.~J.}\ \bibnamefont
  {{Hogan}}}, \bibinfo {author} {\bibfnamefont {S.}~\bibnamefont {{Jha}}},
  \bibinfo {author} {\bibfnamefont {R.~P.}\ \bibnamefont {{Kirshner}}},
  \bibinfo {author} {\bibfnamefont {B.}~\bibnamefont {{Leibundgut}}}, \bibinfo
  {author} {\bibfnamefont {M.~M.}\ \bibnamefont {{Phillips}}}, \bibinfo
  {author} {\bibfnamefont {D.}~\bibnamefont {{Reiss}}}, \bibinfo {author}
  {\bibfnamefont {B.~P.}\ \bibnamefont {{Schmidt}}}, \bibinfo {author}
  {\bibfnamefont {R.~A.}\ \bibnamefont {{Schommer}}}, \bibinfo {author}
  {\bibfnamefont {R.~C.}\ \bibnamefont {{Smith}}}, \bibinfo {author}
  {\bibfnamefont {J.}~\bibnamefont {{Spyromilio}}}, \bibinfo {author}
  {\bibfnamefont {C.}~\bibnamefont {{Stubbs}}}, \bibinfo {author}
  {\bibfnamefont {N.~B.}\ \bibnamefont {{Suntzeff}}}, \ and\ \bibinfo {author}
  {\bibfnamefont {J.}~\bibnamefont {{Tonry}}},\ }\href {\doibase
  10.1086/300499} {\bibfield  {journal} {\bibinfo  {journal} {Astrophys. J.}\
  }\textbf {\bibinfo {volume} {116}},\ \bibinfo {pages} {1009} (\bibinfo {year}
  {1998})},\ \Eprint {http://arxiv.org/abs/astro-ph/9805201} {astro-ph/9805201}
  \BibitemShut {NoStop}%
\bibitem [{\citenamefont {{Perlmutter}}\ and\ \citenamefont {{et
  al.}}(1999)}]{Perlmutter99}%
  \BibitemOpen
  \bibfield  {author} {\bibinfo {author} {\bibfnamefont {S.}~\bibnamefont
  {{Perlmutter}}}\ and\ \bibinfo {author} {\bibnamefont {{et al.}}},\ }\href
  {\doibase 10.1086/307221} {\bibfield  {journal} {\bibinfo  {journal}
  {Astrophys. J.}\ }\textbf {\bibinfo {volume} {517}},\ \bibinfo {pages} {565}
  (\bibinfo {year} {1999})},\ \Eprint {http://arxiv.org/abs/astro-ph/9812133}
  {astro-ph/9812133} \BibitemShut {NoStop}%
\bibitem [{\citenamefont {{Yang}}\ \emph {et~al.}(2018)\citenamefont {{Yang}},
  \citenamefont {{Shahalam}}, \citenamefont {{Pal}}, \citenamefont {{Pan}},\
  and\ \citenamefont {{Wang}}}]{Yang18}%
  \BibitemOpen
  \bibfield  {author} {\bibinfo {author} {\bibfnamefont {W.}~\bibnamefont
  {{Yang}}}, \bibinfo {author} {\bibfnamefont {M.}~\bibnamefont {{Shahalam}}},
  \bibinfo {author} {\bibfnamefont {B.}~\bibnamefont {{Pal}}}, \bibinfo
  {author} {\bibfnamefont {S.}~\bibnamefont {{Pan}}}, \ and\ \bibinfo {author}
  {\bibfnamefont {A.}~\bibnamefont {{Wang}}},\ }\href@noop {} {\bibfield
  {journal} {\bibinfo  {journal} {ArXiv e-prints}\ } (\bibinfo {year}
  {2018})},\ \Eprint {http://arxiv.org/abs/1810.08586} {arXiv:1810.08586
  [gr-qc]} \BibitemShut {NoStop}%
\bibitem [{\citenamefont {{Krauss}}\ and\ \citenamefont
  {{Turner}}(1995)}]{Krauss95}%
  \BibitemOpen
  \bibfield  {author} {\bibinfo {author} {\bibfnamefont {L.~M.}\ \bibnamefont
  {{Krauss}}}\ and\ \bibinfo {author} {\bibfnamefont {M.~S.}\ \bibnamefont
  {{Turner}}},\ }\href {\doibase 10.1007/BF02108229} {\bibfield  {journal}
  {\bibinfo  {journal} {General Relativity and Gravitation}\ }\textbf {\bibinfo
  {volume} {27}},\ \bibinfo {pages} {1137} (\bibinfo {year} {1995})},\ \Eprint
  {http://arxiv.org/abs/astro-ph/9504003} {astro-ph/9504003} \BibitemShut
  {NoStop}%
\bibitem [{\citenamefont {{Weinberg}}(1989)}]{Weinberg89}%
  \BibitemOpen
  \bibfield  {author} {\bibinfo {author} {\bibfnamefont {S.}~\bibnamefont
  {{Weinberg}}},\ }\href {\doibase 10.1103/RevModPhys.61.1} {\bibfield
  {journal} {\bibinfo  {journal} {Reviews of Modern Physics}\ }\textbf
  {\bibinfo {volume} {61}},\ \bibinfo {pages} {1} (\bibinfo {year}
  {1989})}\BibitemShut {NoStop}%
\bibitem [{\citenamefont {{Peebles}}\ and\ \citenamefont
  {{Ratra}}(2003)}]{Peebles03}%
  \BibitemOpen
  \bibfield  {author} {\bibinfo {author} {\bibfnamefont {P.~J.}\ \bibnamefont
  {{Peebles}}}\ and\ \bibinfo {author} {\bibfnamefont {B.}~\bibnamefont
  {{Ratra}}},\ }\href {\doibase 10.1103/RevModPhys.75.559} {\bibfield
  {journal} {\bibinfo  {journal} {Reviews of Modern Physics}\ }\textbf
  {\bibinfo {volume} {75}},\ \bibinfo {pages} {559} (\bibinfo {year} {2003})},\
  \Eprint {http://arxiv.org/abs/astro-ph/0207347} {astro-ph/0207347}
  \BibitemShut {NoStop}%
\bibitem [{\citenamefont {{Padmanabhan}}(2003)}]{Padmanabhan03}%
  \BibitemOpen
  \bibfield  {author} {\bibinfo {author} {\bibfnamefont {T.}~\bibnamefont
  {{Padmanabhan}}},\ }\href {\doibase 10.1016/S0370-1573(03)00120-0} {\bibfield
   {journal} {\bibinfo  {journal} {Physics Reports}\ }\textbf {\bibinfo
  {volume} {380}},\ \bibinfo {pages} {235} (\bibinfo {year} {2003})},\ \Eprint
  {http://arxiv.org/abs/hep-th/0212290} {hep-th/0212290} \BibitemShut {NoStop}%
\bibitem [{\citenamefont {{Jassal}}\ \emph
  {et~al.}(2005{\natexlab{a}})\citenamefont {{Jassal}}, \citenamefont
  {{Bagla}},\ and\ \citenamefont {{Padmanabhan}}}]{Jassal05}%
  \BibitemOpen
  \bibfield  {author} {\bibinfo {author} {\bibfnamefont {H.~K.}\ \bibnamefont
  {{Jassal}}}, \bibinfo {author} {\bibfnamefont {J.~S.}\ \bibnamefont
  {{Bagla}}}, \ and\ \bibinfo {author} {\bibfnamefont {T.}~\bibnamefont
  {{Padmanabhan}}},\ }\href {\doibase 10.1103/PhysRevD.72.103503} {\bibfield
  {journal} {\bibinfo  {journal} {Phys. Rev. D}\ }\textbf {\bibinfo {volume}
  {72}},\ \bibinfo {eid} {103503} (\bibinfo {year} {2005}{\natexlab{a}})},\
  \Eprint {http://arxiv.org/abs/astro-ph/0506748} {astro-ph/0506748}
  \BibitemShut {NoStop}%
\bibitem [{\citenamefont {{Jassal}}\ \emph
  {et~al.}(2005{\natexlab{b}})\citenamefont {{Jassal}}, \citenamefont
  {{Bagla}},\ and\ \citenamefont {{Padmanabhan}}}]{Jassal05a}%
  \BibitemOpen
  \bibfield  {author} {\bibinfo {author} {\bibfnamefont {H.~K.}\ \bibnamefont
  {{Jassal}}}, \bibinfo {author} {\bibfnamefont {J.~S.}\ \bibnamefont
  {{Bagla}}}, \ and\ \bibinfo {author} {\bibfnamefont {T.}~\bibnamefont
  {{Padmanabhan}}},\ }\href {\doibase 10.1111/j.1745-3933.2005.08577.x}
  {\bibfield  {journal} {\bibinfo  {journal} {Mon. Not. R. Astron. Soc}\
  }\textbf {\bibinfo {volume} {356}},\ \bibinfo {pages} {L11} (\bibinfo {year}
  {2005}{\natexlab{b}})},\ \Eprint {http://arxiv.org/abs/astro-ph/0404378}
  {astro-ph/0404378} \BibitemShut {NoStop}%
\bibitem [{\citenamefont {{Samushia}}\ and\ \citenamefont
  {{Ratra}}(2006)}]{Samushia06}%
  \BibitemOpen
  \bibfield  {author} {\bibinfo {author} {\bibfnamefont {L.}~\bibnamefont
  {{Samushia}}}\ and\ \bibinfo {author} {\bibfnamefont {B.}~\bibnamefont
  {{Ratra}}},\ }\href {\doibase 10.1086/508662} {\bibfield  {journal} {\bibinfo
   {journal} {Astrophys. J. Letter}\ }\textbf {\bibinfo {volume} {650}},\
  \bibinfo {pages} {L5} (\bibinfo {year} {2006})},\ \Eprint
  {http://arxiv.org/abs/astro-ph/0607301} {astro-ph/0607301} \BibitemShut
  {NoStop}%
\bibitem [{\citenamefont {{Xia}}\ \emph {et~al.}(2008)\citenamefont {{Xia}},
  \citenamefont {{Li}}, \citenamefont {{Zhao}},\ and\ \citenamefont
  {{Zhang}}}]{Xia08}%
  \BibitemOpen
  \bibfield  {author} {\bibinfo {author} {\bibfnamefont {J.-Q.}\ \bibnamefont
  {{Xia}}}, \bibinfo {author} {\bibfnamefont {H.}~\bibnamefont {{Li}}},
  \bibinfo {author} {\bibfnamefont {G.-B.}\ \bibnamefont {{Zhao}}}, \ and\
  \bibinfo {author} {\bibfnamefont {X.}~\bibnamefont {{Zhang}}},\ }\href
  {\doibase 10.1103/PhysRevD.78.083524} {\bibfield  {journal} {\bibinfo
  {journal} {Phys. Rev. D}\ }\textbf {\bibinfo {volume} {78}},\ \bibinfo {eid}
  {083524} (\bibinfo {year} {2008})},\ \Eprint {http://arxiv.org/abs/0807.3878}
  {arXiv:0807.3878} \BibitemShut {NoStop}%
\bibitem [{\citenamefont {G.-B.~Zhao}(2007)}]{Zhao07}%
  \BibitemOpen
  \bibfield  {author} {\bibinfo {author} {\bibfnamefont {X.-M. Z. B.~F.}\
  \bibnamefont {G.-B.~Zhao}, \bibfnamefont {J.-Q.~Xia}},\ }\href@noop {}
  {\bibfield  {journal} {\bibinfo  {journal} {Int. J. Mod. Phys. D.}\ }\textbf
  {\bibinfo {volume} {16}},\ \bibinfo {pages} {1229} (\bibinfo {year}
  {2007})}\BibitemShut {NoStop}%
\bibitem [{\citenamefont {{Xia}}\ \emph {et~al.}(2006)\citenamefont {{Xia}},
  \citenamefont {{Zhao}}, \citenamefont {{Feng}}, \citenamefont {{Li}},\ and\
  \citenamefont {{Zhang}}}]{Xia06}%
  \BibitemOpen
  \bibfield  {author} {\bibinfo {author} {\bibfnamefont {J.-Q.}\ \bibnamefont
  {{Xia}}}, \bibinfo {author} {\bibfnamefont {G.-B.}\ \bibnamefont {{Zhao}}},
  \bibinfo {author} {\bibfnamefont {B.}~\bibnamefont {{Feng}}}, \bibinfo
  {author} {\bibfnamefont {H.}~\bibnamefont {{Li}}}, \ and\ \bibinfo {author}
  {\bibfnamefont {X.}~\bibnamefont {{Zhang}}},\ }\href {\doibase
  10.1103/PhysRevD.73.063521} {\bibfield  {journal} {\bibinfo  {journal} {Phys.
  Rev. D}\ }\textbf {\bibinfo {volume} {73}},\ \bibinfo {eid} {063521}
  (\bibinfo {year} {2006})},\ \Eprint {http://arxiv.org/abs/astro-ph/0511625}
  {astro-ph/0511625} \BibitemShut {NoStop}%
\bibitem [{\citenamefont {{Kiselev}}(2003{\natexlab{a}})}]{Kiselev03a}%
  \BibitemOpen
  \bibfield  {author} {\bibinfo {author} {\bibfnamefont {V.~V.}\ \bibnamefont
  {{Kiselev}}},\ }\href@noop {} {\bibfield  {journal} {\bibinfo  {journal}
  {ArXiv General Relativity and Quantum Cosmology e-prints}\ } (\bibinfo {year}
  {2003}{\natexlab{a}})},\ \Eprint {http://arxiv.org/abs/gr-qc/0303031}
  {gr-qc/0303031} \BibitemShut {NoStop}%
\bibitem [{\citenamefont {{Toshmatov}}\ \emph {et~al.}(2017)\citenamefont
  {{Toshmatov}}, \citenamefont {{Stuchl{\'{\i}}k}},\ and\ \citenamefont
  {{Ahmedov}}}]{Toshmatov17}%
  \BibitemOpen
  \bibfield  {author} {\bibinfo {author} {\bibfnamefont {B.}~\bibnamefont
  {{Toshmatov}}}, \bibinfo {author} {\bibfnamefont {Z.}~\bibnamefont
  {{Stuchl{\'{\i}}k}}}, \ and\ \bibinfo {author} {\bibfnamefont
  {B.}~\bibnamefont {{Ahmedov}}},\ }\href {\doibase 10.1140/epjp/i2017-11373-4}
  {\bibfield  {journal} {\bibinfo  {journal} {European Physical Journal Plus}\
  }\textbf {\bibinfo {volume} {132}},\ \bibinfo {eid} {98} (\bibinfo {year}
  {2017})}\BibitemShut {NoStop}%
\bibitem [{\citenamefont {{Kiselev}}(2003{\natexlab{b}})}]{Kiselev03}%
  \BibitemOpen
  \bibfield  {author} {\bibinfo {author} {\bibfnamefont {V.~V.}\ \bibnamefont
  {{Kiselev}}},\ }\href@noop {} {\bibfield  {journal} {\bibinfo  {journal}
  {Classical and Quantum Gravity}\ }\textbf {\bibinfo {volume} {20}},\ \bibinfo
  {pages} {1187} (\bibinfo {year} {2003}{\natexlab{b}})},\ \Eprint
  {http://arxiv.org/abs/gr-qc/0210040} {gr-qc/0210040} \BibitemShut {NoStop}%
\bibitem [{\citenamefont {{Fernando}}\ \emph {et~al.}(2015)\citenamefont
  {{Fernando}}, \citenamefont {{Meadows}},\ and\ \citenamefont
  {{Reis}}}]{Fernando15}%
  \BibitemOpen
  \bibfield  {author} {\bibinfo {author} {\bibfnamefont {S.}~\bibnamefont
  {{Fernando}}}, \bibinfo {author} {\bibfnamefont {S.}~\bibnamefont
  {{Meadows}}}, \ and\ \bibinfo {author} {\bibfnamefont {K.}~\bibnamefont
  {{Reis}}},\ }\href {\doibase 10.1007/s10773-015-2601-7} {\bibfield  {journal}
  {\bibinfo  {journal} {International Journal of Theoretical Physics}\ }\textbf
  {\bibinfo {volume} {54}},\ \bibinfo {pages} {3634} (\bibinfo {year}
  {2015})},\ \Eprint {http://arxiv.org/abs/1411.3192} {arXiv:1411.3192 [gr-qc]}
  \BibitemShut {NoStop}%
\bibitem [{\citenamefont {{Oteev}}\ \emph {et~al.}(2016)\citenamefont
  {{Oteev}}, \citenamefont {{Abdujabbarov}}, \citenamefont
  {{Stuchl{\'{\i}}k}},\ and\ \citenamefont {{Ahmedov}}}]{Oteev16}%
  \BibitemOpen
  \bibfield  {author} {\bibinfo {author} {\bibfnamefont {T.}~\bibnamefont
  {{Oteev}}}, \bibinfo {author} {\bibfnamefont {A.}~\bibnamefont
  {{Abdujabbarov}}}, \bibinfo {author} {\bibfnamefont {Z.}~\bibnamefont
  {{Stuchl{\'{\i}}k}}}, \ and\ \bibinfo {author} {\bibfnamefont
  {B.}~\bibnamefont {{Ahmedov}}},\ }\href {\doibase 10.1007/s10509-016-2850-9}
  {\bibfield  {journal} {\bibinfo  {journal} {Astrophys. Space Sci.}\ }\textbf
  {\bibinfo {volume} {361}},\ \bibinfo {eid} {269} (\bibinfo {year}
  {2016})}\BibitemShut {NoStop}%
\bibitem [{\citenamefont {{Abdujabbarov}}\ \emph {et~al.}(2017)\citenamefont
  {{Abdujabbarov}}, \citenamefont {{Toshmatov}}, \citenamefont
  {{Stuchl{\'{\i}}k}},\ and\ \citenamefont {{Ahmedov}}}]{Abdujabbarov17b}%
  \BibitemOpen
  \bibfield  {author} {\bibinfo {author} {\bibfnamefont {A.}~\bibnamefont
  {{Abdujabbarov}}}, \bibinfo {author} {\bibfnamefont {B.}~\bibnamefont
  {{Toshmatov}}}, \bibinfo {author} {\bibfnamefont {Z.}~\bibnamefont
  {{Stuchl{\'{\i}}k}}}, \ and\ \bibinfo {author} {\bibfnamefont
  {B.}~\bibnamefont {{Ahmedov}}},\ }\href {\doibase 10.1142/S0218271817500511}
  {\bibfield  {journal} {\bibinfo  {journal} {International Journal of Modern
  Physics D}\ }\textbf {\bibinfo {volume} {26}},\ \bibinfo {eid} {1750051-239}
  (\bibinfo {year} {2017})}\BibitemShut {NoStop}%
\bibitem [{\citenamefont {{Newman}}\ and\ \citenamefont
  {{Janis}}(1965)}]{Newman65b}%
  \BibitemOpen
  \bibfield  {author} {\bibinfo {author} {\bibfnamefont {E.~T.}\ \bibnamefont
  {{Newman}}}\ and\ \bibinfo {author} {\bibfnamefont {A.~I.}\ \bibnamefont
  {{Janis}}},\ }\href {\doibase 10.1063/1.1704350} {\bibfield  {journal}
  {\bibinfo  {journal} {Journal of Mathematical Physics}\ }\textbf {\bibinfo
  {volume} {6}},\ \bibinfo {pages} {915} (\bibinfo {year} {1965})}\BibitemShut
  {NoStop}%
\bibitem [{\citenamefont {Broccoli}\ and\ \citenamefont
  {Vigan{\`o}}(2018)}]{Broccoli18}%
  \BibitemOpen
  \bibfield  {author} {\bibinfo {author} {\bibfnamefont {M.}~\bibnamefont
  {Broccoli}}\ and\ \bibinfo {author} {\bibfnamefont {A.}~\bibnamefont
  {Vigan{\`o}}},\ }\href {\doibase 10.1103/PhysRevD.98.084007} {\bibfield
  {journal} {\bibinfo  {journal} {Phys. Rev. D}\ }\textbf {\bibinfo {volume}
  {98}},\ \bibinfo {pages} {084007} (\bibinfo {year} {2018})},\ \Eprint
  {http://arxiv.org/abs/1807.08313} {arXiv:1807.08313 [gr-qc]} \BibitemShut
  {NoStop}%
\bibitem [{\citenamefont {{Erbin}}(2017)}]{Erbin17}%
  \BibitemOpen
  \bibfield  {author} {\bibinfo {author} {\bibfnamefont {H.}~\bibnamefont
  {{Erbin}}},\ }\href {\doibase 10.3390/universe3010019} {\bibfield  {journal}
  {\bibinfo  {journal} {Universe}\ }\textbf {\bibinfo {volume} {3}},\ \bibinfo
  {pages} {19} (\bibinfo {year} {2017})},\ \Eprint
  {http://arxiv.org/abs/1701.00037} {arXiv:1701.00037 [gr-qc]} \BibitemShut
  {NoStop}%
\bibitem [{\citenamefont {{Gonzalez de Urreta}}\ and\ \citenamefont
  {{Socolovsky}}(2015)}]{Gonzalez15}%
  \BibitemOpen
  \bibfield  {author} {\bibinfo {author} {\bibfnamefont {E.~J.}\ \bibnamefont
  {{Gonzalez de Urreta}}}\ and\ \bibinfo {author} {\bibfnamefont
  {M.}~\bibnamefont {{Socolovsky}}},\ }\href@noop {} {\bibfield  {journal}
  {\bibinfo  {journal} {ArXiv e-prints}\ } (\bibinfo {year} {2015})},\ \Eprint
  {http://arxiv.org/abs/1504.01728} {arXiv:1504.01728 [gr-qc]} \BibitemShut
  {NoStop}%
\bibitem [{\citenamefont {{Erbin}}\ and\ \citenamefont
  {{Heurtier}}(2015)}]{Erbin15}%
  \BibitemOpen
  \bibfield  {author} {\bibinfo {author} {\bibfnamefont {H.}~\bibnamefont
  {{Erbin}}}\ and\ \bibinfo {author} {\bibfnamefont {L.}~\bibnamefont
  {{Heurtier}}},\ }\href {\doibase 10.1088/0264-9381/32/16/165004} {\bibfield
  {journal} {\bibinfo  {journal} {Classical and Quantum Gravity}\ }\textbf
  {\bibinfo {volume} {32}},\ \bibinfo {eid} {165004} (\bibinfo {year}
  {2015})},\ \Eprint {http://arxiv.org/abs/1411.2030} {arXiv:1411.2030 [gr-qc]}
  \BibitemShut {NoStop}%
\bibitem [{\citenamefont {{Keane}}(2014)}]{Keane14}%
  \BibitemOpen
  \bibfield  {author} {\bibinfo {author} {\bibfnamefont {A.~J.}\ \bibnamefont
  {{Keane}}},\ }\href {\doibase 10.1088/0264-9381/31/15/155003} {\bibfield
  {journal} {\bibinfo  {journal} {Classical and Quantum Gravity}\ }\textbf
  {\bibinfo {volume} {31}},\ \bibinfo {eid} {155003} (\bibinfo {year}
  {2014})},\ \Eprint {http://arxiv.org/abs/1407.4478} {arXiv:1407.4478 [gr-qc]}
  \BibitemShut {NoStop}%
\bibitem [{\citenamefont {{Bambi}}\ and\ \citenamefont
  {{Modesto}}(2013)}]{Bambi13}%
  \BibitemOpen
  \bibfield  {author} {\bibinfo {author} {\bibfnamefont {C.}~\bibnamefont
  {{Bambi}}}\ and\ \bibinfo {author} {\bibfnamefont {L.}~\bibnamefont
  {{Modesto}}},\ }\href {\doibase 10.1016/j.physletb.2013.03.025} {\bibfield
  {journal} {\bibinfo  {journal} {Physics Letters B}\ }\textbf {\bibinfo
  {volume} {721}},\ \bibinfo {pages} {329} (\bibinfo {year} {2013})},\ \Eprint
  {http://arxiv.org/abs/1302.6075} {arXiv:1302.6075 [gr-qc]} \BibitemShut
  {NoStop}%
\bibitem [{\citenamefont {Lombardo}(2004)}]{Lombardo04}%
  \BibitemOpen
  \bibfield  {author} {\bibinfo {author} {\bibfnamefont {D.~J.~C.}\
  \bibnamefont {Lombardo}},\ }\href
  {http://stacks.iop.org/0264-9381/21/i=6/a=009} {\bibfield  {journal}
  {\bibinfo  {journal} {Classical and Quantum Gravity}\ }\textbf {\bibinfo
  {volume} {21}},\ \bibinfo {pages} {1407} (\bibinfo {year}
  {2004})}\BibitemShut {NoStop}%
\bibitem [{\citenamefont {{Azreg-A{\"\i}nou}}(2014{\natexlab{a}})}]{Azreg14}%
  \BibitemOpen
  \bibfield  {author} {\bibinfo {author} {\bibfnamefont {M.}~\bibnamefont
  {{Azreg-A{\"\i}nou}}},\ }\href {\doibase 10.1103/PhysRevD.90.064041}
  {\bibfield  {journal} {\bibinfo  {journal} {Phys. Rev. D}\ }\textbf {\bibinfo
  {volume} {90}},\ \bibinfo {eid} {064041} (\bibinfo {year}
  {2014}{\natexlab{a}})},\ \Eprint {http://arxiv.org/abs/1405.2569}
  {arXiv:1405.2569 [gr-qc]} \BibitemShut {NoStop}%
\bibitem [{\citenamefont
  {{Azreg-A{\"\i}nou}}(2014{\natexlab{b}})}]{Azreg-Ainou14}%
  \BibitemOpen
  \bibfield  {author} {\bibinfo {author} {\bibfnamefont {M.}~\bibnamefont
  {{Azreg-A{\"\i}nou}}},\ }\href {\doibase 10.1140/epjc/s10052-014-2865-8}
  {\bibfield  {journal} {\bibinfo  {journal} {European Physical Journal C}\
  }\textbf {\bibinfo {volume} {74}},\ \bibinfo {eid} {2865} (\bibinfo {year}
  {2014}{\natexlab{b}})},\ \Eprint {http://arxiv.org/abs/1401.4292}
  {arXiv:1401.4292 [gr-qc]} \BibitemShut {NoStop}%
\bibitem [{\citenamefont {{Azreg-A{\"\i}nou}}(2011)}]{Azreg-Ainou11}%
  \BibitemOpen
  \bibfield  {author} {\bibinfo {author} {\bibfnamefont {M.}~\bibnamefont
  {{Azreg-A{\"\i}nou}}},\ }\href {\doibase 10.1088/0264-9381/28/14/148001}
  {\bibfield  {journal} {\bibinfo  {journal} {Classical and Quantum Gravity}\
  }\textbf {\bibinfo {volume} {28}},\ \bibinfo {eid} {148001} (\bibinfo {year}
  {2011})},\ \Eprint {http://arxiv.org/abs/1106.0970} {arXiv:1106.0970 [gr-qc]}
  \BibitemShut {NoStop}%
\bibitem [{\citenamefont {{Toshmatov}}\ \emph {et~al.}(2014)\citenamefont
  {{Toshmatov}}, \citenamefont {{Ahmedov}}, \citenamefont {{Abdujabbarov}},\
  and\ \citenamefont {{Stuchl{\'{\i}}k}}}]{Toshmatov14}%
  \BibitemOpen
  \bibfield  {author} {\bibinfo {author} {\bibfnamefont {B.}~\bibnamefont
  {{Toshmatov}}}, \bibinfo {author} {\bibfnamefont {B.}~\bibnamefont
  {{Ahmedov}}}, \bibinfo {author} {\bibfnamefont {A.}~\bibnamefont
  {{Abdujabbarov}}}, \ and\ \bibinfo {author} {\bibfnamefont {Z.}~\bibnamefont
  {{Stuchl{\'{\i}}k}}},\ }\href {\doibase 10.1103/PhysRevD.89.104017}
  {\bibfield  {journal} {\bibinfo  {journal} {Phys. Rev. D.}\ }\textbf
  {\bibinfo {volume} {89}},\ \bibinfo {eid} {104017} (\bibinfo {year}
  {2014})},\ \Eprint {http://arxiv.org/abs/1404.6443} {arXiv:1404.6443 [gr-qc]}
  \BibitemShut {NoStop}%
\bibitem [{\citenamefont {{Nam}}(2018)}]{Nam18}%
  \BibitemOpen
  \bibfield  {author} {\bibinfo {author} {\bibfnamefont {C.~H.}\ \bibnamefont
  {{Nam}}},\ }\href {\doibase 10.1007/s10714-018-2380-6} {\bibfield  {journal}
  {\bibinfo  {journal} {General Relativity and Gravitation}\ }\textbf {\bibinfo
  {volume} {50}},\ \bibinfo {eid} {57} (\bibinfo {year} {2018})}\BibitemShut
  {NoStop}%
\bibitem [{\citenamefont {{Hayward}}(2006)}]{Hayward06}%
  \BibitemOpen
  \bibfield  {author} {\bibinfo {author} {\bibfnamefont {S.~A.}\ \bibnamefont
  {{Hayward}}},\ }\href {\doibase 10.1103/PhysRevLett.96.031103} {\bibfield
  {journal} {\bibinfo  {journal} {Physical Review Letters}\ }\textbf {\bibinfo
  {volume} {96}},\ \bibinfo {eid} {031103} (\bibinfo {year} {2006})},\ \Eprint
  {http://arxiv.org/abs/gr-qc/0506126} {gr-qc/0506126} \BibitemShut {NoStop}%
\bibitem [{\citenamefont {Misner}\ \emph {et~al.}(1973)\citenamefont {Misner},
  \citenamefont {Thorne},\ and\ \citenamefont {Wheeler}}]{Misner73}%
  \BibitemOpen
  \bibfield  {author} {\bibinfo {author} {\bibfnamefont {C.~W.}\ \bibnamefont
  {Misner}}, \bibinfo {author} {\bibfnamefont {K.~S.}\ \bibnamefont {Thorne}},
  \ and\ \bibinfo {author} {\bibfnamefont {J.~A.}\ \bibnamefont {Wheeler}},\
  }\href@noop {} {\emph {\bibinfo {title} {Gravitation}}},\ Misner73\ (\bibinfo
   {publisher} {W. H. Freeman},\ \bibinfo {address} {San Francisco},\ \bibinfo
  {year} {1973})\BibitemShut {NoStop}%
\bibitem [{\citenamefont {Chandrasekhar}(1998)}]{Chandrasekhar98}%
  \BibitemOpen
  \bibfield  {author} {\bibinfo {author} {\bibfnamefont {S.}~\bibnamefont
  {Chandrasekhar}},\ }\href@noop {} {\emph {\bibinfo {title} {{The mathematical
  theory of black holes}}}}\ (\bibinfo  {publisher} {Oxford University
  Press.},\ \bibinfo {address} {New York},\ \bibinfo {year} {1998})\BibitemShut
  {NoStop}%
\bibitem [{\citenamefont {Wald}(1984)}]{Wald84}%
  \BibitemOpen
  \bibfield  {author} {\bibinfo {author} {\bibfnamefont {R.~M.}\ \bibnamefont
  {Wald}},\ }\href@noop {} {\emph {\bibinfo {title} {General relativity}}},\
  Wald84\ (\bibinfo  {publisher} {The University of Chicago Press},\ \bibinfo
  {address} {Chicago},\ \bibinfo {year} {1984})\BibitemShut {NoStop}%
\bibitem [{\citenamefont {{Fan}}\ and\ \citenamefont {{Wang}}(2016)}]{Fan16}%
  \BibitemOpen
  \bibfield  {author} {\bibinfo {author} {\bibfnamefont {Z.-Y.}\ \bibnamefont
  {{Fan}}}\ and\ \bibinfo {author} {\bibfnamefont {X.}~\bibnamefont {{Wang}}},\
  }\href {\doibase 10.1103/PhysRevD.94.124027} {\bibfield  {journal} {\bibinfo
  {journal} {Phys. Rev. D}\ }\textbf {\bibinfo {volume} {94}},\ \bibinfo {eid}
  {124027} (\bibinfo {year} {2016})},\ \Eprint
  {http://arxiv.org/abs/1610.02636} {arXiv:1610.02636 [gr-qc]} \BibitemShut
  {NoStop}%
\bibitem [{\citenamefont {{Dymnikova}}\ and\ \citenamefont
  {{Galaktionov}}(2015)}]{Dymnikova15}%
  \BibitemOpen
  \bibfield  {author} {\bibinfo {author} {\bibfnamefont {I.}~\bibnamefont
  {{Dymnikova}}}\ and\ \bibinfo {author} {\bibfnamefont {E.}~\bibnamefont
  {{Galaktionov}}},\ }\href {\doibase 10.1088/0264-9381/32/16/165015}
  {\bibfield  {journal} {\bibinfo  {journal} {Classical and Quantum Gravity}\
  }\textbf {\bibinfo {volume} {32}},\ \bibinfo {eid} {165015} (\bibinfo {year}
  {2015})},\ \Eprint {http://arxiv.org/abs/1510.01353} {arXiv:1510.01353
  [gr-qc]} \BibitemShut {NoStop}%
\bibitem [{\citenamefont {{Erbin}}(2015)}]{Erbin15a}%
  \BibitemOpen
  \bibfield  {author} {\bibinfo {author} {\bibfnamefont {H.}~\bibnamefont
  {{Erbin}}},\ }\href {\doibase 10.1007/s10714-015-1860-1} {\bibfield
  {journal} {\bibinfo  {journal} {General Relativity and Gravitation}\ }\textbf
  {\bibinfo {volume} {47}},\ \bibinfo {eid} {19} (\bibinfo {year} {2015})},\
  \Eprint {http://arxiv.org/abs/1410.2602} {arXiv:1410.2602 [gr-qc]}
  \BibitemShut {NoStop}%
\bibitem [{\citenamefont {Bambi}(2017)}]{Bambi17e}%
  \BibitemOpen
  \bibfield  {author} {\bibinfo {author} {\bibfnamefont {C.}~\bibnamefont
  {Bambi}},\ }\href@noop {} {\emph {\bibinfo {title} {Black Holes: A Laboratory
  for Testing Strong Gravity}}}\ (\bibinfo  {publisher} {Springer, Singapore},\
  \bibinfo {year} {2017})\BibitemShut {NoStop}%
\bibitem [{\citenamefont {Carroll}(2003)}]{Carroll03}%
  \BibitemOpen
  \bibfield  {author} {\bibinfo {author} {\bibfnamefont {S.}~\bibnamefont
  {Carroll}},\ }\href
  {http://www.amazon.com/Spacetime-Geometry-Introduction-General-Relativity/dp/0805387323}
  {\emph {\bibinfo {title} {Spacetime and Geometry: An Introduction to General
  Relativity}}}\ (\bibinfo  {publisher} {Benjamin Cummings},\ \bibinfo {year}
  {2003})\BibitemShut {NoStop}%
\bibitem [{\citenamefont {Bardeen}\ and\ \citenamefont
  {Petterson}(1975)}]{Bardeen:1975zz}%
  \BibitemOpen
  \bibfield  {author} {\bibinfo {author} {\bibfnamefont {J.~M.}\ \bibnamefont
  {Bardeen}}\ and\ \bibinfo {author} {\bibfnamefont {J.~A.}\ \bibnamefont
  {Petterson}},\ }\href {\doibase 10.1086/181711} {\bibfield  {journal}
  {\bibinfo  {journal} {Astrophys. J.}\ }\textbf {\bibinfo {volume} {195}},\
  \bibinfo {pages} {L65} (\bibinfo {year} {1975})}\BibitemShut {NoStop}%
\end{thebibliography}%

\end{document}